\documentclass[11pt,a4paper,reqno,centertags]{amsbook} 
\usepackage[T1]{fontenc} 
\usepackage{amsxtra,amsfonts,amssymb,amsthm}
\usepackage{graphicx,epsfig} 
\usepackage[mathscr]{eucal} 
\usepackage{mathlagrange}
\usepackage{mycite} 

\usepackage{feynmf}
\usepackage{a4} 
\usepackage[vcentermath]{youngtab}

\numberwithin{equation}{chapter}

\DeclareSymbolFontAlphabet{\mathbm}{AMSb}

\newcommand{\emt}{\textsc{emt}}
\newcommand{\ope}{\textsc{ope}}
\newcommand{\qft}{\textsc{qft}}

\newcommand{\rhs}{\textsc{rhs}}
\newcommand{\lhs}{\textsc{lhs}}

\newcommand{\supp}{\operatorname{supp}}
\newcommand{\sgn}{\operatorname{sgn}} 
 
\newcommand{\singord}{\operatorname{sing\,ord}} 
\newcommand{\scaledeg}{\operatorname{scal\,deg}} 
 
\newcommand{\End}{\operatorname{End}} 
\newcommand{\Diag}{\operatorname{Diag}} 
\newcommand{\Dist}{\operatorname{Dist}}

\newcommand{\const}{\mathit{const}} 
\newcommand{\I}{\mathrm{int}}
\newcommand{\Ad}{\mathrm{Ad}}

\newcommand{\D}{\ensuremath{\partial}}
\newcommand{\dif}{\ensuremath{\mathrm{d}}}
 
\newcommand{\parz}[2]{\frac{\D#1}{\D#2}}

\newcommand{\vek}[1]{\ensuremath{\mathbf{#1}}} 
\newcommand{\xvek}{\vek{x}}
\newcommand{\yvek}{\vek{y}}

\newcommand{\pvek}{\vek{p}}

\newcommand{\menge}[1]{\mathbb{#1}} 
\newcommand{\eins}{\menge{I}} 
\newcommand{\SLC}{\ensuremath{\mathit{SL}(2,\menge{C})}}
\newcommand{\CC}{\menge{C}}
\newcommand{\RR}{\menge{R}}
\newcommand{\NN}{\menge{N}}
\newcommand{\MM}{\menge{M}}

\newcommand{\fblc}{\ensuremath{\overline{V}_{\pm}}} 
\newcommand{\flc}{\ensuremath{\overline{V}_{+}}} 
\newcommand{\blc}{\ensuremath{\overline{V}_{-}}}

\newcommand{\Lcal}{\ensuremath{\mathcal{L}}} 

\newcommand{\Rd}{\ensuremath{\menge{R}^{d}}} 
\newcommand{\Rdon}{\Rd\setminus\{0\}} 
\newcommand{\Rv}{\ensuremath{\menge{R}^{4}}}

\newcommand{\MMmon}{\ensuremath{\MM^{m}\setminus\{0\}}}

\newcommand{\Do}{\ensuremath{\mathcal{D}}}
\newcommand{\Fo}{\ensuremath{\mathcal{F}}}
\newcommand{\Ho}{\ensuremath{\mathcal{H}}}

\newcommand{\Rcal}{\ensuremath{\mathcal{R}}}
\newcommand{\Ccal}{\ensuremath{\mathcal{C}}}

\newcommand{\Acal}{\ensuremath{\mathcal{A}}} 
\newcommand{\Pcal}{\ensuremath{\mathcal{P}}} 
\newcommand{\Ncal}{\ensuremath{\mathcal{N}}} 

\newcommand{\Cunend}{\ensuremath{\mathscr{C}^{\infty}}} 
\newcommand{\Dd}{\ensuremath{\mathscr{D}}} 
\newcommand{\Sd}{\ensuremath{\mathscr{S}}} 

\newcommand{\Aa}{\ensuremath{\mathfrak{A}}} 
 
\newcommand{\Ba}{\ensuremath{\mathfrak{B}}} 
\newcommand{\Ga}{\ensuremath{\mathfrak{G}}} 

\newcommand{\Ocal}{\ensuremath{\mathscr{O}}}
 
\newcommand{\Lor}{\ensuremath{\Lcal_{+}^{\uparrow}}} 
\newcommand{\Poinc}{\ensuremath{\Pcal_{+}^{\uparrow}}} 

\newcommand{\Ll}{\ensuremath{\mathlagrange{L}}} 
\newcommand{\Kl}{\ensuremath{\mathlagrange{K}}}

\newcommand{\scp}[2]{\ensuremath{\left\langle #1, #2 \right\rangle}} 
 
\newcommand{\W}[2]{\ensuremath{W_{(#1;#2)}}} 
\newcommand{\tR}[2]{\ensuremath{t_{(#1;#2)}}}
\newcommand{\tiltR}[2]{\ensuremath{\overline{t}_{(#1;#2)}}} 
\newcommand{\tRli}[2]{\ensuremath{t_{(#1;#2)}^{\mathrm{linv}}}}

\newcommand{\wick}[1]{\ensuremath{:\! #1 \!:}}

\newcommand{\wpv}{\wick{\vp,\vp}}
\newcommand{\wpvg}{\wick{\vp,\vp}_{g\Ll}}

\newcommand{\Tbar}{\overline{T}}
\newcommand{\K}[1]{\left( #1 \right)}
\newcommand{\Kg}[1]{\left( #1 \right)_{g\Ll}}

\newcommand{\vp}{\varphi}
\newcommand{\xbar}{\ensuremath{\widetilde{x}}}

\newcommand{\Nnull}{\textbf{N0}} 
\newcommand{\Neins}{\textbf{N1}}
\newcommand{\Nzwei}{\textbf{N2}}
\newcommand{\Ndrei}{\textbf{N3}}
\newcommand{\Ndreistrich}{\textbf{N3'}}
\newcommand{\Nvier}{\textbf{N4}}
\newcommand{\Weins}{\textbf{WI 1}}
\newcommand{\Wzwei}{\textbf{WI 2}}

\newcommand{\Peins}{\textbf{P1}}

\newcommand{\Pdrei}{\textbf{P3}}
\newcommand{\Pvier}{\textbf{P4}}

\newcommand{\Deltaret}{\Delta_{\mathrm{ret}}}
\newcommand{\Dret}{D_{\mathrm{ret}}}
\newcommand{\Deltaav}{\Delta_{\mathrm{av}}}
\newcommand{\Dav}{D_{\mathrm{av}}}
\newcommand{\Deltaretav}{\Delta_{\mathrm{ret/av}}}

\newcommand{\Tmn}{\Theta^{\mu\nu}}
\newcommand{\Tmnnc}{\Tmn_{0\,\mathrm{can}}}

\newcommand{\Tmnncg}{\Tmn_{0\,\mathrm{can }g\Ll}}
\newcommand{\Tmncg}{\Tmn_{\mathrm{can}\,g\Ll}}
\newcommand{\Tmncl}{\Theta^{\mathrm{class}\,\mu\nu}}
\newcommand{\Tmnccl}{\Theta_{\mathrm{can}}^{\mathrm{class}\,\mu\nu}}
\newcommand{\Tmnnccl}{\Theta_{0\,\mathrm{can}}^{\mathrm{class}\,\mu\nu}}

\newcommand{\Tmnicl}{\Theta_{\mathrm{imp}}^{\mathrm{class}\,\mu\nu}}
\newcommand{\Tmnig}{\Tmn_{\mathrm{imp}\,g\Ll}}
\newcommand{\Tmnni}{\Tmn_{0\,\mathrm{imp}}}

\newcommand{\Tnn}{\Theta^{0\nu}}

\newcommand{\Tnnnc}{\Tnn_{0\,\mathrm{can}}}
\newcommand{\Tnnncg}{\Tnn_{0\,\mathrm{can}\,g\Ll}}
\newcommand{\Tnncg}{\Tnn_{\mathrm{can}\,g\Ll}}
\newcommand{\Tnnig}{\Tnn_{\mathrm{imp}\,g\Ll}}

\newcommand{\Tnm}{\Theta^{0\mu}}

\newcommand{\Tnmnc}{\Tnm_{0\,\mathrm{can}}}

\newcommand{\Tnc}{\Theta_{0\,\mathrm{can}}}

\newcommand{\Imn}{I^{\mu\nu}}
\newcommand{\Imncl}{I^{\mathrm{class}\,\mu\nu}}
\newcommand{\Imng}{\Imn_{g\Ll}}
\newcommand{\Imnn}{\Imn_0}
\newcommand{\Imnng}{\Imn_{0\,g\Ll}}

\newcommand{\Dmg}{D^\mu_{g\Ll}}

\newcommand{\vpg}{\vp_{g\Ll}}

\newcommand{\phicl}{\phi^{\mathrm{class}}}
\newcommand{\Llcl}{\Ll^{\mathrm{class}}}
\newcommand{\Llcln}{\Ll^{\mathrm{class}}_0}
\newcommand{\Llcli}{\Ll^{\mathrm{class}}_\mathrm{int}}

\newcommand{\ktilde}{\widetilde{k}}

\newcommand{\ptilde}{\widetilde{\pvek}}
\newcommand{\ptilt}{\widetilde{p}}

\newcommand{\Xdot}{\dot{X}} 
\newcommand{\Ydot}{\dot{Y}}  
\newcommand{\Adot}{\dot{A}} \newcommand{\Bdot}{\dot{B}}

\newcommand{\Wg}{W_{g\Ll}}
\newcommand{\Sg}{S_{g\Ll}}
\newcommand{\Klg}{\Kl_{g\Ll}}
\newcommand{\Llg}{\Ll_{g\Ll}}
\newcommand{\Aag}{\Aa_{g\Ll}}

\newcommand{\Png}{P^\nu_{g\Ll}}

\newcommand{\Nk}{N^{(e_k)}}

\newtheorem{thm}{Theorem}

\theoremstyle{definition}
\newtheorem{defi}{Definition}
\newtheorem{exa}{Example}
\newtheorem*{Woperation}{$W$-operation}
\theoremstyle{remark}
\newtheorem*{rem}{Remark}  

\begin{document}
\begin{fmffile}{dissgrafen}

\title{Energy Momentum Tensor and \\Operator
Product Expansion in \\Local Causal Perturbation Theory}

\author{
\begin{center}
        \vfill
    {\large
    Dissertation\\
    zur Erlangung des Doktorgrades\\
    des Fachbereichs Physik\\
    der Universit\"at Hamburg
    \vfill
    vorgelegt von\\
    Dirk Prange\\
    aus Hamburg
    \vfill
    Hamburg\\
    2000}
  \end{center}
}
\maketitle


\vfill
\begin{tabular}{ll}
  Gutachter der Dissertation: & Prof. Dr. K. Fredenhagen\\
                              & Priv.-Doz. Dr. M. D\"utsch\\
                              &\\
  Gutachter der Disputation:  & Prof. Dr. K. Fredenhagen\\
                              & Prof. Dr. G. Mack\\
                              &\\
  Datum der Disputation:      & 12. September 2000\\
  &\\
  Dekan des Fachbereichs Physik und & \\
  Vorsitzender des Promotionsausschusses: & Prof. Dr. F.-W. B\"u\ss er
\end{tabular}
\newpage

\begin{abstract}

We derive new examples for algebraic relations of interacting fields
in local perturbative quantum field theory.  The fundamental building
blocks in this approach are time ordered products of free (composed)
fields. We give explicit formulas for the construction of Poincar{\'e}
covariant ones, which were already known to exist through
cohomological arguments.
  
For a large class of theories the canonical energy momentum tensor is
shown to be conserved. Classical theories without dimensionful
couplings admit an improved tensor that is additionally traceless.  On
the example of $\vp^4$-theory we discuss the improved tensor in the
quantum theory. Its trace receives an anomalous contribution due to
its conservation.
  
Moreover we define an interacting bilocal normal product for scalar
theories. This leads to an operator product expansion of two time
ordered fields.
\\
\\
  \noindent\textsc{Zusammenfassung.}
  Im Rahmen der kausalen St\"orungstheorie leiten wir neue Beispiele
  f\"ur algebraische Relationen wechselwirkender Quantenfelder her.
  Wir geben explizite Formeln f\"ur die Konstruktion Poincar\'e
  kovarianter zeitgeordneter Produkte freier (zusammengesetzter)
  Felder an, welche in diesem Zugang die Grundbausteine bilden. Bisher
  war nur deren Existenz aufgrund kohomologischer Argumente bekannt.
  
  F\"ur eine gro\ss e Klasse von Theorien zeigen wir die Erhaltung des
  kanonischen Energie-Impuls-Tensors. F\"ur klassische Theorien, die
  keine dimensionsbehafteten Kopplungen enthalten, existiert ein
  verbesserter Tensor, der zus\"atzlich spurfrei ist. Am Beispiel der
  $\vp^4$-Theorie untersuchen wir diesen Tensor in der Quantentheorie.
  Als Folge der Erhaltung bekommt die Spur anomale Beitr\"age.
  
  Dar\"uberhinaus geben wir die Definition eines bilokalen
  wechselwirkenden Normalproduktes f\"ur skalare Theorien. Mithilfe
  des Normalproduktes finden wir die Operator-Produkt-Ent\-wick\-lung
  f\"ur das zeitgeordnete Produkt zweier Felder.
\end{abstract}


\setcounter{tocdepth}{1} 
\tableofcontents

%
%
\chapter{Introduction}
\label{ssec:intro}
\hfill
\begin{minipage}{6cm}
\footnotesize \itshape
``There is a theory which states that if ever anyone discovers exactly
what the Universe is for and why it is here, it will instantly
disappear and be replaced by something even more bizarre and
inexplicable \dots'' \footnotemark
\end{minipage}\\[\baselineskip]
\footnotetext{to be continued on page \pageref{pg:end}.}
\addtocounter{footnote}{1}

Today quantum field theory (\qft) provides the best unification of
classical relativistic field theory with the axioms of quantum
mechanics. During the last decades a high agreement between
theoretical predictions and observed data was achieved. The
theoretical results referring to experimental data from scattering
processes are usually derived via perturbation theory around the free
quantum field or around the classical field theory. In these regimes
either the coupling or Planck's constant can be regarded as a small
parameter and therefore perturbation theory seems to apply as a
suitable tool.

On the other hand there is no hope to derive any realistic statement
about the strong coupling regime of elementary particle physics from
the perturbative point of view. This subject is addressed in the
formulation of \qft\ on the lattice where spacetime becomes discrete.
In the last years especially lattice gauge theory benefits from the
increasing computing power that is available to describe the structure
of hadrons.

A quite severe drawback in the formulation of \qft\ is the fact that
up to now no realistic model of an interacting non perturbative and
continuous theory exists in four dimensions. Therefore the
perturbative approach has become the most popular one, (see e.g.\ 
\cite{bk:itzykson}). The quantization of free fields as an operator
relation following the principles of quantum mechanics and special
relativity constitutes the starting point. The interaction is
introduced as a perturbation and the coupling is treated as an
expansion parameter for the interacting Green's functions. These
functions show two different divergencies, ultraviolet and infrared
ones. The former ones reflect the distributional character of the free
field operators whereas the latter ones originate from the long range
forces carried by massless particles. These divergencies are removed
in the various processes of renormalization, of which we mention only
the most general one given by BPHZ(L). Unfortunately, all
investigations concerning the convergence of the perturbation series
yield a negative result.

In contrast to the perturbative formulation of quantum field theory
which relies on the existence of some specific models, an axiomatic
approach to quantum field theory (also called algebraic \qft) was
given by Haag and Kastler \cite{bk:haag}[and references there]. In
their approach the starting point is the existence of a net of
\emph{local} algebras of observables. The principle of locality is
taken into account by the requirement that space like separated observables
commute reflecting the fact that measurements on space like separated
experiments can not influence each other. Under the requirement that
the Poincar\'e group is unitarily implemented and isotony holds the
net is completely fixed.  A representation of this abstractly defined
algebra by operators on a Hilbert space is generated through the
GNS-construction provided one has defined (physically relevant)
states, namely positive normed linear functionals. 

Neither the existence of quantum fields nor of particles serves as an
input and their correlation still is a subject of research
\cite{pap:buchholz-haag1}. Especially the treatment of non
perturbative and model independent effects, like the relation between
spin \& statistics turn out to be customary applications of algebraic
\qft.

An adapted version of perturbative quantum field theory that fits
nicely into the algebraic framework was derived by Bogoliubov, Shirkov
\cite{bk:bogol} and Epstein-Glaser \cite{pap:ep-gl}. Their approach
focuses on the construction of a local $S$-matrix as a formal power
series of a compactly supported coupling by the axioms of causality
and Poincar{\'e} covariance mainly. This $S$-matrix gives rise to a
local observable in the algebraic sense \cite{proc:brun-fred}, namely
the relative $S$-matrix.  From this functional interacting (composed)
fields are derived which serve as a a basis for local observables
constituting the local algebras. By the principles of locality, the
$S$-matrix only has to be known in the region in which the interacting
algebras live and on which the coupling is kept fixed. No IR-divergencies
related to an infinite range of interaction appear. Moreover the
inductive construction takes care of the UV-renormalization in a very
elegant way.  The limit in which the coupling becomes a constant over
all spacetime like in the usual approach is referred to as the
adiabatic limit. Since all objects are defined by formal
power series the question of convergence is not addressed in that
framework.

Local perturbation theory was applied successfully to QED
\cite{bk:scharf}[and references herein]. Blanchard and Seneor
\cite{pap:blanchard} have shown that the adiabatic limit for 
Green's functions and for Wightman functions exists for QED and also
for massless $\varphi^4$-theory. Epstein and Glaser already proved the
existence of the adiabatic limit in massive theories \cite{pap:ep-gl}.

\sloppy
It was shown by D\"utsch \cite{pap:duet2} that the Epstein-Glaser
definition of Green's functions yields the Gell-Man-Low formula in
that limit. But for non Abelian gauge theories the situation is worse.
Therefore one tries to avoid the adiabatic limit in general. In
\cite{pap:brun-fred2} Brunetti and Fredenhagen have shown that a
variation of the interaction outside the localization region only acts
as a unitary transformation on the interacting fields. Moreover they
have given a purely algebraic construction of the adiabatic limit.

\fussy
The elementary building blocks for the $S$-matrices are the time
ordered products of (composed) free fields, i.e.\ Wick monomials. In
their inductive construction ambiguities generically appear
through the process of extending distributions. Restrictions on these
extensions are called \emph{normalization conditions}. The requirement
of Poincar{\'e} covariance forms the most important one. Stora and
Popineau \cite{prep:stora-pop} and D\"utsch et al. \cite{pap:scharf2}
,\cite{bk:scharf}[chapter 4.5] have given a cohomological
existence proof of such an extension. In \cite{prep:wir} we presented
an explicit extension in lowest order and gave a general
result in \cite{prep:prange2} in form of an inductive construction.

Further normalization conditions have been given in
\cite{pap:duet-fred1}. They imply the interacting field equations. In
\cite{phd:boas} Boas has given a generalization of these conditions such that
they can be applied to derivative couplings, too.

In the perturbative construction of \qft's symmetries play a major
role. In classical field theory any symmetry of the Lagrangian gives
rise to a conserved current via the Noether procedure. In the
corresponding \qft\ one tries to preserve as many of these symmetries
as possible. The breaking of a classical symmetry in the quantum
theory is called an anomaly. In \cite{pap:duet-fred1} BRST invariance
was shown to hold for QED and in \cite{phd:boas} the result was
extended to non Abelian theories.  The corresponding current
conservation follows from a Ward identity for time ordered products
involving one (free) current.  These identities can be regarded as
further normalization conditions.

This thesis focuses on two subjects. The first deals with translation
invariance. In classical field theory the energy momentum tensor
(\emt) is derived as the Noether current subject to a translation of
the fields. In our situation translation invariance is broken through
the coupling term and translation invariance only holds where the
coupling is constant. We show that the same conservation equation can
be maintained in local perturbative \qft\ for a quite general theory
without derivative couplings. This equation is a consequence of a Ward
identity which we prove with the methods developed in
\cite{pap:duet-fred1,phd:boas}. Our result coincides with a similar
investigation by Lowenstein \cite{pap:low3} who also has shown the
conservation of the canonical \emt\ for $\vp^4$-theory in the
framework of Zimmermann's normal product quantization
\cite{lect:zimmermann,pap:zimmermann1}.

For classical theories which possess no dimensionful parameters an
improved \emt\ exists which is also traceless \cite{pap:callan1}.  The
improved tensor is derived by addition of a conserved tensor, called
improvement tensor.  On the example of massless $\vp^4$-theory we show
that in the perturbative local formulation there still exists a
conserved improvement tensor by a corresponding Ward identity. But the
improved \emt\ inevitably produces the well known trace anomaly
\cite{pap:coleman} by virtue of the compatibility of both Ward
identities.  The anomaly is only defined up to a real parameter
related to some normalization choice. Our result is in accordance with
Zimmermann's analysis \cite{proc:zimmermann} who also has given a
derivation of the anomaly in terms of normal products.

The second focus is on the derivation of an operator product expansion
in local perturbation theory.  In \cite{lect:zimmermann,
  pap:zimmermann2} Zimmermann has generalized his concept of local
normal products to the case of bilocal normal products which allow for
restricting the coordinates to the same value (the same way which is
allowed in Wick or normal ordering of free fields). With the help of
these objects he found an operator product expansion of the time
ordered product of two scalar fields as suggested by Wilson
\cite{pap:wil1}.  We give a similar approach here. The definition of a
time ordered product with a bilocal Wick product in one entry allows
for an operator product expansion of the time ordered product of the
corresponding interacting fields. We perform the construction for
$\varphi^4$-theory. With these interacting normal products we suggest
a definition of a ground state in analogy to the free field case. We
show that the corresponding two point function is positive in the
sense of formal power series as defined in \cite{pap:duet-fred1}.

The thesis is organized as follows. In chapter~\ref{chap:free} we
review the process of quantization of free scalar fields which are the
fields that our thesis mainly deals with. 

Chapter~\ref{chap:timeordered} introduces the notion of time ordered
products. We use the formulation of Boas \cite{phd:boas} where
auxiliary variables are used instead of Wick products. The inductive
construction in the form \cite{proc:brun-fred,pap:brun-fred2} is
explained. Then the normalization conditions
\cite{pap:duet-fred1,phd:boas} follow. We cite the solution of the
Poincar\'e covariant extension from \cite{prep:wir,prep:prange2}.

\sloppy
Interacting fields as formal power series according to Bogoliubov and
Epstein-Glaser \cite{bk:bogol,pap:ep-gl} are introduced in
chapter~\ref{chap:intfields}.

\fussy
The \emt\ is discussed in chapter~\ref{chap:emt}. The canonical tensor
is shown to be conserved locally for a quite general theory without
derivative couplings. Its charge is shown to define the interacting
momentum operator. The improved tensor is studied on the example of
massless $\vp^4$-theory. The trace anomaly is derived.

In the last chapter~\ref{chap:ope} we study the \ope\ for the time
ordered product of two interacting scalar fields. The expansion
follows from a suitable definition of a bilocal time ordered product.

At the end we give a conclusion and some outlook.


\chapter{Canonical quantization of free scalar fields}
\label{chap:free}

We review the process of quantization for the (bosonic) scalar free
field. Our presentation of the subject follows the lecture notes
\cite{lect:fred2}. It emphasizes the possibility to define the algebra
of the quantum field first in a purely algebraic manner. A
representation of this algebra by (unbounded) operators on a
Hilbert space is derived via the GNS-reconstruction once a state on
the algebra is given. With the usual two point function defining a
quasi free state, the Hilbert space becomes the usual symmetric Fock space.

Minkowski space is denoted by $\MM$ and the
scalar product is $x\cdot y = xy = \eta_{\mu\nu}x^\mu y^\nu$.

\section{Free quantum field algebra}
We consider the free scalar field $\vp$ of mass $m$ which solves the
Klein-Gordon equation:
\begin{equation}
  \label{eq:kg}
D\vp \doteq  (\square+m^2)\vp = 0.
\end{equation}

As a hyperbolic differential equation the Klein-Gordon equation
possesses a fundamental solution $\Delta\in\Dd'(\MM)$ with causal 
support, $\supp \Delta \in \flc \cup \blc $, which solves the equation
(in the weak sense):
\begin{equation}
  \label{eq:kgD}
 \Delta(Df)=0, \quad \forall f\in \Dd(\MM)
\end{equation}
and has the following initial values:
\begin{align}
  \label{eq:ivD}
 \Delta(0,\xvek)&=0,
& \D_0\Delta(0,\xvek)&=-\delta(\xvek).
\end{align}
Then any solution of the differential equation is completely
determined by the initial values at time $t$:
\begin{align}
  \label{eq:initial}
  f(t,\xvek)&=\phi_0(\xvek) & \D_0f(t,\xvek)&=\phi_1(\xvek),
\end{align}
according to
\begin{equation}
  \label{eq:solcauchy}
  f(x^0,\cdot)=\Delta(x^0-t)\ast\phi_1+(\D_0\Delta)(x^0-t)\ast\phi_0.
\end{equation}
The distribution $\Delta$ has a unique decomposition into advanced and 
retarded Green's functions:
\begin{align}
\Delta&=\Deltaret-\Deltaav, \label{eq:funda1} \\
D\Delta_{\mathrm{ret,av}}&=\delta,\quad  
\supp \Delta_{\substack{\mathrm{ret}\\ \mathrm{av}}} 
\in \fblc, \label{eq:funda2}
\end{align}

The \emph{algebra of observables} $\Aa$ is defined as an abstract algebra
generated by $\varphi(f), f\in\Dd(\MM)$ and the following
conditions:
\begin{enumerate}
\item $f\mapsto\varphi(f)$ is linear,
\item $\varphi(Df)=0$,
\item $\varphi(f)^\ast= \varphi(\overline{f})$,
\item $[\varphi(f),\varphi(g)]= i \scp{\Delta\ast g}{f}.$
\end{enumerate}
The $\ast$-operation is an algebra involution. The brackets
$\scp{t}{f}=t(f), f\in\Dd(\MM)$ denote the evaluation of the funtional
$t\in\Dd'(\MM)$ and the $\ast$ means convolution. The algebra $\Aa$
is uniquely determined as a $^*$-algebra with unit. 
If the supports of the
test functions are contained in a bounded (usually causally complete)
space time region $\Ocal$ one talks about the \emph{local} algebra of
observables $\Aa(\Ocal)$. 

The commutation relation is frequently written as
\begin{equation}
  \label{eq:comm}
  [\vp(x),\vp(y)]=i\Delta(x-y).
\end{equation}


\section{Representations of the observable algebras}

\label{sec:fock}
A \emph{state} $\omega$ on the observable algebra $\Aa$ is a linear
functional $\omega: \Aa\mapsto\CC$ with the properties:
\begin{align}
\omega(\eins)&=1, \label{eq:normiert} \\ 
\omega(A^\ast A)&\geq 0. \label{eq:positiv}
\end{align}
If a state on $\Aa$ is given one gets a representation of
$\Aa$ by operators on a Hilbert space via the GNS
construction. Since the algebra is generated by $\varphi(f)$ the state 
is already determined by the $n$-point functions:
\begin{equation}
  \label{def:npoint}
  \omega_n(f_1,\dots,f_n)=\omega(\varphi(f_1)\dots\varphi(f_n)).
\end{equation}
A special class of states is given by the \emph{quasi free} states: All 
higher $n$-point functions are completely determined by the 2-point
function $\omega_2$. Because of \eqref{eq:positiv} the 2-point
function has to fulfil the positivity condition:
\begin{equation}
  \label{eq:positivity}
\int\dif x\,\dif y\,\omega_2(x,y)\overline{f(x)}f(y)\geq 0.
\end{equation}


\section{The Fock space}
From the notions above the construction of a representation space  is
just an application of the GNS theorem. We start by choosing a
suitable 2-point function, which is given as usual by the positive
frequency part of the commutator function. This completely defines a
quasi free state on the algebra, called the vacuum state $\omega_0$.
 
We begin by introducing some abbreviations. The mass shell is defined
by the hyperboloid
\begin{equation}
  \label{def:massshell}
  H_m=\{p\in\Rv, p^2=m^2, p^0>0\}.
\end{equation}
We denote a four vector on the mass shell by
\begin{align}
H_m\ni\ptilt &\doteq (E_\pvek,\pvek), \label{def:ptilt} \\
E_\pvek&\doteq\sqrt{\pvek^2+m^2}. \label{def:Ep}
\end{align}
The invariant volume measure on the mass shell is given by
\begin{equation}
\dif\ptilde\doteq\frac{\dif^3\pvek}{(2\pi)^3 2E_\pvek}.
\label{def:ptilde} 
\end{equation}
With these abbreviations the commutator function is found to be
\begin{equation}
  \label{def:Delta}
\Delta(x)=-2\int\dif\ptilde\,\sin(\ptilt x).
\end{equation}
The verification of this expression only requires to check if it
fulfils the Klein-Gordon equation \eqref{eq:kg} and the right initial
values given by \eqref{eq:ivD}.The positive frequency part of $\Delta$
is denoted by $\Delta_+$. It is given by:
\begin{equation}
  \label{def:Deltaplus}
\Delta_{+}(x)=\int\dif\ptilde\, e^{-i\ptilt x}.
\end{equation}
Corresponding to our state $\omega_0$ we define the 2-point function by 
\begin{equation}
  \label{def:zweipkt1}
\omega_0(\varphi(f)\varphi(g))= i\scp{f}{\Delta_+\ast g}.
\end{equation}
Bearing in mind that we are working with distributions we can write
this as
\begin{equation}
  \label{def:zweipkt2}
\omega_2(x,y)= i\Delta_+(x-y).
\end{equation}
Because of \eqref{eq:comm} and \eqref{eq:normiert} the commutator is
given by the asymmetric part of the 2-point function, hence:
\begin{equation}
  \label{eq:comm-zweipkt}
  i\Delta(x)=\Delta_+(x)-\Delta_+(-x).
\end{equation}
Computing
\begin{equation}
  \label{eq:positivcheck1}
  \omega_2(\overline{f},f)
=\int\dif y\, \dif x \, \omega_2(x,y)\overline{f(x)}f(y)
=\int\dif\ptilde\,|\widehat{f}(\ptilt)|^2\geq 0,
\end{equation}
we find that the positivity condition \eqref{eq:positivity} is
fulfiled. Therefore $\omega_2$ defines a positive semi definite scalar
product on $\Dd(\MM)$. This generalizes to a positive semidefinite
scalar product $\scp{\cdot}{\cdot}$ on $\Dd(\MM^n)$ via
\begin{equation}
  \label{def:scalarprodn}
\scp{f}{g}\doteq\int\dif x_1\dots \dif x_n\dif y_1\dots\dif y_n
\prod_{i=1}^n \omega_2(x_i,y_i)
\overline{f(x_1,\dots,x_n)}g(y_1,\dots,y_n).
\end{equation}
Now consider the space of sequences of test functions
$(\Phi_n)_{n\geq0}$ with: $\Phi_0\in\CC,\Phi_n \in \Dd(\MM^n)$, and
$\Phi_n$ is invariant under any permutation of its arguments. It is
equipped with a positive semidefinite scalar product
\begin{equation}
  \label{eq:scalarfock}
  \scp{\Phi}{\Psi}\doteq\sum_{n=0}^\infty\scp{\Phi_n}{\Psi_n},
\end{equation}
where the scalar product on the \rhs\ is given by
\eqref{def:scalarprodn} for $n>0$ and
$\scp{\Phi_0}{\Psi_0}\doteq\Phi_0\Psi_0$. The space 
\begin{equation}
\label{def:fockspace}
\Fo(\Dd(\Rv))
\doteq \{\Phi, \Phi_n\in\Dd(\MM^n), \scp{\Phi}{\Phi}<\infty\},
\end{equation}
is the symmetric Fock space . To define an action of $\varphi$ on
$\Fo(\Dd(\MM))$, we have to define an action on a dense subspace,
which consists of finite sequences only, denoted by
\begin{equation}
\Do\doteq\{\Phi\in\Fo,\exists m\in\NN, \Phi_n=0, \forall n>m\}
\subset\Fo.
\label{def:DefibereichD}
\end{equation}
The field operator is decomposed into
creation and annihilation operators according to
$\phi(f)=a(f)+a^\ast(f)$.  These operators act in Fock space in the 
following way:
\begin{align}
  (a(f)\Phi)_n(x_1,\dots,x_n)
&=\sqrt{n+1}\int\dif x\, \dif y \, f(x) \omega_2(x,y)
\Phi_{n+1}(y,x_1,\dots,x_n) \label{def:annihilator}\\
(a^\ast(f)\Phi)_n(x_1,\dots,x_n)
&=\begin{cases}
0,& n=0,\\
\frac{1}{\sqrt{n}}\sum_{k=1}^n f(x_k) \Phi_{n-1}(x_1,\dots,\not
k,\dots,x_n), & n\not=0.
\end{cases}
\label{def:creator}
\end{align}
This automatically implements the commutation relations
\eqref{eq:comm}. Dividing out the ideal $\Ncal$ that is
generated by the null space of the scalar product produces a pre
Hilbert space $\Ho=\Fo/\Ncal$. Because of \eqref{eq:positivcheck1} 
\Ncal\ consists of all test functions $f$ whose Fourier transform%
\footnote{All conventions and symbols are explained in
  appendix~\ref{app:convention}} 
$\widehat{f}(p_1,\dots,p_n)$ vanish
if at least one momentum $p_i$ is on the mass shell, $p_i\in H_m$.

The representative $\Omega$ of the
class $(1,0,0,\dots)\in\Fo$ is called the vacuum vector. It defines 
a state $\omega_0$ on $\Aa$
according to $\omega_0(A)=(\Omega,A\Omega)$. The scalar product
$(\cdot,\cdot)$ now is the positive definite one on the classes of
$\Ho$. The state $\omega_0$ has the two point function $\omega_2$.  

The relations \eqref{def:annihilator}, \eqref{def:creator} uniquely
fix the higher $n$-point functions by $\omega_2$ according to
\begin{align}
  \label{def:npktung}
\omega_{2n+1}&=0, \\
\omega_{2n}(x_1,\dots,x_{2n})
&=\sum_{\text{pairings of }\{1,\dots,n\}}
\prod_{\text{pairs }i<j} \omega_2(x_i,x_j). 
\label{def:npktg}
\end{align}
Hence $\omega_0$ is a quasi free state. 

We define a unitary representation of the proper orthochronous
Poincar\'e group on $\Fo(\Dd(\MM))$ by
\begin{equation}
  \label{def:poincfock}
  (U(L)\Phi)_n(x_1,\dots,x_n)
  =\Phi_n\left(L^{-1}x_1,\dots,L^{-1}x_n\right),
  \quad\forall L\in\Poinc,
\end{equation}
with $Lx=\Lambda x + a$, $L^{-1} x = \Lambda^{-1} (x - a)$,
$L=(a,\Lambda)$. That $U$ is unitary follows easily from the fact that
$\omega_2$ is invariant: $\omega_2(Lx,Ly) = \omega_2(x,y)$. Then the
field operator transforms according to
\begin{equation}
  \label{eq:poinctrafofield}
  U(L)\vp(x)U(L)^{-1}=\vp(Lx).
\end{equation}

\section{A remark on the commutator functions}
\label{sec:commutator}
Let us look how the different relatives of the commutator functions
look like in our convention. With the support properties and our
definition \eqref{eq:funda1}, \eqref{eq:funda2} we have:
\begin{align}
\Deltaret(x)&=\theta(x^0)\Delta(x), \label{def:Deltaret}\\
\Deltaav(x)&=-\theta(-x^0)\Delta(x)=\Deltaret(-x). \label{def:Deltaav}
\end{align}
Inserting the representation \eqref{def:Delta} one finds:
\begin{equation}
  \label{eq:Deltaretav}
\Deltaretav(x) = - \frac{1}{(2\pi)^4} \int \dif^4 p
\frac{e^{-ipx}}{p^2-m^2\pm i\epsilon p^0}.
\end{equation}
Another very important distribution emerges from the time ordering of
the 2-point function, namely the Feynman propagator.
\begin{align}
 i\Delta^F(x)&\doteq\theta(x^0)\Delta_+(x)+\theta(-x^0)\Delta_+(-x) \\
&=i\Deltaret(x)-\Delta_+(-x)\\
&=i\Deltaav(x)-\Delta_+(x) \\
&=i\Delta^F(-x)\\
&=-\frac{i}{(2\pi)^4} \int \dif^4 p
\frac{e^{-ipx}}{p^2-m^2+i\epsilon}.
\label{def:Deltaf}
\end{align}
Because of \eqref{eq:funda2}, $\Delta^F$ is a Green's function, too:
\begin{equation}
  \label{eq:greenpropagator}
(\square+m^2)\Delta^F=\delta.
\end{equation}
An explicit configuration space expression of all these distributions
can be found in \cite{bk:scharf}[chapter 2.3].


\chapter{Time ordered products}
\label{chap:timeordered}


In the last chapter we have discussed the free scalar quantum field,
obeying the Klein-Gordon equation. We introduce Wick polynomials of
this field, that are composed operators and allow to define
interactions, currents and the energy momentum tensor, for example.
The introduction of a perturbative interaction into the theory
requires the definition of time ordered products ($T$-products) of
Wick polynomials.  The naive ansatz for a time ordering
prescription of $n$ Wick monomials would be%
\footnote{If there are also fermions and ghost fields present, this
  ansatz requires a modification involving the sign of the
  permutation \cite{phd:boas}.}
\begin{multline}
  \label{eq:timeansatz}
  T(W_1(x_1)\dots W_n(x_n))= \\
=\sum_{\pi\in S_n}\theta(x^0_{\pi(1)}-x^0_{\pi(2)})\dots
\theta(x^0_{\pi(n-1)}-x^0_{\pi(n)})
W_{\pi(1)}(x_{\pi(1)})\dots W_{\pi(n)}(x_{\pi(n)}).
\end{multline}
Unfortunately the operators are distribution valued and the $\theta$
function is not continuous at $0$. Therefore the above products are
not a priori well defined. As long as the $W_i$ are linear in the
fields \eqref{eq:timeansatz} still works but already at
the level of quadratic Wick monomials in the fields this naive ansatz
breaks down leading to the well know ultra violet divergencies in
generic (perturbative) quantum field theories. On the other hand we
see that \eqref{eq:timeansatz} gives a well defined expression as long
as no points coincide.

In the definition of time ordered products the arguments are Wick
products (or linear combinations of them). Hence different (looking)
Wick monomials may be related through free field equations, e.g.\ 
$\square\varphi$ and $-m^2\varphi$ represent the same object. It turns out to
be useful to solve this degeneration by introducing an abstract
algebra $\Ba$ of auxiliary variables that is freely generated, as was
shown by Boas \cite{phd:boas}. Then the time ordering becomes a
map from $\Ba^n$ to operator valued distributions on $\Do$.
Hence, commutators and free field equations are formulated for first
order T-products. Moreover this language is adapted to deal with
couplings containing derivated fields, like Yang-Mills for example
\cite{phd:boas}. We present the algebra $\Ba$ in section~\ref{sec:aux}.

A very elegant solution for the definition of time ordering has been
given by Epstein and Glaser \cite{pap:ep-gl}, following ideas of
Bogoliubov and Shirkov \cite{bk:bogol} and St\"uckelberg. A more
accessible way was suggested by Stora \cite{lect:stora} and worked out
in detail by Brunetti-Fredenhagen
\cite{proc:brun-fred,pap:brun-fred2,lect:fred2}.  We review their
solution in section \ref{sec:axiom},\ref{sec:inductive}. Starting
point is a set of axioms from which the causality -- decoding the time
ordering -- is the most important one. Then, for every order the
products are determined up to the total diagonal (all points coincide)
by the products of lower order. Therefore, one is left with an
extension problem that can be solved in distribution theory.

As is well known in quantum field theory this solution is not unique
in general. Therefore one forces the time ordering to respect certain
symmetry relations. Moreover it has turned out to be useful to demand two
more properties for the time ordered products: One deals with the case
that one argument is a field (and imply the equations of
motion for the interacting fields, defined in the next chapter). The
other one relates the normalization of the operators to a
normalization of vacuum expectation values \cite{pap:duet-fred1}.  All
these conditions together are called normalization conditions. They are
presented in section~\ref{sec:normalization}.

\sloppy A very important normalization condition is given by the property of
Poincar\'e covariance. Epstein and Glaser gave a proof of the
existence of such a $T$-product for massive fields \cite{pap:ep-gl}.
Later Stora-Popineau \cite{prep:stora-pop} and  D\"utsch et al.
\cite{pap:scharf2}, \cite{bk:scharf}[chapter 4.5] found a
cohomological existence proof that applies to arbitrary fields. In
\cite{prep:wir} we have worked out their solution into an
explicit form in lowest order perturbation theory. An inductive
construction for higher orders was given in \cite{prep:prange2}.
These preprints are subject of section~\ref{sec:poincare}.

\fussy

\section{The algebra of auxiliary variables}
\label{sec:aux}

This section follows Boas \cite{phd:boas}. Since our work focuses on
scalar fields we formulate this section for bosonic fields only
and comment on the changes that are relevant if fermionic (or ghost)
fields are present. The necessary modifications can be found in
\cite{phd:boas}.  

The algebra $\Ba$ is defined as a freely generated $*$-algebra adapted
to the fields that our quantum theory is formulated with. Assume our
model contains $r$ fields $\vp_1, \dots, \vp_r$. The $\varphi_i$ are
called \emph{basic generators}. Additionally we have to consider
spacetime derivatives, denoted by $\varphi_{i,\mu_1\mu_2,\dots}$. The
elements of the set
\begin{equation}
  \label{def:boasgenerators}
 \Ga\doteq\{\varphi_i,\varphi_{i,\mu_1},\varphi_{i,\mu_1\mu_2}, 
\dots, i=1,\dots,r \} 
\end{equation}
are the \emph{generators} of $\Ba$. We define $\Ba$ as the
unital free commutative algebra%
\footnote{Fermions (and ghosts) give rise to a charge and additional
  (anti-) commutators. The corresponding equivalence relation has to
  be divided out.}  
generated by the elements of $\Ga$.  There is a natural definition of
the derivation with respect to the generators
according to%
\footnote{The derivative becomes graded for fermions (and ghosts).}
\begin{align}
  \frac{\D\varphi_i}{\D\varphi_j}&=\delta_i^j\eins,
  \label{eq:boasdervi1}\\
\frac{\D}{\D\varphi_j}(AB)
&=\frac{\D A}{\D\varphi_j}B+A\frac{\D B}{\D\varphi_j},
\quad\forall A,B\in\Ba,\ \varphi_i,\varphi_j\in\Ga.
\label{eq:boasderiv2}
\end{align}
Assume our free quantum field operators transform under the Lorentz
group according to%
\footnote{We assume summation over double indices.}
\begin{equation}
  \label{eq:lorentzfields}
  U(0,\Lambda)\vp_i(x)U(0,\Lambda)^{-1}
=D{\left(\Lambda^{-1}\right)_i}^j\vp_j(\Lambda x), 
\forall\Lambda\in\Lor,
\end{equation}
where $D$ is a finite dimensional representation of $\Lor$ and $U$ is
a unitary representation of of $\Poinc$ on $\Do$.  The Lorentz group
acts on $\Ba$ as an algebra homomorphism, i.e. a linear mapping which
satisfies
\begin{equation}
  \label{eq:boaslorentz}
  D(\Lambda)\left(\prod_i\varphi_i\right)
=\prod_i D(\Lambda)(\varphi_i),\quad
\Lambda\in\Lor,\ \varphi_i\in\Ga.
\end{equation}
Hence we only need to specify the action on the generators:
\begin{align}
  D(\Lambda)(\vp_i)&={D(\Lambda)_i}^j\vp_j, \label{eq:boaslorentz1} \\
D(\Lambda)\left(\vp_{i,\mu_1\dots\mu_n}\right)
&={\Lambda_{\mu_1}}^{\nu_1}\dots{\Lambda_{\mu_n}}^{\nu_n}
{D(\Lambda)_i}^j\vp_{j,\nu_1\dots\nu_n}. \label{eq:boaslorentz2}
\end{align}
At the end we have additionally a $\ast$-involution acting on $\Ba$
according to
\begin{equation}
  \label{eq:boasstar}
  (aAB)^\ast=\overline{a}B^\ast A^\ast.
\end{equation}
For one scalar field this is obviously trivial
\begin{align}
  \label{eq:boasstargen}
  \varphi^\ast&=\varphi, &\text{and}& &
\varphi_{,\nu_1\dots\nu_n}^*&=\varphi_{,\nu_1\dots\nu_n}.
\end{align}
Let us mention that the goal of this algebra is that a field and its
derivatives are treated as independent objects, thus allowing to
uniquely define a derivative with respect to a generator. This
symbolic derivation is the same one uses in classical mechanics for
deriving the Euler-Lagrange equations, for example.

Now we discuss the mapping of elements of $\Ba$ to Wick
polynomials on $\Do$. This is obtained by the time ordering
map $T$ of one argument:
\begin{equation}
  \label{def:Teins}
  T: \Ba\mapsto\Dist_1(\Do),
\end{equation}
where $\Dist_1$ denotes the space of operator valued distributions on
$\Do$, namely the set of linear continuous maps $\Dd(\MM) \mapsto
\End(\Do)$.  The mapping $T$ is $\CC$-linear. But it is no algebra
homomorphism since there exists no product because of the
distributional character of the images. On the generators the map is
defined as follows:
\begin{align}
  T(\vp_i)(x)&\doteq\vp_i(x) \label{def:Teinsmap1}\\
T(\vp_{i,\nu_1\dots\nu_n})(x)&\doteq
\D_{\nu_1}\dots\D_{\nu_n}\vp_i(x). \label{def:Teinsmap2}
\end{align}
The free fields are quantized according to
\begin{equation}
  \label{eq:commutatorT}
  [T(\varphi_i)(x),T(\varphi_j)(y)]=i\Delta_{ij}(x-y),\ 
\forall\varphi_i,\varphi_j\in\Ga\subset\Ba.
\end{equation}
Since the indices $i,j$ also represent higher generators the
corresponding commutator contributions are given by
\begin{equation}
  \label{def:highercommutator}
  \Delta_{i,\mu_1\dots\mu_n\ j,\nu_1\dots\nu_m}
=(-)^m\D_{\mu_1}\dots\D_{\mu_n}\D_{\nu_1}\dots\D_{\nu_m}\Delta_{ij}.
\end{equation}
The equations of motion read:
\begin{equation}
  \label{eq:motionT}
  D_{ij}T(\vp_j)=0,
\end{equation}
where $D$ is a hyperbolic partial differential operator. 
Then we define the mapping of general elements of $W\in\Ba$ through
the implicit formula:
\begin{align}
  [T(W)(x), T(\varphi_i)(y)]
&=iT\left(\frac{\D W}{\D\varphi_j}\right)(x)\Delta_{ji}(x-y),
\label{eq:boascommutator1}\\
\omega_0\left(T(W)(x)\right)&=0.\label{eq:boascommutator2}
\end{align}
As was shown by Boas this fixes the $T(W)$ uniquely: Let
$W=\prod_i\varphi_i$ be a monomial, then
$T(W)=\wick{\prod_i\varphi_i}$. The colons denote \emph{Wick} (or
\emph{normal}) \emph{ordering} which is defined by the recursion
\begin{align}
  \wick{\eins}&=\eins, \label{def:wick1}\\
\wick{\varphi_i(x)}&=\varphi_i(x), \label{def:wick2}\\
\wick{\varphi_{i_1}(x_1)\dots\varphi_{i_n}(x_n)}
&=\wick{\varphi_{i_1}(x_1)\dots\varphi_{i_{n-1}}(x_n-1)}\varphi_{i_n}(x_n)+
\notag\\
&\quad-
\sum_{i=1}^{n-1}\omega_{2i_ii_n}(x_i,x_n)
\wick{\varphi_{i_1}(x_1)\dots\not i\dots\varphi_{i_{n-1}}(x_{n-1})}.
\label{def:wick3}
\end{align}
Since \eqref{def:npktg} is just the summation of this recursion we
have
\begin{equation}
\omega_0\left(\wick{\varphi_{i_1}(x_1)\dots\varphi_{i_n}(x_n)}\right)
=0.
\label{eq:wickvakuumnull}
\end{equation}
The normal ordering allows to restrict the distributional operators to 
any sub manifold of coinciding points. This provides us with composed
fields like
\begin{equation}
  \label{eq:examplewick}
  \wick{\varphi(x)^2}, \wick{\varphi(x)^{18}\square\varphi(x)}, 
\wick{\D_\mu\varphi(x)\D_\nu\varphi(x)}, \dots
\end{equation}
From the definition \eqref{def:wick1}--\eqref{def:wick3} follows, that 
derivations commute with the Wick ordering and the free field equations
hold inside the Wick colons. 

Because of the free field equations the map $T$ defines a non faithful 
representation of $\Ba$ in $\Dist_1(\Do)$. 


\section{The axioms for time ordered products}
\label{sec:axiom}

We have related the symbolic algebra $\Ba$ to the vector space of Wick 
polynomials by the $T$ operation of one argument. Now we formulate the 
axioms that make $T$ a time ordered product of $n$ arguments that
reduces to the naive ansatz \eqref{eq:timeansatz} for non coincident
points. 
 
Let us denote elements of $\Ba$ by $W_i$ such that under the map
$T:W_i\mapsto T(W_i)$ represents the Wick polynomials whose time
ordering should be defined. The space of distributions on $\Dd(\MM^n)$
with values in $\End(\Do)$ is denoted by $\Dist_n(\Do)$. We require the
T-products to fulfil the following axioms:
\begin{itemize}
\item[\bf P1.]{\bf Well-posedness.}\sloppy
The time ordered products of $n$ symbols denoted by $T(W_1, \dots,
W_n)(x_1, \dots, x_n)$ are multi linear%
\footnote{We also allow for the mapping $(\Cunend\otimes\Ba)^n \mapsto
  \Dist_n(\Do)$ according to $T(f_1W_1,\dots,f_nW_n)(x_1,\dots,x_n) =
  f(x_1) \dots f(x_n) T(W_1,\dots,W_n)(x_1,\dots,x_n)$, which for
  $f_i=1, \forall i$, reduces to the above case.}
strongly continuous maps \mbox{$\Ba^n \mapsto \Dist_n(\Do)$}.

\item[\bf P2.]{\bf Symmetry.} \fussy
The time ordered products are invariant under
  any permutation of their arguments,%
  \footnote{If fermions or ghost are present there is an
    additional (-1)-factor corresponding to the sign of the permutation,
    see \cite{phd:boas}.}
  \begin{equation}
    \label{eq:Tsymmetry}
T\left(W_{\pi(1)},\dots, W_{\pi(n)}\right)(x_{\pi(1)},\dots,x_{\pi(n)})
=T(W_1,\dots, W_n)(x_1,\dots,x_n).
  \end{equation}
$\forall \pi\in S_n$. 

\item[\bf P3.]{\bf Causality.} Assume the points $x_1,\dots,x_n$ can
  by separated by a space like hypersurface, such that $x_i\gtrsim
  x_j, \forall i=1,\dots,k, j=k+1,\dots, n$, with the notation:
  $x\gtrsim y \Leftrightarrow y\not\in\flc(x)$.
  Then the $T$-products factorize:
  \begin{multline}
    \label{eq:Tcausal}
    T(W_1,\dots, W_n)(x_1,\dots,x_n)= \\
    =T(W_1,\dots,W_k)(x_1,\dots,x_k)
    T(W_{k+1},\dots,W_n)(x_{k+1},\dots,x_n).
  \end{multline}

\item[\bf P4.]{\bf Translation covariance.} Under a translation the
  $T$-product transforms according to
  \begin{equation}
    \label{eq:Ttranslation}
    (\Ad U(\eins,a))T(W_1,\dots,W_n)(x_1,\dots,x_n)
=T(W_1,\dots,W_n)(x_1+a,\dots,x_n+a)).
  \end{equation}
 
\end{itemize}

We use the abbreviation $T(I)(x_I) \doteq T\left(W_i,i\in I\right)
  \left(x_i,i\in I\right)$.  The causality condition \Pdrei\ implies
  that space like separated time ordered products commute, since
  $I\sim J \Leftrightarrow (I\gtrsim J) \wedge (I\lesssim J)$ implies
\begin{equation}
  \label{eq:Tcommute}
  T(I\cup J)(x_{I\cup J})=T(I)(x_I)T(J)(x_J)=T(J)(x_J)T(I)(x_I).
\end{equation}

We have demanded multi linearity in \Peins. Therefore it is important
that the arguments of the $T$-products are from our algebra $\Ba$
where no equations of motion hold. Otherwise all time ordered products
containing the Wick polynomial
$\wick{\varphi\square\varphi}+\wick{m^2\varphi^2}$ would be zero for
example.  Although this choice provides for a well defined time
ordering prescription we encounter a situation where one needs a non
zero definition, i.e.\ in the energy momentum tensor.


\section{The inductive construction}
\label{sec:inductive}

The last section has stated the axioms for time ordered
products. Now we formulate an inductive construction: under the
assumption that all products are know up to order $n-1$ we show how
to derive $T$ in order $n$. This procedure goes back to Epstein and
Glaser \cite{pap:ep-gl}. They determined a causal distribution through 
$T$-products of lower order and developed a procedure for a causal
splitting into a retarded and advanced supported part -- analogous to
the decomposition of $\Delta$ into $\Deltaret$ and $\Deltaav$ but also 
applicable to more singular distributions. Then the time ordered
product can be expressed by the splitting solution.

Later, Stora has suggested a method to derive the time ordered
products directly, without the detour using the advanced and retarded 
distributions \cite{lect:stora}. Brunetti-Fredenhagen
\cite{proc:brun-fred,pap:brun-fred2} have provided a complete
analysis (and moreover generalized the whole construction for the
treatment of scalar fields on curved spacetime) 
based on that idea which we review here. 

\subsection{The induction start} 
\label{subsec:start}
If we have no arguments for our $T$-product we set
\begin{equation}
  \label{def:Tleer}
  T(\emptyset)=\eins \in \End(\Do).
\end{equation}
The $T$-products of one argument were already defined in the last
section according to
\begin{equation}
  \label{eq:Teins}
  T(W)(x)=\wick{W(x)}\in \Dist_1(\Do).
\end{equation}
We proceed with the 

\subsection{Recursion to higher orders}
\label{subsec:recursion}
Assume that all time ordered products have been constructed and
satisfy the axioms {\bf P1--P4}. Let us use the abbreviation
$N\doteq\{1,\dots,n\}$. Then we define the
  sets:
  \begin{equation}
    \label{def:CI}
    \Ccal_I\doteq\left\{
(x_1,\dots,x_n)\in\MM^n|x_i\not\in\blc(x_j), \forall i \in I, j 
\in I^c\right\},
  \end{equation}
and $I^c\doteq N\setminus I$ is the complement in $N$. Let
$\Diag_n\doteq\{(x_1,\dots,x_n)\in\MM^n|x_1=\dots=x_n\}$ be the diagonal
in $\MM^n$, then 
\begin{equation}
  \label{eq:CIvereinigung}
  \bigcup_{\substack{I\subset N \\ I\not=\emptyset\\ I\not=N}}
=\MM^n\setminus\Diag_n.
\end{equation}
Then, on any $\Ccal_I$ we set
\begin{equation}
  \label{def:TI}
  T_I(N)(x_N)\doteq T(I)(x_I)T(I^c)(x_{I^c}).
\end{equation}
Now one has to glue together all $T_I$ to a distribution on
$\MM^n \setminus \Diag_n$. But since different $\Ccal_I$ can overlap one
has to check, that the compatibility condition%
\footnote{In the following we frequently omit the arguments $x_I$ 
  since they are already determined by the sets $I$ in $T(I)$.}
\begin{equation}
  \label{def:Tcompatible}
  T_{I}\restriction_{\Ccal_I\cap\Ccal_J}
  =T_J\restriction_{\Ccal_I\cap\Ccal_J},
\end{equation}
holds, if $\Ccal_I\cap\Ccal_J\not=\emptyset$. 
But this follows from causality (\Pdrei) in lower orders: Since
$J\gtrsim J^c$ and $I\gtrsim I^c$ we have
\begin{align}
T_I(N)&=T(I)T(I^c)\label{eq:Tcompatible1}\\
&=T(I\cap J) T(I\cap J^c) T(I^c\cap J) T(I^c\cap J^c)
\label{eq:Tcompatible2}\\
&=T(I\cap J) T(I^c\cap J) T(I\cap J^c) T(I^c\cap
J^c)\label{eq:Tcompatible3}
\\
&=T(J)T(J^c)\label{eq:Tcompatible4}\\
&=T_J(N)\label{eq:Tcompatible5},
\end{align}
where the two inner $T$-products in \eqref{eq:Tcompatible3} commute since
$I^c \cap J \sim I \cap J^c$. Now we take a locally finite smooth
partition of unity $\{f_I\}_{I\in N, I\not=\emptyset, I\not= N}$ on $\MM^n
\setminus \Diag_n$ subordinate to $\{\Ccal\}_{I\in
  N,I\not=\emptyset,I\not=N}$:
\begin{equation}
  \label{def:fI}
  \sum_{\substack{I\subset N \\ I\not=\emptyset\\ I\not=N}}f_I=1
\text{ on } \MM^n\setminus\Diag_n, \supp f_I\subset\Ccal_I.
\end{equation}
Then we define:
\begin{equation}
  \label{def:nullT}
^0T(N)\doteq\sum_{\substack{I\subset N \\ I\not=\emptyset\\ I\not=N}}
f_IT_I(N).
\end{equation}
It can be verified that this definition does not depend on the choice
of the partition of unity $\{f_I\}$ and that $^0T$ is a well defined
operator valued distribution on $\Dd(\MM^n)$ satisfying \Peins\ --
\Pvier. 

By construction ${^0T}$ is a linear combination of numerical translation
invariant distributions multiplied with certain Wick
products:
\begin{equation}
  \label{eq:nullT}
{^0T}(N)(x_N)
={^0t}(x_1-x_2,\dots,x_{n-1}-x_n)\wick{V_1(x_1)\dots V_n(x_n)}.
\end{equation}
As was shown by Epstein-Glaser, these products always exist
\cite{pap:ep-gl}[Theorem 0]. Hence the definition of $T$ reduces to
finding an extension of the numerical distribution $^0t$ to $\Diag_n$, 
which in difference coordinates translates into the problem of
extending a distribution to the origin. This problem is addressed in
the next subsection.

\subsection{Extension of distributions to the origin} 
\label{subsec:extension}
The solution of this problem requires the introduction of a quantity
that measures the singularity of the distribution at the origin
\cite{bk:steinmann}.
\begin{defi}
A distribution \( t\in\Dd'(\Rd) \) has 
scaling degree \( s \) at \( x=0 \), if
\begin{equation}
s=\inf\{s'\in\RR|\lambda^{s'}T(\lambda x)
\overset{\lambda \searrow 0}{\longrightarrow} 0 \text{ in the sense 
of distributions} \}.
\end{equation}
We set \( \scaledeg(t)\doteq s \) and define 
\(\singord(t):=[s]-n \), the singular order.%
\footnote{\( [s] \) is the largest integer that is smaller than or 
equal to \( s \).} 
\end{defi}
The definition also holds if \( t\in\Dd'(\Rdon) \). One easily finds
that differentiation increases the scaling degree while multiplication 
with $x$ decreases it:
\begin{align}
  \scaledeg(x^\beta t)&=\scaledeg(t)-|\beta|,\\
\scaledeg(\D^\beta t)&=\scaledeg(t)+|\beta|, \\
\scaledeg(ft)\leq\scaledeg(t),
\end{align}
for all $t\in\Dd'(\Rd)$ and $f\in\Dd(\Rd)$.
Now, the solution of the extension problem depends on the sign 
of the singular order.
\begin{thm}
\label{thm:fortsetzung1} 
Let \( ^{0}t\in\Dd'(\Rdon) \) with scaling degree \( s<n \). 
Then there exists a unique \( t\in\Dd'(\Rd) \) with scaling degree 
\( s \) and \( t(f)={^{0}t}(f) \) for all \( 
f\in\Dd(\Rdon) \).
\end{thm}
Otherwise we introduce the 
\begin{Woperation}
Let \( \Dd^{\omega}(\Rd) \) be the subspace of test functions 
vanishing up to order \( \omega \) at \( 0 \). Define
\begin{gather}
\W{\omega}{w}:\Dd(\Rd)\mapsto\Dd^{\omega}(\Rd), 
\quad f\mapsto\W{\omega}{w}f, \notag \\
\left(\W{\omega}{w}f\right)(x)=f(x)-w(x)\sum_{|\alpha|\leq\omega} 
\frac{x^{\alpha}}{\alpha!}\left(\D^{\alpha}\frac{f}{w}\right)(0),
\label{def:W}
\end{gather}
with \( w\in\Dd(\Rd), w(0)\not=0 \).
\end{Woperation}
Now we can discuss the general case.
\begin{thm}
\label{thm:fortsetzung2} 
Let \( ^0t\in\Dd'(\Rdon) \) with scaling degree \(s\geq n\).
Given \( w\in\Dd(\Rd) \) with \( w(0)\not=0 \) and constants \(
c^{\alpha}\in\CC  \) for all multi indices \(
\alpha, |\alpha|\leq\omega \) , then there is exactly one distribution \(
t'\in\Dd'(\Rd) \) with scaling degree \( s \) and following
properties:
\begin{enumerate}
\item \( \scp{t'}{f}=\scp{^0t}{f}\quad\forall  
f\in\Dd(\Rdon) \), \label{erst}
\item \( \scp{t'}{wx^{\alpha}}=c^{\alpha} \), \label{zweit}
\end{enumerate}
with \( t' \) given by:
\begin{equation}
\scp{t'}{f}=\scp{t}{\W{\omega}{w}f}+\sum_{|\alpha|\leq\omega} 
\frac{c^{\alpha}}{\alpha!}\left(\D^{\alpha}\frac{f}{w}\right)(0).
\label{def:T'}
\end{equation}
Here \( t \) is the unique extension by
theorem~\ref{thm:fortsetzung1}, \( \W{\omega}{w} \) is given by
(\ref{def:W}) and \( \omega \) is the singular order of \( ^0t
\).
\end{thm}
With these theorems the extension can be done right away. In the case
of non negative singular order we notice that an ambiguity appears,
namely all choices of the constants $c^\alpha$ yield a well defined
solution. The normalization condition in the next section 
restricts this freedom further. Before we proceed to these conditions
we introduce the


\subsection{Anti time ordered products}
\label{subsec:antitime}
If one wants to define interacting perturbative fields one could
equally well have started with the definition of anti chronological
products, that give a meaning to the expression 
\begin{multline}
  \label{eq:antitimeansatz}
  \overline{T}(W_1,\dots,W_n)(x_1,\dots,x_n)= \\
=\sum_{\pi\in S_n}\theta\left(x^0_{\pi(1)}-x^0_{\pi(2)}\right)\dots
\theta\left(x^0_{\pi(n-1)}-x^0_{\pi(n)}\right)
W_{\pi(n)}\left(x_{\pi(n)}\right)\dots 
W_{\pi(1)}\left(x_{\pi(1)}\right)
\end{multline}
in the case of coinciding points, corresponding to equation
\eqref{eq:timeansatz}. The causality property that encodes the wanted
anti chronological factorization reads according to \Pdrei:
\begin{equation}
  \label{eq:anticausal}
  \Tbar(N)(x_N)=\Tbar(I^c)(x_{I^c})\Tbar(I)(x_I),
\text{ if } I\gtrsim I^c.
\end{equation}
It turns out in the next chapter that the functional of the anti
chronological products is the inverse $S$-matrix hence we have
\begin{equation}
  \label{eq:TmalTbar}
\sum_{I\in N}(-)^{|I|}\Tbar(I)(x_I)T(I^c)(x_{I^c})
=\sum_{I\in N}(-)^{|I^c|}T(I)(x_I)\Tbar(I^c)(x_{I^c})=0.
\end{equation}
This equation allows a recursive definition of the $\Tbar$-products of 
order $n$ by lower order $\Tbar$- and all order $T$-products:
\begin{align}
  \Tbar(N)\K{y_N}
&=-\sum_{\substack{I\subset N \\ I\not=\emptyset}}
(-)^{|I|}T(I)\K{x_I}\Tbar\K{I^c}\K{x_{I^c}}, 
\label{def:Tbarrekurs1}\\
&=-\sum_{I\subsetneq N}
(-)^{|I^c|}\Tbar(I)\K{x_I}T\K{I^c}\K{x_{I^c}}.
\label{def:Tbarrekurs2}
\end{align}
Explicit solution of the recursion gives:
\begin{equation}
  \label{def:Tbar}
  \Tbar(N)(x_N)=\sum_{P\in\text{Part}(N)}(-)^{|P|+|N|}
\prod_{Q\in P} T(Q)(x_Q).
\end{equation}


\section{Normalization conditions}
\label{sec:normalization}
As was seen in the last section the extension procedure that has to
be applied to the numerical distribution $^0t$ in order to define the
$T$-products everywhere produces an ambiguity related to the constants 
$c^\alpha$ from theorem~\ref{thm:fortsetzung2}. We formulate a
set of conditions, that apply to these constants in a way that certain 
properties for the complete $T$-products are fulfiled. These
conditions were introduced in \cite{pap:duet-fred1} and afterwards
generalized to the case, when derivated fields are present in
\cite{phd:boas}. 

The first normalization condition deals with the possible occurrence of
discrete symmetries of $^0T$. Assume, there is a finite group $G$
acting invariantly on $^0T$ according to $:{^0T}\mapsto{^0T}_a$, and
${^0T}_a={^0T}_\eins$ for all $a\in G$. The extension of ${^0T}_a$
is denoted by $T_a$. We require the extension to be invariant
under $G$, too:
\begin{equation}
T_a(W_1,\dots,W_n)(x_1,\dots,x_n)
=T_\eins(W_1,\dots,W_n)(x_1,\dots,x_n), \quad\forall W_i\in\Ba, a\in G.
\tag{\Nnull}
\end{equation}
Obviously this condition can always be fulfiled by choosing
\begin{equation}
  \label{eq:Nnull}
  T(W_1,\dots,W_n)(x_1,\dots,x_n)
=\frac{1}{|G|}\sum_{a\in G}T_a(W_1,\dots,W_n)(x_1,\dots,x_n)
\end{equation}
This condition already establishes the correct $P,C,T$-transformation
properties of the $T$-products, see (\cite{bk:scharf}). Since this
condition is quite trivial, we only talk of a $T$-product if
it is fulfiled.

The next condition implements the conservation of Poincar\'e
covariance. Let $\Poinc$ act on $\End(\Do)$ through the
representation $U$. Then we demand:
\begin{multline}
  (\Ad U(L))T(W_1,\dots,W_n)(x_1,\dots,x_n)=\\
=T\K{D\K{\Lambda^{-1}}(W_1),\dots,D\K{\Lambda^{-1}}}(W_n))
(Lx_1,\dots,Lx_n),
\tag{\Neins}
\end{multline}
for all $L=(a,\Lambda)\in\Poinc$ and the action of the representation
$D$ on $\Ba$ is defined in
\eqref{eq:boaslorentz1},\eqref{eq:boaslorentz2}. Let us remark that
our current formulation is redundant since \Ndrei\ already contains
the axiom \Pvier\, namely for $\Lambda=\eins$. Nevertheless we 
stick to this formulation for the following reason: It was already
remarked by Epstein-Glaser \cite{pap:ep-gl} that translation
covariance is a crucial condition for the causal construction. In that
case their theorem~0 always guarantees a solution in the form
\eqref{eq:nullT}. In contrast to \Pvier\ normalization condition
\Neins\ is not necessary for performing the inductive construction. 

Moreover Brunetti-Fredenhagen \cite{proc:brun-fred,pap:brun-fred2}
have shown that the inductive construction can also be performed if
\Pvier\ is replaced by a weaker assumption formulated with the
techniques of micro local analysis. Their approach also applies for
curved spacetime. Since there is no symmetry in a generic spacetime a
normalization condition like \Neins\ reflecting the symmetry of the
Minkowski space has to be abandoned.

It was shown by Epstein and Glaser \cite{pap:ep-gl} that \Neins\ can
be fulfiled. In the next section we derive a procedure for an
explicit construction.

The anti time ordered products were introduced in the last
chapter. Then unitarity is fulfiled if
\begin{equation}
  \tag{\Nzwei}
T(W_1,\dots,W_n)(x_1,\dots,x_n)^*
=\Tbar(W_n^*,\dots,W_1^*)(x_n,\dots,x_1),\quad\forall W_i\in\Ba.
\end{equation}
The $*$ on the \lhs\ is the adjoint on $\Do$. On the \rhs\ the order
of the symbols is reversed. It can be rearranged using \Pdrei. Epstein
and Glaser have shown that \Nzwei\ can always be satisfied: Assume it
holds to order $n-1$ together with \Neins. For every normalization
$T'=T(W_1,\dots,W_n)$ that fulfils \Neins\ we can form $T=\frac{1}{2}
(T'+ {\Tbar'}^*)$ which satisfies \Nzwei. Then \Neins\ is
fulfiled automatically since $U$ is unitary: $U^\ast=U^{-1}$. 

If we commute the $T$-product with a generator we require (compare to
\eqref{eq:boascommutator1}):
\begin{multline}
  \tag{\Ndrei}
  \left[T(W_1,\dots,W_n)(x_1,\dots ,x_n),\vp_i(y)\right] = \\
  = i \sum_{k=1}^n 
T\left(W_1,\dots,\parz{W_k}{\vp_j},\dots,W_n\right)
(x_1,\dots,x_n) \Delta_{ji}(x_k-y),
\end{multline}
for every $W_i \in \Ba$ and $\vp_i(y) = T(\vp_i)(y), \vp_i \in\Ga$.
Since the center of $\End(\Do)$ only contains multiples of the identity
\cite{bk:scharf}, \Ndrei\ determines the $T$-product up to a
$\CC$-number distribution by the $T$-products of the sub monomials.
This distribution is given by the vacuum expectation value. It was
shown by Boas that \Ndrei\ can always be satisfied (together with
\Neins\ and \Nzwei) and is equivalent to the \emph{causal Wick
  expansion}:
\begin{multline}
\label{eq:causalwick}
  T(W_1,\dots,W_n)(x_1,\dots,x_n) = \\
=  \sum_{\gamma_1,\dots,\gamma_n} \omega_0
  \left(
    T\left(W_1^{(\gamma_1)},\dots,W_n^{(\gamma_n)}\right)(x_1,\dots, x_n)
  \right) 
  \frac{\wick{\vp^{\gamma_1}(x_1) \cdots \vp^{\gamma_n}(x_n)}}
{\gamma_1!\cdots \gamma_n!} 
\end{multline}
Here the $\gamma_i \in \NN^r$ are multi indices with one entry 
for each of the $r$ generators in $\Ga$, i.e. 
\begin{equation}
  \gamma_i = ((\gamma_i)_1,\dots , (\gamma_i)_r) \quad \in \NN^r
\end{equation}
The $W^{(\gamma_i)}$ are derivatives with respect to the generators
\begin{equation}
  W^{(\gamma_i)} \doteq \parz{^{|\gamma_i|}W}{^{(\gamma_i)_1} 
  \vp_1\cdots \D^{(\gamma_i)_r} \vp_r} ,
\end{equation}
where $|\gamma_i| = \sum_{k=1}^r (\gamma_i)_k$. The $\vp^{\gamma_i}$ 
are defined by
\begin{equation}
  \vp^{\gamma_i}(x) 
\doteq T\left(\prod_{k=1}^r \vp_k^{(\gamma_i)_k}\right)(x),
\end{equation}
and
\begin{equation}
  (\gamma_i)! \doteq \prod_{k=1}^r (\gamma_i)_k !. 
\end{equation}

The following normalization condition is a partial differential
equation concerning $T$-products with only one generator. 
\begin{multline}
\tag{\Nvier}
D^x_{ij}T(W_1,\dots,W_n,\vp_j)(x_1,\dots,x_n,z) = \\
  = i\sum_{k=1}^n 
T\left(W_1\dots\parz{W_k}{\vp_i}\dots W_n\right)(x_1,\dots ,x_n)
 \,\delta (x_k-z)  ,
\end{multline}
where $W_i\in \Ba$ and $ \vp_i\in \Ga$. This differential
equation can be solved by
\begin{multline}
\label{eq:nvierintegrated}
    T(W_1,\dots,W_n\vp_i)(x_1,\dots ,x_n,z) = \\
\begin{split}
&= i \sum_{k=1}^{n} \Delta^F_{ij}(z-x_k) 
 T\left(W_1,\dots,\parz{W_k}{\vp_j},\dots,W_n\right)(x_1,\dots ,x_n)+ \\
    &\quad +\sum_{\gamma_1\cdots\gamma_n}\omega_0
    \left(
      T\left(W_1^{(\gamma_1)},\cdots, W_n^{(\gamma_n)}\right)
      (x_1, \dots, x_n)
    \right)
    \frac{\wick{\vp^{\gamma_1}(x_1) \dots \vp^{\gamma_n}(x_n)
    \vp_i(z)}}{\gamma_1!\dots\gamma_n!} .
  \end{split}
\end{multline}
All proofs of the compatibility statements can be found in
\cite{phd:boas}. Moreover it is shown that the normalization conditions
also hold in the presence of derivated fields in the $W_i$. This can
be accomplished by a suitable adaption of the differential operator
$D_{ij}$, the commutator functions $\Delta_{ij}$ and the Feynman
propagators $\Delta^F_{ij}(x-y) = \omega_0\K{T(\vp_i,\vp_j)(x,y)}$.

As we have seen in the last subsection the definition of $T$ requires 
an extension of the numerical distributions of the decomposition
\eqref{eq:causalwick}. We do not want to increase the scaling degree
of these distribution in the renormalization process. Then the scaling 
degree of the time ordered numerical distributions is already
determined by the dimensions of the symbols from $\Ba$:
\begin{equation}
  \label{eq:scaledegT}
  \scaledeg \omega_0\K{T(W_1,\dots,W_n)}\Bigr\vert_{\Diag_n}=
\sum_{i=1}^n \dim W_i,
\end{equation}
for all monomials $W_i\in\Ba$.  Since in $\Ba$ every $W$ can be
decomposed uniquely into basic generators and derivatives we have
according to this decomposition:
\begin{equation}
  \label{eq:dimW}
  \dim W=\sum_{i\in\Ga_b}\dim \vp_i + \# \D, 
\end{equation}
and the dimension of the bosonic (fermionic) fields is one (3/2).


\section{Poincar\'e covariance}
\label{sec:poincare}

In the last section we have demanded Poincar\'e covariance to hold for 
the $T$-products by condition \Nzwei. In the decomposition of the
products according to the causal Wick expansion \eqref{eq:causalwick}
we see that covariance properties have to be fulfiled by the numerical 
distributions only, since the Wick products already transform
correctly under Poincar\'e transformations. Since we are working in an 
inductive procedure we require covariance to be fulfiled in lower
orders and have to provide an extension of the numerical distributions 
according to theorem~\ref{thm:fortsetzung2} that respects these
properties. Therefore the solution of this problem reduces to finding
a suitable set of constants $c^\alpha$. 

Epstein-Glaser \cite{pap:ep-gl} gave an existence proof for the
\emph{central solution} of the subtraction procedure in case of a
massive theory. It corresponds to the choice $w=1$ and $c^\alpha=0$ in
theorem~\ref{thm:fortsetzung2} which obviously preserves Lorentz
covariance. For the massless case they suggested an averaging over a
maximal compact subgroup of the complexified Lorentz group.

Another proof was given by Stora and Popineau
\cite{prep:stora-pop}[unpublished] and D{\"u}tsch et al.
\cite{pap:scharf2}. A detailed representation can be found in the book
of Scharf \cite{bk:scharf}. It is based on cohomological arguments. We
review their analysis here, adapted to our case.

In \cite{prep:wir} we have explicitly calculated the constants in
lowest order perturbation theory for scalar fields. This solution may
also apply in the case of special symmetry in higher orders
\cite{pap:pinter1}. A general solution for higher orders and
arbitrary covariance could be derived only recursively in
\cite{prep:prange2}. This section reviews the content of these
preprints. 


\subsection{The subtraction procedure}
We remind the reader of the subtraction operator $W$ \eqref{def:W}
which was necessary to define the distributional extension. The
$W$-operation is simplified if we require $w(0)=1$ and
$\D^{\alpha}w(0)=0$, for $0<|\alpha|\leq\omega$ (this was our
assumption in \cite{prep:wir}). A test function with these properties
can be derived from an arbitrary test function by application of the
following $V$-operation ($\D^\mu w^{-1}$ means $\D^{\mu}(w^{-1})$):
\begin{equation}
V_{\omega}:\Dd(\Rd)\mapsto\Dd(\Rd),\quad (V_{\omega}w)(x)\doteq 
w(x)\sum_{|\mu|\leq \omega}\frac{x^{\mu}}{\mu!}\D^{\mu}w^{-1}(0),
\label{def:V}
\end{equation}
where $w(0)\not=0$ is still assumed.   We can write $W$ as
\begin{equation}
  \label{eq:varW}
\left(\W{\omega}{w}f\right)(x)
=f(x)-\sum_{|\alpha|\leq\omega} 
\frac{x^{\alpha}}{\alpha!}V_{\omega-|\alpha|}w\,\D^{\alpha}f(0).
\end{equation}
Let us denote the extension corresponding to this subtraction by
\tR{\omega}{w}:
\begin{equation}
  \label{def:tR}
  \scp{\tR{\omega}{w}}{f}\doteq\scp{^0t}{\W{\omega}{w}f}.
\end{equation}
Adding any polynomial in derivatives of $\delta$ up to order $\omega$
produces another extension $\tiltR{\omega}{w}$:
\begin{align}
\scp{\tiltR{\omega}{w}}{f}&=\scp{\tR{\omega}{w}}{f}
+\sum_{|\alpha|\leq\omega} 
\frac{a^{\alpha}}{\alpha!}\D^{\alpha}f(0), \\
\intertext{or rearranging the coefficients} 
&=\scp{\tR{\omega}{w}}{f}
+\sum_{|\alpha|\leq\omega} 
\frac{c^{\alpha}}{\alpha!}\D^{\alpha}\left(f w^{-1}\right)(0)
\label{def:tiltR}
\end{align}
Since $\W{\omega}{w}( w x^\alpha) = \W{\omega}{w} (x^\alpha
V_{\omega-|\alpha|}w) = 0$ for $|\alpha| \leq \omega$, $c$ resp.\ $a$ 
are given by
\begin{align}
a^\alpha&=\scp{\tiltR{\omega}{w}}{x^\alpha V_{\omega-|\alpha|}w}, & 
c^\alpha&=\scp{\tiltR{\omega}{w}}{x^\alpha w}.
\label{eq:cundaaust}
\end{align}
They are related through:
\begin{align}
 a^\alpha&=c^\alpha\sum_{|\mu|\leq\omega-|\alpha|}
\frac{c^{\mu}}{\mu!}\D_{\mu}w^{-1}(0), &
 c^\alpha&=a^\alpha\sum_{|\mu|\leq\omega-|\alpha|}
\frac{a^{\mu}}{\mu!}\D_{\mu}w(0), &
1\leq|\alpha|\leq\omega, \notag \\
a^0&=\sum_{|\mu|\leq\omega}
\frac{c^{\mu}}{\mu!}\D_{\mu}w^{-1}(0), &
 c^0&=\sum_{|\mu|\leq\omega} \frac{a^{\mu}}{\mu!}\D_{\mu}w(0).
\label{eq:causa}
\end{align}
The equation for $a$ follows from the Leibnitz rule in
\eqref{def:tiltR}, while the equation for $c$ is derived from
\eqref{eq:cundaaust}.

\subsection{The $G$-covariant extension}
\label{sec:g}
We begin defining the notion of a $G$-co\-vari\-ant distribution.  So
let $G$ be a linear transformation group on \Rd\ i.e.  $x\mapsto gx$,
$g\in G$. Then
\begin{equation}
        x^{\alpha}\mapsto{g^{\alpha}}_{\beta}x^{\beta}=(gx)^{\alpha}
        \label{}
\end{equation}
denotes the corresponding tensor representation.  $G$ acts on
functions in the following way:
 \begin{equation}
 (gf)(x)\doteq f(g^{-1}x),
 \label{def:gphi}
 \end{equation}
so that \Dd\ is made a $G$-module.  We further have
\begin{align}
g(fh)&=(gf)(gh),\label{eq:gprod} \\
x^{\alpha}\D_{\alpha}(g^{-1}f)
&=(gx)^{\alpha}g^{-1}(\D_{\alpha}f), \label{eq:1} \\
x^{\alpha}\D_{\alpha}(g^{-1}f)(0)
&=(gx)^{\alpha}\D_{\alpha}f(0)
\forall g\in G, f,h \in \Dd(\Rd) \label{eq:2}
\end{align}
Now assume we have a distribution ${^0t}\in\Dd'(\Rdon,V)$ taking
values in a finite vector space $V$ that serves as a representation
space for the group $G$. The distribution transforms 
covariantly under the Group $G$ as a density, i.e.
\begin{equation}
{^0t}(gx)|\det g|=D(g){^0t}(x),
\label{eq:tcov}
\end{equation}
where $D$ is the corresponding representation.  That means:
\begin{equation}
\scp{{^0t}}{gf}=\scp{D(g){^0t}}{f}\doteq D(g)\scp{{^0t}}{f}. 
\label{def:tcov}
\end{equation}
We now investigate the covariance properties in the extension process.
We compute:
\begin{multline}
D(g)\scp{\tR{\omega}{w}}{g^{-1}f} 
-\scp{\tR{\omega}{w}}{f}= \\
\begin{split}
  &=D(g)\scp{{^0t}}{\W{\omega}{w}g^{-1}f} -\scp{{^0t}}{\W{\omega}{w}f} \\
&\stackrel{\eqref{eq:varW}}{=}
D(g)\scp{{^0t}}{g^{-1}f-\sum_{|\alpha|\leq\omega}
  \frac{x^{\alpha}}{\alpha!}V_{\omega-|\alpha|}w\,\D_{\alpha}(g^{-1}f)(0)}
-\scp{{^0t}}{\W{\omega}{w}f} \\
&\stackrel{(\ref{eq:gprod},\ref{eq:2})}{=}
D(g)\scp{{^0t}}{g^{-1}\biggl(f- \sum_{|\alpha|\leq\omega}
  \frac{x^{\alpha}}{\alpha!}\left(gV_{\omega-|\alpha|}w\right)
  \D_{\alpha}f(0)\biggr)} -\scp{{^0t}}{\W{\omega}{w}f}\\
&\stackrel{(\ref{def:tcov})}{=} \sum_{|\alpha|\leq\omega}
\scp{{^0t}}{x^{\alpha}(\eins-g)(V_{\omega-|\alpha|}w)}
\frac{\D_{\alpha}f(0)}{\alpha!} \\
&\doteq\sum_{|\alpha|\leq\omega}b^{\alpha}(g)
\frac{\D_{\alpha}f(0)}{\alpha!}.
\end{split}
\label{def:balpha}
\end{multline}
Then (\ref{def:balpha}) defines a map from $G$ to a finite dimensional
complex vector space. Now we follow \cite{prep:stora-pop},
\cite{bk:scharf}[chapter 4.5]: Applying two transformations
\begin{align}
b^{\alpha}(g_{1}g_{2})
&=\scp{{^0t}}{x^{\alpha}(\eins-g_{1}g_{2})(V_{\omega-|\alpha|}w)} \\
&=\scp{{^0t}}{x^{\alpha}\left((\eins-g_{1})
+g_{1}(\eins-g_{2})\right)(V_{\omega-|\alpha|}w)} \\
&=b^{\alpha}(g_{1})+|\det g_{1}|\scp{{^0t}(g_{1}x)} 
{(g_{1}x)^{\alpha}(\eins-g_{2})(V_{\omega-|\alpha|}w)},
\end{align} 
and omitting the indices we see
$b(g_{1}g_{2})=b(g_{1})+D(g_{1})g_{1}b(g_{2}), $ which is a 1-cocycle
for $ b(g) $.  Its trivial solutions are the 1-coboundaries
\begin{equation}
        b(g)=(\eins-D(g)g)a,
\label{def:cobound}
\end{equation}
and these are the only ones if the first cohomology group of $G$ is
zero. In that case we can restore $G$-covariance by adding the
following counter terms:
\begin{equation}
\scp{\tR{\omega}{w}^{G-\mathrm{cov}}}{f} 
\doteq\scp{\tR{\omega}{w}}{f} 
+\sum_{|\alpha|\leq\omega}
\frac{1}{\alpha!}a^{\alpha}(w)\D_{\alpha}f(0).
 \label{def:tRGcov}
 \end{equation} 
The task is to determine $a$ from (\ref{def:cobound}) and (\ref{def:balpha}):
\begin{equation}
\scp{{^0t}}{x^{\alpha}(\eins-g)(V_{\omega-|\alpha|}w)}
=\bigl[(\eins-D(g)g)a\bigr]^\alpha
\label{def:a}
\end{equation}

\subsection{Bosonic Lorentz covariance}
\label{subsec:tensor}
The first cohomology group of $\Lcal_{+}^{\uparrow}$ vanishes
\cite{bk:scharf}[chapter 4.5 and references there]. We determine $ a $
from the last equation. The most simple solution appears in the case
of Lorentz invariance in one coordinate. This situation was
completely analyzed in \cite{prep:wir} for $\D_{\alpha}w(0) =
\delta_{\alpha}^{0}$.  The following two subsections generalize the
results to arbitrary $w,\ w(0)\not=0$.

\subsubsection{Lorentz invariance in $\MM$}
\label{subsubsec:Linv4d}
If we expand the index $\alpha$ into Lorentz indices $\mu_1, \dots, 
\mu_n$, \eqref{def:a} is symmetric in $\mu_1, \dots, \mu_n$ and 
therefore $a$ is, too. We just state our result from 
\cite{prep:wir} which is modified by the generalization of $w$:
\begin{multline}
  a^{(\mu _1 \ldots \mu_n)} = \frac{ (n-1)!!}{(n+2)!!}
  \sum_{s=0}^{\left[\frac{n-1}{2} \right] }
  \frac{(n-2s)!!}{(n-2s-1)!!} \eta^{(\mu_1 \mu _2} 
  \ldots \eta^{\mu_{2s-1}\mu _{2s}} \times \\ 
  \times \left\langle 
  {^0t}, (x^2)^s x ^{\mu _{2s+1}} \ldots x ^{\mu_{n-1}} 
  \left(x^2\D^{\mu_{n})}-x^{\mu_{n})}x^{\beta}\D_{\beta}\right)
  V_{\omega-n}w\right\rangle,
\label{eq:amun}
\end{multline}
if we choose the fully contracted part of $ a $ to be zero in case 
of $ n $ being even. We used the notation
\begin{align*}
b^{(\mu_{1}\dots\mu_{n})}&=\frac{1}{n!}\sum_{\pi\in S_{n}} 
b^{\mu_{\pi(1)}\dots\mu_{\pi(n)}}, &
b^{[\mu_{1}\dots\mu_{n}]}&=\frac{1}{n!}\sum_{\pi\in S_{n}} 
\sgn(\pi)b^{\mu_{\pi(1)}\dots\mu_{\pi(n)}},
\end{align*}
for the totally symmetric resp.\ antisymmetric part of a tensor.


\subsubsection{Dependence on $w$}
\label{subsubsec:scale} 
Performing a functional derivation of the Lorentz invariant extension
with respect to $w$, only Lorentz invariant counter terms appear.
\begin{defi}
The functional derivation is given by:
\begin{equation}
\scp{\frac{\delta}{\delta h}F(h)}{f}\doteq
\left.\frac{\dif}{\dif\lambda}F(h+\lambda f)\right\vert_{\lambda=0}. 
\end{equation}
\end{defi}
This definition implies the following functional derivatives:
\begin{align}
\scp{\frac{\delta}{\delta w}\tR{\omega}{w}(f)}{h}
&=-\sum_{|\alpha|\leq\omega}\frac{1}{\alpha!}
\scp{\tR{\omega}{w}}{x^\alpha h}
\D_\alpha\left(f w^{-1}\right)(0),\label{eq:derivtR} \\
\scp{\frac{\delta}{\delta w}\scp{s}{V_\omega w}}{h}
&=\sum_{|\alpha|\leq\omega}\frac{1}{\alpha!}
\scp{s}{\W{\omega}{w}(x^\alpha h)}
\D_\alpha w^{-1} (0), \label{eq:derivS}
\end{align}
for any distribution $s\in\Dd'(\Rd)$ and $f,h\in\Dd(\Rd)$. 
\begin{proof} 
We show how to derive the first relation. Inserting the definition
we find:
\begin{multline}
\left.\frac{\dif}{\dif\lambda}
\tR{\omega}{w+\lambda h}(f)\right\vert_{\lambda=0}=\\
\begin{split}
  &=\scp{^0t}{-h\sum_{|\alpha|\leq\omega}
    \frac{x^\alpha}{\alpha!}\D_\alpha\left(f w^{-1}\right)(0)
    +w\sum_{|\alpha|\leq\omega}\frac{x^\alpha}{\alpha!}
    \D_\alpha\left(f h w^{-2}\right)(0)}, \\
  &=\sum_{|\alpha|\leq\omega}\frac{1}{\alpha!}
  \Biggl\langle{^0t},-h x^\alpha\D_\alpha\left(f w^{-1}\right)(0)+\\
  &\qquad +w x^\alpha\D_\alpha\left(f w^{-1}\right)(0)
  \sum_{|\nu|\leq\omega-|\alpha|}\frac{x^\nu}{\nu!}  \D_\nu\left(h
    w^{-1}\right)(0) \Biggr\rangle \\
  &=-\sum_{|\alpha|\leq\omega}\frac{1}{\alpha!}
  \scp{^0t}{x^\alpha\W{\omega-|\alpha|}{w}h}
  \D_\alpha\left(f w^{-1}\right)(0) \\
  &=-\sum_{|\alpha|\leq\omega}\frac{1}{\alpha!}
  \scp{\tR{\omega}{w}}{x^\alpha h}\D_\alpha\left(f w^{-1}\right)(0),
\end{split}
\end{multline}
where we used Leibnitz rule and rearranging of  the summation of the
second term in the second line and the
the relation $x^\alpha\W{\omega-|\alpha|}{w}f = 
\W{\omega}{w}(x^\alpha f)$ on the last line. The second equation 
follows from a similar calculation.
\end{proof}
We calculate the dependence of $a$ on $w$. With \eqref{eq:derivS} we
get:
\begin{multline}
\scp{\frac{\delta}{\delta w} a^{(\mu_1\dots\mu_n)}(w)}{f}
= \frac{(n-1)!!}{(n+2)!!}
  \sum_{s=0}^{\left[\frac{n-1}{2} \right] }
  \frac{(n-2s)!!}{(n-1-2s)!!}  \eta^{(\mu_1\mu_2} \dots
  \eta^{\mu _{2s-1} \mu _{2s}} \times \\ 
  \times\sum_{|\beta|\leq\omega-n}\frac{\D_\beta w^{-1}(0)}{\beta!}
  \scp{^0t}{(x^2)^s x^{\mu_{2s+1}} \dots x ^{\mu_{n-1}} 
  \left(x^2\D^{\mu _{n})} 
  -x^{\mu_{n})}x^{\beta}\D_{\beta}\right)\W{\omega-n}{w}(x^\beta f)}.
\label{eq:damun}
\end{multline}
Since $\W{\omega-n}{w}(x^\beta h)$ is sufficient regular, we can put
the $x$'s and derivatives on the left and the same calculation like in
\cite{prep:wir} applies. The result is
\begin{multline}
  \scp{\frac{\delta}{\delta w}a^{\mu_{1}\cdots\mu_{n}}(w)}{f}=
  \sum_{|\beta|\leq\omega-n}\frac{\D_\beta w^{-1}(0)}{\beta!}
\Biggl[
\scp{\tR{\omega}{w}}{x^{\mu_{1}}\cdots x^{\mu_{n}}x^\beta f}
 + \\
-\begin{cases}
 0,&n\text{ odd}, \\
 \frac{2(n-1)!!}{(n+2)!!}  
 \scp{\tR{\omega}{w}}{(x^2)^\frac{n}{2}x^\beta f}
 \eta^{(\mu_1\mu_2}\cdots\eta^{\mu_{n-1}\mu_n)}, 
 &n\text{ even.}
\end{cases}
\Biggr]
\end{multline}
Using this result and \eqref{eq:derivtR} we find:
\begin{align}
  \scp{\frac{\delta}{\delta w}\scp{\tRli{\omega}{w}}{f}}{h}
  &=-\sum_{\substack{n=0\\ n\text{ even}}}^{\omega}\frac{d_{n}}{n!}
  \square^{\frac{n}{2}}f(0), \\
  d_{n}&\doteq
  \frac{2(n-1)!!}{(n+2)!!}\sum_{|\beta|\leq\omega-n}\frac{1}{\beta!}
\scp{\tR{\omega}{w}}{(x^{2})^{\frac{n}{2}}x^\beta h}
\D_\beta w^{-1}(0), 
\end{align}
where we set $ d_{0}=1 $.


\subsection{General Lorentz covariance}
If the distribution ${^{0}t}$ depends on more than one variable,
${^{0}t}x^{\alpha}$ is not symmetric in all Lorentz indices in
general.  Since $x^{\alpha}$ transforms like a tensor, it is natural
to generalize the discussion to the case, where ${^{0}t}$ transforms
like a tensor, too. Assume rank$( {^0t} )=r$, then $D(g)g$ is the
tensor representation of rank $p=r+n,n=|\alpha|$, in (\ref{def:a}).
From now on we omit the indices. So if $t\in\Dd(\MMmon)$, we
denote by \xbar\ -- formerly $x^{\alpha}$ -- a tensor of rank $n$
built of $x_{1},\dots,x_{m}$.

To solve \eqref{def:a} we proceed like in 
\cite{prep:wir}. Since the equation holds for all $g$ we solve 
for $a$ by using Lorentz transformations in the infinitesimal 
neighbourhood of $\eins$. If we take 
$\theta_{\alpha\beta}=\theta_{[\alpha\beta]} $ as six coordinates these transformations read:
\begin{equation}
g \approx\eins+\frac{1}{2}\theta_{\alpha\beta}l^{\alpha\beta},
\label{eq:gbyone}
\end{equation}
with the generators
\begin{equation}
{(l^{\alpha\beta})^{\mu}}_{\nu}
=\eta^{\alpha\mu}\delta^{\beta}_{\nu}-\eta^{\beta\mu}\delta^{\alpha}_{\nu}.
\label{def:generator}
\end{equation}
Then, for an infinitesimal transformation one
finds from \eqref{def:a}:
\begin{gather}
B^{\alpha\beta}\doteq 
2\biggl\langle{^0t},\xbar\sum_{j=1}^{m}
x^{[\alpha}_j\D^{\beta]}_j(V_{\omega-n}w)\biggr\rangle 
=(l^{\alpha\beta}\otimes\dots\otimes\eins+\dots 
+\eins\otimes\dots\otimes l^{\alpha\beta})a, 
\label{eq:atensor}
\end{gather}
$\alpha,\beta$ being Lorentz four-indices. In \cite{prep:wir}
our ability to solve that equation heavily relied on the given 
symmetry, which is in general absent here. Nevertheless we can find 
an inductive construction for $a$, corresponding to equation 
(29) in \cite{prep:wir}. 

We build one Casimir operator on the r.h.s. (the other one is always 
zero, since we are in a $(1/2,1/2)^{\otimes p}$ representation).
\subsubsection*{The case $p=1$} 
Just to remind that $p$ is the rank of $\xbar t$, this occurs if
either $t$ is a vector and $\xbar=1, (n=0)$, or $t$ is a scalar and
$\xbar=x_{1},\dots,x_{m}$.  (\ref{eq:atensor}) gives:
\begin{equation}
\frac{1}{2}l_{\alpha\beta}B^{\alpha\beta}
=\frac{1}{2}l_{\alpha\beta}l^{\alpha\beta}a 
=-3\eins a,
\label{eq:cas1}
\end{equation}
since the Casimir operator is diagonal in the irreducible $(1/2,1/2)$ 
representation. 

\subsubsection*{The case $p=2$}
We get
\begin{equation}
\frac{1}{2}(l_{\alpha\beta}\otimes\eins
+\eins\otimes l_{\alpha\beta})B^{\alpha\beta}
=(-6\eins+l_{\alpha\beta}\otimes l^{\alpha\beta})a.
\label{eq:acas2}
\end{equation}
Since $a$ is a tensor of rank 2, let us introduce the projector onto the
symmetric resp. antisymmetric part and the trace:
\begin{align}
{P_{S}^{\mu\nu}}_{\rho\sigma}
&=\frac{1}{2}(\delta^{\mu}_{\rho}\delta^{\nu}_{\sigma}
+\delta^{\nu}_{\rho}\delta^{\mu}_{\sigma}), &
{P_{A}^{\mu\nu}}_{\rho\sigma}
&=\frac{1}{2}(\delta^{\mu}_{\rho}\delta^{\nu}_{\sigma}
-\delta^{\nu}_{\rho}\delta^{\mu}_{\sigma}), \label{def:P} &
{P_\eta^{\mu\nu}}_{\rho\sigma}
&=\frac{1}{4}\eta^{\mu\nu}\eta_{\rho\sigma}, \\
P^2&=P, &
P_S+P_A&=\eins, &
P_S-P_A&=\tau, 
\end{align}
where $\tau$ denotes the permutation of the two indices. 
Using (\ref{def:generator}), we find
\begin{equation}
\frac{1}{2}l_{\alpha\beta}\otimes l^{\alpha\beta}=4P_{\eta}-\tau.
\label{eq:genten}
\end{equation}
Now we insert (\ref{eq:genten}) into (\ref{eq:acas2}). The trace part
is set to zero again. Acting with $P_A$ and $P_S$ on the resulting
equation gives us two equations for the antisymmetric and symmetric
part respectively. This yields:
\begin{equation}
a=-\frac{1}{16}(P_S+2P_A)(l_{\alpha\beta}\otimes\eins
+\eins\otimes l_{\alpha\beta})B^{\alpha\beta}.
\label{eq:a2}
\end{equation}

\subsubsection*{Inductive assumption}
Now we turn back to equation \eqref{def:a}. We note that any
contraction commutes with the (group) action on the rhs. Hence, if we
contract \eqref{eq:atensor}, we find on the rhs:
\begin{multline*}
\eta_{ij}(l^{\alpha\beta}\otimes\dots\otimes\eins+\dots 
+\eins\otimes\dots\otimes l^{\alpha\beta})a= \\
(l^{\alpha\beta}\otimes\dots\otimes\eins+\dots
+\not i+\dots+\not j+\dots
+\eins\otimes\dots\otimes l^{\alpha\beta})(\eta_{ij}a),
\end{multline*}
where $i,j$ denote the positions of the corresponding indices.
Therefore the rank of \eqref{eq:atensor} is reduced by two and we can
proceed inductively.  With the cases $p=1, p=2$ solved, we assume that
all possible contractions of $a$ are known.

\subsubsection*{Induction step}
Multiplying (\ref{eq:atensor}) with the generator and contracting the
indices yields:
\begin{multline}
\biggl(3p\eins+2\sum_{\tau\in S_p}\tau\biggr)a 
=-\frac{1}{2}(l_{\alpha\beta}\otimes\dots\otimes\eins+\dots 
+\eins\otimes\dots\otimes l_{\alpha\beta})B^{\alpha\beta}
+8\sum_{i<j\leq p}{P_\eta}_{ij}a.
\label{eq:acasp}
\end{multline}
The transposition $\tau$ acts on $a$ by permutation of the
corresponding indices. For a general $\pi\in S_p$ the action on $a$ is
given by: $\pi
a^{\mu_1\dots\mu_p}=a^{\mu_{\pi^{-1}(1)}\dots\mu_{\pi^{-1}(p)}}$. In
order to solve this equation we consider the representation of the
symmetric groups. We give a brief summary of all necessary ingredients
in appendix~\ref{app:permrep}. So let $k_\tau\doteq\sum_{\tau\in
  S_p}\tau$ be the sum of all transpositions of $S_p$.  Then $k_\tau$ is
in the center of the group algebra $\Acal_{S_p}$.  It can be
decomposed into the idempotents $e_{(m)}$ that generate the irreducible
representations of $S_p$ in $\Acal_{S_p}$.
\begin{equation}
k_\tau=h_\tau\sum_{(m)}\frac{1}{f_{(m)}}\chi_{(m)}(\tau)e_{(m)}.
\label{eq:decompktau}
\end{equation}
The sum runs over all partitions $(m)=(m_1,\dots,m_r), \sum_{i=1}^r
m_i=p, m_1\geq m_2\geq\dots\geq m_r$ and $h_\tau=\frac{1}{2} p(p-1)$
is the number of transpositions in $S_p$. $\chi_{(m)}$ is the
character of $\tau$ in the representation generated by $e_{(m)}$ which
is of dimension $f_{(m)}$. We use \eqref{eq:decompktau}, the
orthogonality relation $e_{(m)}e_{(m')} =\delta_{(m)(m')}$ and the
completeness $\sum_{(m)}e_{(m)}=\eins$ in (\ref{eq:acasp}). The
expression in brackets on the l.h.s may be orthogonal to some
$e_{(m)}$. The corresponding $e_{(m)}a$ contribution is any
combinations of $\eta$'s and $\epsilon$'s --$\epsilon$ being the
totally antisymmetric tensor in four dimensions -- transforming
invariantly and thus can be set to zero. We arrive at
\begin{multline}
a=\sum_{\substack{(m)\\c(m)\not=0}}\frac{e_{(m)}}{c(m)}
\left(
-\frac{1}{2}(l_{\alpha\beta}\otimes\dots\otimes\eins+\dots 
+\eins\otimes\dots\otimes l_{\alpha\beta})B^{\alpha\beta}
+8\sum_{i<j\leq p}{P_\eta}_{ij}a
\right), \\
c(m)\doteq 3p+p(p-1)\frac{\chi_{(m)}(\tau)}{f_{(m)}}
=3p+\sum_{i=1}^r\left(
b_i^{(m)}(b_i^{(m)}+1)-a_i^{(m)}(a_i^{(m)}+1)
\right),
\label{eq:atenrec}
\end{multline}
with $a=(a_1,\dots,a_r), b=(b_1,\dots,b_r)$ denoting the
characteristics of the frame $(m)$, see appendix~\ref{app:permrep}. 
Let us take $p=4$ as an example:
\[
\begin{array}{cccc}
\text{idempotent} & \text{Young frame} & \text{dimension} & 
\text{character} \\
& & & \\
e_{(4)}       & \yng(4)       & f_{(4)}=1       & \chi_{(4)}(\tau)=1  \\
& & & \\
e_{(3,1)}     & \yng(3,1)     & f_{(3,1)}=3     & \chi_{(3,1)}(\tau)=1  \\
& & & \\
e_{(2,2)}     & \yng(2,2)     & f_{(2,2)}=2     & \chi_{(2,2)}(\tau)=0  \\
& & & \\
e_{(2,1,1)}   & \yng(2,1,1)   & f_{(2,1,1)}=3   & \chi_{(2,1,1)}(\tau)=-1 \\
& & & \\
e_{(1,1,1,1)} & \yng(1,1,1,1) & f_{(1,1,1,1)}=1 & \chi_{(1,1,1,1)}(\tau)=-1
\end{array}
\]
We find for (\ref{eq:acasp})
\begin{equation}
a=\frac{1}{48}(2e_{(4)} + 3e_{(3,1)} +4e_{(2,2)} +6e_{(2,1,1)})\times
\text{r.h.s}(\ref{eq:acasp}).
\end{equation}
We see that no $e_{(1,1,1,1)}$ appears in that equation. It corresponds to the one dimensional
$sgn$-representation of $S_4$, so $e_4a\propto\epsilon$.

\subsection{Spinorial Lorentz covariance}
\label{subsec:spinor}
This subsection uses the conventions of \cite{bk:sexl-urb}.  Any
finite dimensional representation of $\Lcal_{+}^{\uparrow}$ can be
reduced to tensor products of \SLC\ and $\overline{\SLC}$ and direct
sums of these.  A two component spinor $\Psi$ transforms according to
\begin{equation}
\Psi^{A}={g^{A}}_{B}\Psi^{B},
\label{eq:spintrafo}
\end{equation}
where $g$ is a $2\times2$-matrix in the \SLC\ representation of
$\Lcal_{+}^{\uparrow}$.  For the complex conjugated representation we
use the dotted indices, i.e.
\begin{equation}
\overline{\Psi}^{\Xdot}={\overline{g}^{\Xdot}}_{\Ydot}\overline{\Psi}^{\Ydot},
\label{eq:ccspintrafo}
\end{equation}
with $ {\overline{g}^{\Xdot}}_{\Ydot}=\overline{{g^{X}}_{Y}} $ in the $ 
\overline{\SLC} $ representation.  The indices are lowered and raised 
with the $ \epsilon $-tensor.
\begin{gather}
\epsilon_{AB}=\overline{\epsilon}_{\Adot\Bdot}
\doteq\epsilon_{\Adot\Bdot}, \label{eq:eps} \\
\epsilon^{AB}\epsilon_{AC}
=\epsilon^{BA}\epsilon_{CA}
=\delta^{B}_{C}.
\end{gather}
We define the Van-der-Waerden symbols with the help of the Pauli 
matrices $\sigma_{\mu}$ and ${\widetilde{\sigma}}_{\mu}\doteq\sigma^{\mu}$:
\begin{align}
{\sigma_{\mu}}^{A\Xdot}&\doteq\frac{1}{\sqrt{2}}(\sigma_{\mu})^{AX}, &
{\sigma_{\mu}}_{A\Xdot}&\doteq\frac{1}{\sqrt{2}}
({{\widetilde{\sigma}}_{\mu}}^{T})_{AX}.
\label{def:vdw}
\end{align}
They satisfy the following relations
\begin{align}
{\sigma_{\mu}}^{A\Xdot}{\sigma_{\nu}}_{A\Xdot}
&=\eta_{\mu\nu} &
{\sigma_{\mu}}_{A\Xdot}{\sigma^{\mu}}_{B\Ydot}
&=\epsilon_{AB}\epsilon_{\Xdot\Ydot}
\label{eq:vdw}
\end{align}
With the help of these we can build the infinitesimal spinor
transformations
\begin{equation}
g\approx\eins+\frac{1}{2}\theta_{\alpha\beta}S^{\alpha\beta},
\label{eq:spininf}
\end{equation}
with the generators
\begin{equation}
{(S^{\alpha\beta})^{A}}_{B}
={\sigma^{[\alpha}}^{A\Xdot}{\sigma^{\beta]}}_{B\Xdot}.
\label{eq:spingen}
\end{equation}
Note that the $ \sigma $'s are hermitian: $
\overline{{\sigma_{\mu}}^{A\Xdot}}={\sigma_{\mu}}^{X\Adot} $.  Again
we define the projectors for the tensor product. But we have only two
irreducible parts:
\begin{align}
{{P_S}^{AB}}_{CD}&=\frac{1}{2}
(\delta^A_C\delta^B_D+\delta^A_D\delta^B_C), &
{{P_\epsilon}^{AB}}_{CD}&=\frac{1}{2}
\epsilon^{AB}\epsilon_{CD}, \\
P^2_S&=P_S,\quad P^2_\epsilon=P_\epsilon, & 
P_S+P_\epsilon&=\eins.
\end{align}
We get the following identities:
\begin{align}
S^{\alpha\beta}S_{\alpha\beta}
&=\overline{S}^{\alpha\beta}\overline{S}_{\alpha\beta}
=-3\eins, \label{eq:sgen1} \\
S^{\alpha\beta}\otimes S_{\alpha\beta}
&=\overline{S}^{\alpha\beta}\otimes\overline{S}_{\alpha\beta}
=4P_\epsilon-\eins \label{eq:sgen2}, \\
S^{\alpha\beta}\otimes\overline{S}_{\alpha\beta}
&=\overline{S}^{\alpha\beta}\otimes S_{\alpha\beta}
=0. \label{eq:sgen3}
\end{align}

In order to have (\ref{def:a}) in a pure spinor representation we have
to decompose the tensor $\xbar$ into spinor indices according to
\begin{equation}
x^{A\Xdot}\doteq x^\mu{\sigma_{\mu}}^{A\Xdot}.
\end{equation}
Assume $t\widetilde{x}$ transforms under the $u$-fold tensor product of
\SLC\ times the $v$-fold tensor product of $\overline{\SLC}$ then,
for infinitesimal transformations, (\ref{def:a}) yields:
\begin{equation}
B^{\alpha\beta}
=(S^{\alpha\beta}\otimes\dots\otimes\eins+\dots
+\eins\otimes\dots\otimes\overline{S}^{\alpha\beta})a,
\end{equation}
with $B^{\alpha\beta}$ from equation \eqref{eq:atensor} in the
corresponding spinor representation.  The sum consists of $u$ summands
with one $S^{\alpha\beta}$ and $v$ summands with one
$\overline{S}^{\alpha\beta}$ with $u,v>n$.  Multiplying again with the
generator and contracting the indices gives twice the Casimir on the
r.h.s. Inserting (\ref{eq:sgen1}-\ref{eq:sgen3}) yields:
\begin{multline}
(S_{\alpha\beta}\otimes\dots\otimes\eins+\dots
+\eins\otimes\dots\otimes\overline{S}_{\alpha\beta})B^{\alpha\beta} \\
=\biggl(
-3(u+v)\eins
+2\sum_{1\leq i<j\leq u}(4P_{\epsilon_{ij}}-\eins) +
2\sum_{1\leq i<j\leq v}(4P_{\overline{\epsilon}_{ij}}-\eins) 
\biggr)a.
\end{multline}
The sum over $u$ runs over $\frac{u}{2}(u-1)$ possibilities and
similar for $v$, so we find the induction:
\begin{multline}
a=\frac{1}{u(u+2)+v(v+2)}\Biggl[
-(S_{\alpha\beta}\otimes\dots\otimes\eins+\dots
+\eins\otimes\dots\otimes\overline{S}_{\alpha\beta})B^{\alpha\beta}+
\\
8\biggl(
\sum_{1\leq i<j\leq u}P_{\epsilon_{ij}} 
+\sum_{1\leq i<j\leq v}P_{\overline{\epsilon}_{ij}} 
\biggr)a \Biggr].
\label{eq:spinind}
\end{multline}
It already contains the induction start for $a^{(AB)}, a^{(XY)}$ and 
$a^{A\Xdot}$. 

\subsection{General covariant BPHZ subtraction}
\label{subsec:bphz}
\sloppy
We have derived a Lorentz covariant renormalization that applies for a
general choice of $w$. Therefore, choosing $e^{ip\cdot}$ as
test function provides for a covariant renormalization in momentum
space. The choice $w=e^{iq\cdot}$ corresponds to subtraction at
momentum $q$ \cite{pap:prange1}. Hence $q=0\Leftrightarrow w=1$
represents BPHZ subtraction.

\fussy
But this choice leads to infrared divergencies in massless
theories. Lowenstein and Zimmermann \cite{pap:low-zim1} have introduced
an alternative scheme (called BPHZL) that makes use of an auxiliary
mass and requires additional subtractions with respect to a parameter
which scales this mass. This also produces a covariant renormalization
in momentum space. We compare their results with ours in two examples.

We shrink the distribution space to $\Sd'$ since we are 
dealing with Fourier transformation. Let $x,q,p\in\MM^m$, 
\begin{align}
\widehat{\tR{\omega}{e^{iq\cdot}}}(p)
&\doteq\scp{\tR{\omega}{e^{iq\cdot}}} {e^{ip\cdot}} \\
&=\scp{^{0}t}{e^{ip\cdot}-\sum_{|\alpha|\leq\omega} 
\frac{(p-q)^\alpha}{\alpha!}\D^q_{\alpha}e^{iq\cdot}}.
\label{eq:BPHZ}
\end{align}
It is normalized at the subtraction point $q$, i.e.:
$\D_{\alpha}\widehat{\tR{\omega}{e^{iq\cdot}}}(q) = 0,\ |\alpha| \leq
\omega$.  This is always possible for $q$ totally space like,
$(\sum_{j\in I}q_{j})^2<0, \forall I\subset\{1,\dots,m\}$
\cite{pap:ep-gl,priv:duetsch}. If we use the results from above we obtain 
a covariant
BPHZ subtraction at momentum $q$ by adding
$\sum_{|\alpha|\leq\omega} \frac{i^{|\alpha|}}{\alpha!} a^\alpha
p_\alpha$ to (\ref{eq:BPHZ}), according to equation \eqref{def:tRGcov}.
For $|\beta|\geq\omega+1$, ${^0t}x^\beta$ is a well defined distribution
on $\Sd$ and so is $\D_\beta\widehat{^0t}$.

\subsubsection{Lorentz invariance on $\MM$} 
We have
\begin{align}
V_k e^{iq\cdot}&=e^{iq\cdot}\sum_{m=0}^k\frac{1}{m!}(-ipx)^m, & 
\D_\sigma V_k e^{iq\cdot} &=iq_\sigma e^{iq\cdot}\frac{1}{k!}(-ipx)^k.
\end{align}
Inserting this into (\ref{eq:amun}) we find:
\begin{multline}
a^{(\mu _1\dots\mu_n)}= \\
=\frac{i^n(-)^{\omega+1}}{(\omega-n)!}\frac{ (n-1)!!}{(n+2)!!}
q_{\sigma_1}\dots q_{\sigma_{\omega-n}} 
\sum_{s=0}^{\left[\frac{n-1}{2} \right] }
\frac{(n-2s)!!}{(n-2s-1)!!}
\left(q_\rho\D^\rho\D^{(\mu_1}-q^{(\mu_1}\square\right)
\times \\ 
\times
\eta^{\mu_2\mu_3}\dots\eta^{\mu_{2s}\mu_{2s+1}} 
\D^{\mu_{2s+2}}\dots\D^{\mu_n)}\square^s
\D^{\sigma_1}\dots\D^{\sigma_{\omega-n}}\widehat{{^0t}}(q).
\label{eq:amunq}
\end{multline}
Let us consider two examples namely the fish and the setting sun from
massless scalar field theory. We compare the results to BPHZL
\cite{pap:low-zim1}. 
\begin{exa}
  The fish graph corresponds to $^{0}t=\frac{i^{2}}{2}D_{F}^{2}
  \Rightarrow \omega=0$. It needs no counter term, since it is Lorentz
  invariant. We translate the result of BPHZL to coordinate space:
  \begin{equation}
    \label{eq:fishBPHZL}
    \widehat{\left(D^F\right)^2_{\text{BPHZL}}}(p)
=\scp{\left(D^F\right)^2-\left(\Delta^F\right)^2}{e^{ip\cdot}}
+\scp{\left(\Delta^F\right)^2}{\W{0}{1}e^{ip\cdot}}.
  \end{equation}
\end{exa}
\begin{exa}
Take the setting sun in massless scalar field theory: 
$^{0}t=\frac{i^{3}}{6}D_{F}^{3}$ $\Rightarrow \omega=2$.
\begin{align*}
a^\mu&=-\frac{i}{3}(q_\sigma q_\rho \D^\rho \D^\sigma \D^\mu
-q^\mu q_\sigma\D^\sigma\square)\widehat{^{0}t}(q), \\
a^{\mu\nu}&=\frac{1}{4}(q_\rho \D^\rho \D^\mu \D^\nu 
-q^{(\mu}\D^{\nu)}\square)\widehat{^{0}t}(q),
\end{align*}
and adding $ip_{\mu}a^{\mu} - \frac{1}{2}p_{\mu}p_{\nu}a^{\mu\nu}$ 
restores Lorentz invariance of the setting sun graph subtracted at
$q$. Renormalization according to BPHZL gives:
\begin{align}
    \widehat{\left(D^F\right)^3_{\text{BPHZL}}}(p)
&=\scp{\left(D^F\right)^3-\left(\Delta^F\right)^3}
{\W{1}{1}e^{ip\cdot}}+
\notag \\
&\quad+\scp{\left(\Delta^F\right)^3}{\W{2}{1}e^{ip\cdot}}.
    \label{eq:setsunBPHZL}
\end{align}
\end{exa}
\subsection{General induction}
We only have to evaluate $B^{\alpha\beta}$ with $w=e^{iq\cdot}$ and
plug the result into the induction formulas \eqref{eq:spinind} and
\eqref{eq:atenrec}.
\begin{equation}
B^{\alpha\beta}
=2i^{n}(-)^{\omega+1}\sum_{j=1}^{m}\sum_{|\gamma|=\omega-n}
\frac{q^{\gamma}}{\gamma!}q_{j}^{[\alpha}\D_{j}^{\beta]}
\D_{\gamma}\widetilde{\D}\,\widehat{^{0}t}(q).
\end{equation}
Here, $q_{j}$ are the $m$ components of $q$ hence $\gamma$ is a $4m$
index and $\alpha,\beta$ are four indices. The tensor (spinor)
structure of $\widetilde{\D}$ is given by $\widetilde{x}$ in
\eqref{eq:atensor}.
 

\chapter{Local perturbative interacting fields}
\label{chap:intfields}


In the previous chapters we have introduced free quantum fields, Wick
products of these fields and finally time ordered products of 
Wick polynomials. It turns out that this provides a complete frame for 
the definition of interacting (perturbative) fields.%
\footnote{Here and in the following the name ``field'' also includes 
  composed objects, which are Wick polynomials in the free case.}

The $S$-matrix serves as the generating functional for time ordered
products of the interaction by smearing with a test function of compact
support. After coupling additional fields into the $S$-matrix one
obtains interacting fields and time ordered products of them
by Bogoliubov's formula \cite{bk:bogol}.

This situation provides a setting for the introduction of local
observable algebras: We choose a causally complete bounded spacetime
region $\Ocal$ on which the coupling is assumed to be constant. It
was shown by Brunetti and Fredenhagen \cite{proc:brun-fred} that the
interacting fields on this region only change by a unitary
transformation if the interaction is changed outside $\Ocal$.
Therefore, algebraic relations are left invariant.

Let us emphasize that the occurrence of IR-singularities is a
consequence of performing the \emph{adiabatic limit} where the test
function of the interaction tends to a constant. Since we avoid this
limit it is always possible to construct the local algebras.
Especially for asymptotic free theories like Yang-Mills for example
\cite{phd:boas} this is an advantage. Here perturbation theory can be
regarded as a valid approximation only for short distances. The
involved fields on these scales do not correspond to 
asymptotic particles that can be observed in experiments.

Although our work only deals with perturbative fields we remark that
Bogoliubov's formula for interacting fields also applies for the non
perturbative case. 


\section{The $S$-matrix}
For a symbol $\Ll\in\Ba$ describing the interaction
$\wick{\Ll} \in \Dist_1(\Do)$ of our quantum theory
we build the $S$-matrix:
\begin{equation}
  \label{def:smatrix}
  S(g\Ll)
=\sum_{n=0}^\infty \frac{i^n}{n!}
\int\dif x_1\dots \dif x_n\, T(g\Ll,\dots,g\Ll)(x_1,\dots,x_n), 
\end{equation}
where we allowed for the use of $g\Ll$ as an argument in the
$T$-product as explained in \Peins. We also consider the case of a sum
of different couplings, such that $g\Ll=g\cdot\Ll=\sum_i^s g_i\Ll_i$.
The $S$-matrix is the generating functional for all $T$-products of
the $\Ll_i$:
\begin{equation}
  \label{eq:functionalS}
  T\left(\Ll_{i_1},\dots,\Ll_{i_n}\right)(x_1,\dots,x_n)
=\frac{\delta^n}{i^n\delta g_{i_1}(x_1)\dots\delta g_{i_n}(x_n)}
S(g\Ll)\Bigr\vert_{g_1=\dots=g_s=0}.
\end{equation}
The inverse of $S$ is given by 
\begin{equation}
  \label{def:invsmatrix}
  S(g\Ll)^{-1}=\sum_{n=0}^\infty\frac{(-i)^n}{n!}
\int\dif x_1\dots\dif x_n\, \Tbar(g\Ll,\dots,g\Ll)(x_1,\dots,x_n)
\end{equation}
which follows from \eqref{eq:TmalTbar} by the inversion of a formal
power series. Then $S$ is a unitary operator on $\Do$:
\begin{equation}
  \label{eq:unitaryS}
  S(g\Ll)^{-1}=S(g\Ll)^*,
\end{equation}
because of normalization condition \Nzwei. Obviously $S^{-1}$ is the
generating functional of the $\Tbar$-products:
\begin{equation}
  \label{eq:functionalSinv}
\Tbar\left(\Ll_{i_1},\dots,\Ll_{i_n}\right)(x_1,\dots,x_n)
=\frac{\delta^n}{(-i)^n\delta g_{i_1}(x_1)\dots\delta g_{i_n}(x_n)}
S(g\Ll)^{-1}\Bigr\vert_{g_1=\dots=g_s=0}.
\end{equation}
Because of the causal factorization of the $T$-products the $S$-matrix 
fulfils the following causal factorization:
\begin{equation}
  \label{eq:factorS}
  S(fW+gL+hV)=S(fW+gL)S(gL)^{-1}S(gL+hV), 
\end{equation}\sloppy
$\forall f,g,h \in\Dd(\MM), W,L,V \in \Ba$, if $\supp f \cap
\blc(\supp h) =\emptyset$ as was shown in
\cite{pap:ep-gl}. Especially in the case $g=0$ this leads to
\begin{equation}
  \label{eq:Scausal}
  S(fW+hV)=S(fW)S(hV),
\end{equation}
which is the causal factorization of the $T$-products lifted to the
functionals. 

\fussy
\section{Interacting fields}
\label{sec:intfields}
We now couple additional source terms into the local $S$-matrix
defined above. This generates a new functional called the
\emph{relative} $S$-matrix according to:
\begin{equation}
  \label{def:relativeS}
  \Sg(hW)\doteq S(g\Ll)^{-1}S(g\Ll+hW).
\end{equation}
The relative $S$-matrix also satisfies the causal factorization
equations \eqref{eq:factorS},\eqref{eq:Scausal}. Especially the latter 
one reads 
\begin{equation}
\Sg(h_1W_1+h_2W_2)=\Sg(h_1W_1)\Sg(h_2W_2)=\Sg(h_2W_2)\Sg(h_1W_1),
\label{eq:relScommute}
\end{equation}
if $\supp h_1 \sim \supp h_2$ and $W_1,W_2\in\Ba$. Hence the relative
$S$-matrices commute for space like coupled sources
\cite{pap:brun-fred2} and may therefore serve as generating functionals
for local fields. Then the \emph{interacting field} $\Wg$
corresponding to the symbol $W\in\Ba$ is defined by Bogoliubov's
formula \cite{bk:bogol,pap:ep-gl}:
\begin{equation}
  \label{def:Wg}
  \Wg(x)\doteq\frac{\delta}{i\delta h(x)}
\Sg(hW)\Bigr\vert_{h=0}.
\end{equation}
Expanding the interacting field \eqref{def:Wg} into a power series
results in
\begin{equation}
  \label{eq:Wg}
  \Wg(x)=\sum_{n=0}^\infty\frac{i^n}{n!}
\int \dif y_1\dots\dif y_n R(g\Ll,\dots,g\Ll;W)(y_1,\dots,y_n;x),
\end{equation}
where the \emph{retarded products} ($R$-products) are given by
\begin{align}
  \label{def:R}
  R(W_1,\dots,\W_n;W)(y_1,\dots,y_n;x) &\doteq \sum_{I\subset
    N}(-)^{|I|}\Tbar(I)(y_I)T(I^c,W)(y_{I^c},x).
  \\
  \intertext{With the help of equation \eqref{eq:TmalTbar}, $x$ can
    also be put in the $\Tbar$-products, yielding} 
  &=\sum_{I\subset N}(-)^{|I|}\Tbar(I,W)(y_I,x)T(I^c)(y_{I^c}).
\end{align}
The word retarded encodes the support properties of the $R$-products:
\begin{multline}
  \label{eq:suppR}
  \supp R(W_1,\dots,W_n;W)(y_1,\dots,y_n;x) \subset\\
\subset \{(y_1,\dots,y_n,x)\in 
\MM^{n+1}, y_i\in\blc(x), \forall i=1,\dots,n\}.
\end{multline}
This can be seen immediately from the causality property \Pdrei\
\cite{pap:ep-gl}. 


\subsection{Properties of the interacting fields}
\label{subsec:propertyintfield}
The properties of the time ordered products \Peins\ -- \Pvier\ and the
normalization conditions \Neins\ -- \Nvier\ have immediate
consequences for the interacting fields defined above. Because of
\eqref{eq:TmalTbar} and the definition of $R$ we find that 
\begin{equation}
  \label{eq:einsgl}
  \eins_{g\Ll}=\eins.
\end{equation}
Since the interaction $\Ll$ is assumed to be a Lorentz scalar, the
normalization condition \Neins\ implies the conservation of the
Poincar\'e transformation properties:
\begin{equation}
  \label{eq:poincareintfield}
  (\Ad U(L))\Wg(x)=\K{D\K{\Lambda^{-1}}(W)}_{Lg\Ll}(Lx), 
\forall L=(a,\Lambda)\in\Poinc,
\end{equation}
where $D$ is the representation of $\Lor$ according to
\eqref{eq:boaslorentz1},\eqref{eq:boaslorentz2}. $\Poinc$ acts on
$\Dd(\MM)$ as a group homomorphism according to $Lf(x) = f(L^{-1}x),
\, f \in \Dd(\MM)$.
The $^*$-involution on the interacting fields is given by
\begin{equation}
  \label{eq:intfieldstar}
  (\Wg)^*=\K{W^*}_{g\Ll},
\end{equation}
where on the \lhs\ $^*$ is the adjoint on $\Do$ and on the \rhs\ it is
given by \eqref{eq:boasstar},\eqref{eq:boasstargen}. Since the
generating functionals commute for spacelike separated sources
according to \eqref{eq:relScommute} the interacting fields are local:
\begin{equation}
  \label{eq:intlocal}
  \left[\Wg(x),V_{g\Ll}(y)\right]=0,\text{ if } x\sim y.
\end{equation}

Normalization condition \Nvier\ further implies the interacting
equations of motion for the interacting
basic generator.
\begin{equation}
  \label{eq:intmotion}
  D_{ij}\vp_{j\,g\Ll}=-\K{\parz{\Ll}{\vp_i}}_{g\Ll},\quad
\vp_j\in\Ga_b.
\end{equation}

The commutator of two interacting fields was calculated in
\cite{pap:duet-fred1}:
\begin{multline}
  \label{eq:intcomm}
\left[\Wg(x),V_{g\Ll}(y)\right]=
-\sum_{n=0}^\infty\frac{i^n}{n!}\int
\dif y_1\dots\dif y_n \Bigl(R\K{g\Ll,\dots,g\Ll,W;V}\K{y_N,x;y}+\\
-R\K{g\Ll,\dots,g\Ll,V;W}\K{y_N,y;x}\Bigr).
\end{multline}

\subsection{Iterating the interaction}
\label{subsec:iterate}

Bogoliubov's formula not only defines the interacting fields it
also provides for well defined time ordered products of these by multiple
functional differentiation \cite{pap:ep-gl}. We have
\begin{multline}
  T\Kg{W_1,\dots,W_n}(x_1,\dots,x_n)=\\
  \begin{split}
    &=\frac{\delta^n}{i^n\delta h_1(x_1)\dots\delta h_n(x_n)}
    \Sg\left(\sum_{j=1}^n h_j
      W_j\right)\biggr\vert_{h_1=\dots=h_n=0} \\
\label{eq:Tint}
&=\sum_{m=0}^\infty\frac{i^m}{m!}  
\int \dif y_1\dots \dif y_m \times \\
&\quad\times
R(g\Ll,\dots,g\Ll;W_1,\dots,W_n) (y_1,\dots,y_m;x_1,\dots,x_n).
\end{split}
\end{multline}
The retarded products are given by
\begin{equation}
\label{def:Rn}
  R(V_1,\dots,V_m;W_1,\dots,W_n)(y_M;x_N)
\sum_{I\subset M}(-)^{|I|}\Tbar(I)(y_I)T(I^c,N)(y_{I^c},x_N),
\end{equation}
where we used our short hand notation, denoting $N=\{1,\dots,n\}$ and
$M=\{1,\dots,m\}$. They have retarded support, too:
\begin{multline}
  \label{eq:suppRn}
\supp R(V_1,\dots,V_m;W_1,\dots,W_n)(y_1,\dots,y_m;x_1,\dots,x_n)
\subset \\ 
\subset
\biggl\{(y_1,\dots,y_m,x_1,\dots,x_n)\in 
\MM^{m+n}, y_i\in\bigcup_{j=1}^n\blc(x_j), \forall i=1,\dots,m
\biggr\}.
\end{multline}
Following \cite{prep:duet-fred2} we now study the iteration of the
causal construction. We build the $S$-matrix of a new interaction
$\Klg$ as a formal power series in $h\in\Dd(\MM)$:
\begin{align}
  \label{def:intsmatrix}
  S\left(h\Klg\right)&=\sum_{n=0}^\infty\frac{i^n}{n!}
  \int\dif x_1\dots\dif x_n\, T\Kg{h\Kl,\dots,h\Kl}(x_1,\dots,x_n), \\
  \intertext{which is according to \eqref{eq:Tint} and the definition
    of a generating functional} &=\Sg(h\Kl).
\end{align}
Then the corresponding relative $S$-matrix is given by
\begin{align}
  S_{h\Klg}(f\Wg)&=S(h\Klg)^{-1}S(h\Klg+f\Wg)\\
&=\Sg(h\Kl)^{-1}\Sg(h\Kl+fW)\\
&=S(g\Ll+h\Kl)^{-1}S(g\Ll+h\Kl+fW)\\
&=S_{g\Ll+h\Kl}(fW). \label{eq:Sintrelative}
\end{align}
This allows to define the interacting $(\Wg)_{h\Klg}$-field and
$T$-products of them, corresponding to the field $\Wg$ by the
Bogoliubov formula:
\begin{align}
  (\Wg)_{h\Klg}(x)&=\frac{\delta}{i\delta f(x)} 
S_{h\Klg}(f\Wg)\Bigr\vert_{f=0}\\
&=W_{g\Ll+h\Kl}(x),
\end{align}
because of \eqref{eq:Sintrelative}. Especially in the case $\Kl=\Ll$
and $h=-g$ the interacting interacting field just returns the free 
field.

Now we introduce the general setting which is used in the following
chapters. The interaction Lagrangian $\Ll$ is assumed to
lead to a renormalizable quantum field theory, hence $\dim \Ll \leq
4$. Our test function%
\footnote{If we also consider a sum of couplings  $g$ becomes a
  vector.}  
$g$ is assumed to be constant on an open bounded causally
complete spacetime region $\Ocal$, see figure~\ref{fig:setting}. 
\begin{figure}[h]
  \begin{center}
    \input{settingklein.pstex_t}
    \caption{Observable algebra $\Aa(\Ocal)$.}
    \label{fig:setting}
  \end{center}
\end{figure}

Then we construct the interacting fields $\Wg$ in that region
according to
\begin{equation}
  \label{eq:intfieldO}
  \Wg(f) \forall f\in\Dd(\MM)\text{ with } \supp f\subset\Ocal,
W\in\Ba.
\end{equation}
The algebra of these fields is our observable algebra $\Aag(\Ocal)$.
It was shown in \cite{proc:brun-fred} that any change of the
interaction outside the closure of $\Ocal$ leads to a unitary
transformation of $\Sg(W)$ independent of $W$ and hence of all interacting
fields. Therefore the interaction fixes $\Aag(\Ocal)$ up to unitary
equivalence.


\chapter{The energy momentum tensor}
\label{chap:emt}

In classical field theory any symmetry of the Lagrangian generates a
conserved current by the Noether procedure. If the Lagrangian is not
invariant but only shifts by a divergence the same procedure still
applies. The current which is associated to a translation of the
fields is the energy momentum tensor (\emt). If we have a localized
interaction translation invariance is obviously broken and the \emt\ is
conserved only where the localization function is constant.  But this
is already enough in view of our interacting observable algebras.

The Lagrangian possesses a further symmetry, namely scale invariance,
if no dimensionful couplings are present. Callan, Coleman and Jackiw
have shown \cite{pap:callan1} that in this situation an
\emph{improved} \emt\ can be defined by addition of a conserved
improvement tensor. This improved tensor is also traceless.
Contraction of this tensor with $x$ defines the conserved
\emph{dilatation} current reflecting scale invariance of the
Lagrangian.

We pursue another way of defining the improved tensor on the example
of the massless scalar field theory: The equations of motion do not
fix the Lagrangian unambiguously. We find the improved tensor as the
\emt\ of an improved Lagrangian. A similar derivation was also given
by Kasper \cite{pap:kasper}. Since the improvement tensor is strictly
conserved, the improved \emt\ is only conserved up to the breaking
term of the canonical one, related to the localization of the
interaction. This term also causes a breaking of the dilatation
current.  We discuss this classical situation in
section~\ref{sec:canemtcl}.

With the classical preliminaries we study the quantum theory.
In section~\ref{sec:canemtq} we analyze the canonical \emt\ for a
family of theories, where the free field equation is at least of
second order and the interaction contains no derivated fields. We find
that exactly the same conservation equation like for the classical
fields can be fulfiled if we impose a further normalization condition
called \emph{Ward identity}. The Ward identity requires a suitable
normalization of $T$-pro\-ducts involving the canonical free energy
momentum tensor (as a symbol $\in\Ba$).  We show that the Ward
identity can always be satisfied in section~\ref{sec:ward1} by using
the inductive method of D\"utsch and Fredenhagen in
\cite{pap:duet-fred1}.  Our result coincides with a similar result
that was derived for the energy momentum tensor in the framework of
Zimmermann's normal product quantization \cite{pap:zimmermann1} by
Lowenstein \cite{pap:low3} and also by Zimmermann
\cite{proc:zimmermann}. We show that the interacting momentum operator
(as the corresponding charge) implements the right commutation
relation with the interacting fields.

In their paper \cite{pap:coleman} Coleman and Jackiw argued that the
trace of the improved energy momentum tensor generates an anomaly in
the perturbative interacting quantum theory. Lowenstein has verified
this statement for Zimmermann's normal products in \cite{pap:low3}.
Later, Zimmermann has given a derivation of this anomaly in
\cite{proc:zimmermann}. He verified a conjecture by Minkowski
\cite{pap:minkowski} that it can be normalized to be proportional to
the $\beta$-function of the Callan-Symanzik equation. This statement
already contains a result of Schroer \cite{pap:schroer1} that the
anomaly vanishes if the coupling is a zero of the
$\beta$-function. A more comprehensive result, also covering possible
conformal anomalies, was given by Kraus and Sibold
\cite{pap:kraus-sibold1,pap:kraus-sibold3} using the framework of
algebraic renormalization.

In accordance to these results we show in section~\ref{sec:impemtq}
that a conserved (up to the expected $\D g$ breaking) improved \emt\ 
inevitably leads to the trace anomaly in $\vp^4$-theory. The
definition of the improvement tensor requires a new relation between
interacting fields induced by a corresponding Ward identity which is
proved in section~\ref{sec:ward2}. The simultaneous validity of both
Ward identities forces the trace anomaly to be present. The \emt\ and
its improved counterpart both define the same momentum operator. The
breaking of dilatations given by the trace leads to anomalous
contributions to the dimension of the interacting fields described in
section~\ref{sec:anomal}. We mention that dilatations were also quite
recently studied in local perturbation theory by Grigore
\cite{pap:grigore1}. But his main focus is on the $S$-matrix whereas
we focus on the interacting fields.


\section{The energy momentum tensor in classical field theory}
\label{sec:canemtcl}
We discuss the \emt\ in a classical field theory. The Lagrangian 
depends on the fields $\phicl_j, j=1,\dots,r$ and their first and
second derivatives. We assume that it also depends on $x$ explicitely
via a coupling term $-g\Llcli$ which is assumed to contain no
derivated fields:
$\Llcl=\Llcl(\phicl_l,\phicl_{l,\mu},\phicl_{l,\mu\nu},x)$. Then the
Euler-Lagrange equations read:
\begin{equation}
  \label{eq:eulerlagrange}
  \D_\mu\D_\nu\parz{\Llcl}{\phicl_{l,\mu\nu}}
-\D_\mu\parz{\Llcl}{\phicl_{l,\mu}}
+\parz{\Llcl}{\phicl_l}=0.
\end{equation}
\sloppy
The \emt\ is the current associated to a spacetime translation of the
fields: \mbox{$\phicl(x)\rightarrow\phicl(x+a)$}.
By the Noether procedure we find the \emt\ to be:
\begin{align}
  \Tmncl&=\parz{\Llcl}{\phicl_{l,\mu}}\D^\nu\phicl_l
-\K{\D_\rho\parz{\Llcl}{\phicl_{l,\mu\rho}}}\D^\nu\phicl_l+
\notag\\
&\quad+\parz{\Llcl}{\phicl_{l,\mu\rho}}\D_\rho\D^\nu\phicl_l
-\eta^{\mu\nu}\Llcl.
  \label{def:emtclass}
\end{align}
\fussy
In this situation the ``conservation'' equation reads:
\begin{equation}
\D_\mu\Tmncl=\D^\nu g\Llcli.
\label{eq:Tmnclcons}
\end{equation}
In the following we investigate two specific models:

\subsection{The general first order model}
In this subsection we restrict ourselves to the case that there are no
twice derivated fields present. The free Lagrangian is quadratic in
the fields $\phicl_j,\phicl_{j,\mu}, j=1,\dots,r$. The interaction is
given by the term above (containing no derivated fields).
\begin{align}
\Llcl&=\Llcln-g\Llcli\\
  \label{def:Lcl}
\Llcln&=\frac{1}{2}
K^{\mu\nu}_{jl}\D_{\mu}\phicl_j\D_{\nu}\phicl_l
+\frac{1}{2}G^{\mu}_{jl}\D_{\mu}\phicl_j \phicl_l
-\frac{1}{2}M_{jl}\phicl_j\phicl_{l}
\end{align}
with $K^{\mu\nu}_{jl}, G^{\mu}_{jl}, M_{lj} \in \CC$.  The $K,G$ and
$M$ are supposed to possess the following symmetries: $ K^{\mu\nu}_{jl}
= K^{\nu\mu}_{lj} = K^{\mu\nu}_{lj}, G^{\mu}_{jl} = -G^{\mu}_{lj}$ and
$ M_{jl} = M_{lj} $. The Euler-Lagrange equations read
\begin{equation}
  \label{eq:fieldeqcl}
(K^{\mu\nu}_{jl}\D_{\mu}\D_{\nu}+G^{\mu}_{jl}\D_{\mu}+M_{jl})\phicl_{l}
\doteq D_{jl}\phicl_{l}=-g\parz{\Llcli}{\phicl_j}.
\end{equation}
The \emt\ defined by \eqref{def:emtclass} is called the
\emph{canonical} \emt. It is given by
\begin{align}
\Tmnccl&=\Tmnnccl-\eta^{\mu\nu}g\Llcli\label{def:Tmnccl}\\
\Tmnnccl&=K^{\mu\rho}_{lk}\D_{\rho}\phicl_{l}\D^{\nu}\phicl_{k}
+\frac{1}{2}G^{\mu}_{lk}\D^{\nu}\phicl_{l}\phicl_{k}+ \notag\\
&\quad-\frac{1}{2}\eta^{\mu\nu}\left(
K^{\rho\sigma}_{lk}\D_{\rho}\phicl_{l}\D_{\sigma}\phicl_{k}
+G^{\rho}_{lk}\D_{\rho}\phicl_{l}\phicl_{k}
-M_{lk}\phicl_{l}\phicl_{k}\right).
\label{def:Tmnnccl}
\end{align}
In section~\ref{sec:canemtq} we show that the ``conservation''
equation \eqref{eq:Tmnclcons} can also be fulfiled in the
interacting quantum field theory.

\subsection{The massless $(\phicl)^4$-theory}
\label{subsec:impemtcl}
If no dimensionful couplings are pre\-sent, it is always possible to
construct a conserved and traceless \emt. This tensor is called the
\emph{improved} \emt\ and was first introduced by Callan, et.\ al.\ in
\cite{pap:callan1}.  We derive this tensor in $(\phicl)^4$-theory as the
\emt\ of an improved Lagrangian making use of the ambiguity
in the definition of the Lagrangian. A derivation of this kind was
already performed by Kasper \cite{pap:kasper}. 

The equations of motion read:
\begin{equation}
  \label{eq:motiona}
  \square\phicl=-g\parz{\Llcli}{\phicl}.
\end{equation}
Since a total derivative in the Lagrangian does not change the
equations of motion they originate from both following expressions:
\begin{align}
\Llcl_{\mathrm{can}}
&=\frac{1}{2}\D_\rho\phicl\D^\rho\phicl-g\Llcli, \label{eq:Lcancl}\\
\Llcl_{\mathrm{imp}}
&=\frac{1}{6}\D_\rho\phicl\D^\rho\phicl
-\frac{1}{3}\phicl\square\phicl-g\Llcli.\label{eq:Limpcl}
\end{align}
The corresponding \emt's are called the canonical and the improved
one, respectively. The first one already follows from the last
subsection:
\begin{align}
\Tmnccl&=\D^\mu\phicl\D^\nu\phicl
-\frac{1}{2}\eta^{\mu\nu}\D_\rho\phicl\D^\rho\phicl
+g\eta^{\mu\nu}\Llcli,\\
\Tmnicl
&=\frac{2}{3}\D^\mu\phicl\D^\nu\phicl
-\frac{1}{3}\phicl\D^\mu\D^\nu\phicl+\label{def:Tmnicl1} \\
&\quad-\frac{1}{6}\eta^{\mu\nu}\D_\rho\phicl\D^\rho\phicl
+\frac{1}{12}\eta^{\mu\nu}\phicl\square\phicl\label{def:Tmnicl2}\\
&=\Tmnccl-\frac{1}{3}\Imncl\label{def:Tmnicl3},\\
\intertext{where we have introduced the conserved improvement tensor}
\Imncl
&=\D^\mu\phicl\D^\nu\phicl+\phicl\D^\mu\D^\nu\phicl+\notag\\
&\quad
-\eta^{\mu\nu}\D_\rho\phicl\D^\rho\phicl
-\eta^{\mu\nu}\phicl\square\phicl\\
\label{def:Imncl}
&=\frac{1}{2}
\K{\D^\mu\D^\nu-\eta^{\mu\nu}\square}(\phicl)^2.
\end{align}
Contracting the indices, we find $\eta_{\mu\nu}\Tmnicl = 0$.  The
improved tensor gives rise to the dilatation current:
\begin{equation}
  \label{def:dilcurrent}
  D^{\mathrm{class}\,\mu}\doteq x_\nu\Tmnicl.
\end{equation}
Its conservation equation reads:
\begin{equation}
  \label{eq:consdilat}
  \D_\mu D^{\mathrm{class}\,\mu}
=\eta_{\mu\nu}\Tmnicl+x_\nu\D_\mu\Tmnicl
=x^\mu\D_\mu g \Llcli.
\end{equation}
\sloppy
The dilatation current is the Noether current corresponding to the
scaling \mbox{$\phicl(x)\rightarrow e^{d\alpha}\phicl(e^\alpha x)$} with
$d=1$ of the improved Lagrangian \eqref{eq:Limpcl}. On the other hand
we can derive the dilatations from the canonical Lagrangian
\eqref{eq:Lcancl}. In this case we find:
\begin{equation}
  \label{eq:classdilat}
  \widetilde{D}^{\mathrm{class}\,\mu}=x_\nu\Tmnccl+\phicl\D^\mu\phicl,
\end{equation}
\fussy
with the same conservation equation. The zero component of the
difference is a divergence w.r.t.\ the space coordinates and therefore
does not contribute to the charge:
\begin{equation}
  \widetilde{D}^{\mathrm{class}\,0}
  -D^{\mathrm{class}\,0}
  =\frac{1}{3}\D_j(x^{[j}\D^{0]})(\phicl)^2.
\end{equation}
In section~\ref{sec:impemtq} the corresponding \qft\ is considered.


\section{The canonical quantum energy momentum tensor}
\label{sec:canemtq}
We now discuss the perturbative quantum fields which are associated to
the canonical \emt. While for the free theory all equations from
classical field theory can be carried over due
to the Wick ordering procedure, the interacting quantum fields require
a special normalization which is implied by a Ward identity. This
identity enables to conserve the classical structure also in the
quantum theory.


\subsection{Free quantum theory}
\label{subsec:freeemt}
The classical fields from the last section may now serve as the
symbols from our auxiliary variable algebra $\Ba$.  For our
corresponding quantum fields $\vp_j$ we assume the same equations of
motion to be satisfied
\begin{equation}
  \label{eq:motion}
  D_{jl} T(\vp_l)=0.
\end{equation}
By investigation of the $K,G$ and $M$ we find that this covers a lot
of equations like Klein-Gordon and Dirac equation. The commutator
\begin{equation}
  \label{eq:commutator}
  \left[T(\vp_j)(x),T(\vp_k)(y)\right]
=i\Delta_{jk}(x-y),
\end{equation}
may therefore be an anticommutator in the case of Fermi fields. Now we
regard $\Tmnnc\in\Ba$ from \eqref{def:Tmnnccl} as a symbol
($\phicl\rightarrow\vp$):
\begin{multline}
\Tmnnc=K^{\mu\rho}_{lk}\D_{\rho}\vp_{l}\D^{\nu}\vp_{k}
+\frac{1}{2}G^{\mu}_{lk}\D^{\nu}\vp_{l}\vp_{k}+ \\
\quad-\frac{1}{2}\eta^{\mu\nu}\left(
K^{\rho\sigma}_{lk}\D_{\rho}\vp_{l}\D_{\sigma}\vp_{k}
+G^{\rho}_{lk}\D_{\rho}\vp_{l}\vp_{k}
-M_{lk}\vp_{l}\vp_{k}\right).
\label{def:Tmnn}
\end{multline}
Then $T(\Tmnnc)=\wick{\Tmnnc}$ defines the free canonical quantum
\emt. Since the equations of motion hold inside the Wick ordering we
maintain the conservation
\begin{equation}
  \D_\mu\wick{\Tmnnc}=0.
\label{eq:conesmtcanfree}
\end{equation}

\subsection{Interacting quantum theory}
\label{subsec:canemtqint}
Applying the framework of perturbative interacting fields introduced
in the last chapter we investigate the consequences of switching on an
interaction. We assume that the interaction $\Ll$ contains no
derivated fields. Corresponding to the free canonical \emt\ we
construct the interacting counterpart.  According to \eqref{def:Wg} we
have:
\begin{equation}
  \label{def:intemtcan}
  \Tmnncg(x)\doteq
\sum_{n=0}^\infty\frac{i^n}{n!}
\int \dif y_1\dots \dif y_n
R(g\Ll,\dots,g\Ll;\Tmnnc)(y_1,\dots,y_n;x).
\end{equation}
But this is only the part corresponding to the free fields. The total
tensor $\Tmncg$  receives another contribution from the interaction
(cp. \eqref{def:Tmnccl}):
\begin{equation}
  \label{def:Tmng}
  \Tmncg\doteq\Tmnncg+\eta^{\mu\nu}g\Llg.
\end{equation}
Since $g$ is of compact support global translation invariance is
broken. Hence we expect the conservation equation to
be satisfied that takes account of the non-invariance of the coupling
function (cp. \eqref{eq:Tmnclcons}):
\begin{equation}
  \label{eq:consemtg}
  \D_\mu\Tmncg=\D^\nu g\, \Llg.
\end{equation}
If this equation is true, we have local conservation on $\Ocal$:%
\footnote{We have assumed $g\restriction_\Ocal = \const
  \Rightarrow \D_\mu g\restriction_\Ocal=0$.}
\begin{equation}
  \label{eq:consemtint}
  \Tmncg(\D_\mu f)=0,\ \forall f \text{ with } \supp f\subset\Ocal.
\end{equation}
Equation \eqref{eq:consemtg} is the main statement. We now give a
formulation in terms of the perturbative contributions. It comes out
that the conservation can be completely discussed on the level of
$T$-products. The corresponding equation is a Ward identity involving
the free canonical \emt.

Inserting the definition of $\Tmncg$ from \eqref{def:Tmng} we see that
\eqref{eq:consemtg} is equivalent to
\begin{equation}
  \label{eq:consemtcang}
  \D_\mu\Tmnncg=-g\,\D^\nu\Llg.
\end{equation}
We expand this into the formal power series in the coupling. The \rhs\ 
becomes
\begin{multline}
-g\D^\nu\Llg(x)=\\
\begin{split}
&=-g(x)\sum_{n=0}^\infty\frac{i^n}{n!}\int \dif y_1 \dots \dif y_n\,
\D^\nu_x R\left(\Ll,\dots,\Ll;\Ll\right)(y_1,\dots,y_n;x) 
\,g(y_1)\dots g(y_n) \\
&=i\sum_{n=0}^\infty\frac{i^{n+1}}{n!}\int \dif y_1 \dots \dif
y_{n+1}\, g(y_1)\dots g(y_{n+1})
\times\\
&\quad\times
\frac{1}{n+1}\sum_{k=1}^{n+1}
\D^\nu_x R\left(\Ll,\dots,\not k,\dots,\Ll;\Ll\right)
(y_1,\dots,\not k,\dots,y_{n+1};x) \delta(y_k-x)\\
&= i \sum_{n=1}^\infty\frac{i^n}{n!}\int \dif y_1 \dots \dif y_n
\, g(y_1)\dots g(y_n)\times\\
&\quad\times
\sum_{k=1}^{n}
\D^\nu_x R\left(\Ll,\dots,\not k,\dots,\Ll;\Ll\right)
(y_1,\dots,\not k,\dots,y_n;x) \delta(y_k-x).\\
\end{split}
\end{multline}
The expansion of $\Tmnncg$ was already given at the beginning. Then
\eqref{eq:consemtcang} is fulfiled if 
\begin{multline}
\D^x_\mu R\left(\Ll,\dots,\Ll;\Tmnnc\right)(y_1,\dots,y_n;x)=\\
=i\sum_{k=1}^{n}
\D_x^\nu R\left(\Ll,\dots,\not k,\dots,\Ll;\Ll\right)
(y_1,\dots,\not k,\dots,y_n;x) \delta(y_k-x)
\label{eq:RWI}
\end{multline}
is satisfied to all orders. The $R$-products are completely
determined in terms of the $T$-products \eqref{def:R}. We show that 
it is sufficient to prove the following Ward identity:%
\footnote{The index on the derivative on the \rhs\ refers to the respective
  $y$-coordinate.}
\begin{multline}
\D^x_\mu T\left(W_1,\dots,W_n,\Tmnnc\right)(y_1,\dots,y_n,x)=\\
=i\sum_{k=1}^{n} \delta(y_k-x)
\D_{k}^\nu T\left(W_1,\dots,W_n\right)
(y_1,\dots,y_n), 
\tag{\Weins}
\end{multline}
for all $W_i\in\Ba$ which are (not necessarily proper) sub monomials of the
coupling $\Ll$. We therefore prove the statement for all
$W_i\in\Ba$ that contain no derivated fields and have $\dim\leq 4$.

The Ward identity \Weins\ can be integrated to the functional
equation:
\begin{multline}
  \D^x_\mu\frac{\delta}{i\delta f_{\mu\nu}(x)}
  S\biggl(g\Ll+f\cdot\Tnc+\sum_{j=1}^s h_j W_j\biggr)\biggr\vert_{f=0}=\\
  =-\biggl(g(x)\D^\nu_x\frac{\delta}{i\delta g(x)}
  +\sum_{j=1}^s h_j(x)\D^\nu_x\frac{\delta}{i\delta h_j(x)}\biggr)
  S\biggl(g\Ll+\sum_{j=1}^s h_j W_j\biggr).
\label{eq:WIeinsfunct}
\end{multline}
Multiplying with $S(g\Ll)^{-1}$ from the left and expanding in powers
of the coupling yields \eqref{eq:RWI}. Using 
\begin{equation}
\frac{\delta}{i\delta
  f} S(g\Ll+f\cdot\Tnc)^{-1}\biggr\vert_{f=0} = 
-S(g\Ll)^{-1}
\frac{\delta}{i\delta f} S(g\Ll+f\cdot\Tnc)\biggr\vert_{f=0} 
S(g\Ll)^{-1} 
\end{equation}
we find that \eqref{eq:WIeinsfunct} also holds for the inverse
functional. This implies the corresponding Ward identity for the
$\Tbar$-products to have a minus sign. A simple calculation also shows
that \eqref{eq:WIeinsfunct} implies
\begin{multline}
  \D^x_\mu\frac{\delta}{i\delta f_{\mu\nu}(x)}
  S_{g\Ll+f\cdot\Tnc}(hW)\Bigr\vert_{f=0}=\\
  =-\biggl(g(x)\D^\nu_x\frac{\delta}{i\delta g(x)}
  +h(x)\D^\nu_x\frac{\delta}{i\delta h(x)}\biggr)
  S_{g\Ll}(hW).
\label{eq:ward1R2funct}
\end{multline}
Expanding this equation in $n$'th order $g$ and first order $h$
results in the following identity for the $R$-products:
\begin{align}
\D_\mu^{x_1} R\K{N,\Tmnnc;W}\K{y_N,x_1;x_2}
&=i\sum_{k\in N}\delta(y_k-x_1)\D^\nu_k R(N;W)\K{y_N;x_2}+
\notag\\
&\quad+i\delta(x_2-x_1)\D^\nu_{x_2} R(N;W)\K{y_N;x_2}.
  \label{eq:ward1R2}
\end{align}
The next section gives a proof of \Weins.

\section{Proof of the Ward identity}
\label{sec:ward1}
D{\"u}tsch and Fredenhagen have presented a very general framework for
proving a Ward identity of this kind in \cite{pap:duet-fred1}. It was
generalized by Boas \cite{phd:boas} in the presence of derivated
fields. This is our situation here and we apply their methods. The
strategy is as follows: A possible violation of the Ward identity is
called an \emph{anomaly}. Since all $T$-products are supposed to fulfil
the normalization conditions \Nnull\ -- \Nvier\ we perform a double
induction, one over $n$ and one over the degree (number of generators)
of the Wick monomials. We  assume that the anomaly in both lower
orders is zero.
\begin{itemize}
\item {\bf Step 1.} The commutator of the anomaly with the free fields
  vanishes. Therefore it can appear only in the vacuum sector.
\item {\bf Step 2.} If one $W_i$ is a generator (which is the lowest
  degree sub monomial), the anomaly vanishes due to \Nvier. 
\item {\bf Step 3.} Because of \Neins\ and the induction on $n$, the
  anomaly is a Poincar\'e covariant $\CC$-number distribution with
  support on the total diagonal. We show that it vanishes by an
  appropriate normalization, i.e.\ adding a $\delta$-polynomial with
  the right symmetry properties (\Nnull).%
\footnote{If the current itself is a sub monomial of the coupling, this 
  may lead to non trivial conditions like in \cite{pap:duet-fred1}.}
\end{itemize}
 
To save some space we use the short hand notations from above. We
define the anomaly $a$ by:
\begin{equation}
  a^{\nu}\K{x,y_N}\doteq 
  \D^x_\mu T\K{\Tmnnc,N}\K{x,y_N} -i\sum_{k=1}^{n}\delta(y_k-x)
  \D_{k}^\nu T(N)\K{y_N}.
\label{eq:anomaly}
\end{equation}
 
\subsection*{Step 1}
We commute the anomaly with the free fields.%
\footnote{We use the symbols $\vp_{l,\mu}$ and $\D_\mu\vp_l$
  synonymously.}  We  use a double induction, one on $n$ and the
other on the degree of the Wick sub monomials. Using (\Ndrei) we need
the sub monomials of $\Tmnnc$:
\begin{align}
\frac{\D\Tmnnc}{\D\vp_{j}} &=-\frac{1}{2}G^{\mu}_{jl}\D^{\nu}\vp_{l} 
+\frac{1}{2}\eta^{\mu\nu}G^{\rho}_{jl}\D_{\rho}\vp_{l} 
+\eta^{\mu\nu}M_{jl}\vp_{l} \label{eq:Tmncd1}\\
\frac{\D\Tmnnc}{\D\vp_{j,\rho}}, &=K^{\mu\rho}_{jl}\D^{\nu}\vp_{l} 
+2\eta^{\nu(\rho}K^{\mu)\sigma}_{jl}\D_{\sigma}\vp_{l} 
+\eta^{\nu(\rho}G^{\mu)}_{jl}\vp_{l}.\label{eq:Tmncd2}
\end{align}
We explicitely distinguished between the basic generators and the
first order ones. Here and in the following the sums only run
over the basic generators therefore. Equations
\eqref{eq:Tmncd1},\eqref{eq:Tmncd2} are linear in the fields. We demand
$T$-products containing once derivated basic generators to fulfil the
following normalization:%
\footnote{Because of \eqref{eq:nvierintegrated} this means $\omega_0
  (T(\vp_{j,\mu},\vp_k)(x,y)) = \D_\mu^x
  \omega_0(T(\vp_j,\vp_k)(x,y))$. If the fields $\vp_j,\vp_k$ are
  bosonic with mass dimension 1 this is automatically fulfiled because
  of the negative singular order. In the case of two Fermi fields with
  mass dimension $\frac{3}{2}$ and non vanishing anticommutator, e.g.\
  $\psi,\overline{\psi}$ we have:
  $\omega_0(T(\psi_{,\mu},\overline{\psi})(x,y)) = i \D_\mu S^F(x-y) +
  c\gamma_\mu \delta(x-y)$. The normalization \eqref{eq:oncederiv}
  requires $c=0$.
}
 \begin{equation}
 \D_{\mu}^{x}T(\vp_{j},N)(x,y_{N})=T\K{\D_{\mu}\vp_{j},N}(x,y_{N}).
 \label{eq:oncederiv}
 \end{equation}
This translates into 
\begin{equation}
\D_{\mu}\vp_{j\,g\Ll}(x)=\K{\D_{\mu}\vp_{j}}_{g\Ll}
\label{eq:DvpjgL}
\end{equation}
for the interacting fields.  We calculate the following relevant term:
\begin{multline}
\label{eq:hilfseq1}
\D_{\mu}^{x}\biggl\{ 
T\left(\frac{\D\Tmnnc}{\D\vp_{j}},N\right)\K{x,y_{N}}\Delta_{ji}(x-z)
+T\K{\parz{\Tmnnc}{\vp_{j,\rho}},N}\K{x,y_{N}}
\D_{\rho}\Delta_{ji}(x-z)\biggr\}= 
\\
\begin{split} 
&=\D_{\mu}^{x}\biggl\{ \Delta_{ji}(x-z)\left[ 
-\frac{1}{2}G^{\mu}_{jl}\D^{\nu}_{x} 
+\frac{1}{2}\eta^{\mu\nu}G^{\rho}_{jl}\D_{\rho}^{x} 
+\eta^{\mu\nu}M_{jl}\right]+ \\
&\qquad+\D_{\rho}\Delta_{ji}(x-z)\biggl[ K^{\mu\rho}_{jl}\D^{\nu}_{x} 
+\eta^{\nu\rho}K^{\mu\sigma}_{jl}\D_{\sigma}^{x} 
-\eta^{\mu\nu}K^{\rho\sigma}_{jl}\D_{\sigma}^{x}+ \\
&\qquad+\frac{1}{2}\eta^{\nu\rho}G^{\mu}_{jl} 
-\frac{1}{2}\eta^{\mu\nu}G^{\rho}_{jl}\biggr] \biggr\} 
T(\vp_{l},N)\K{x,y_{N}} \\
&=\Bigl( -\D_{\mu}\Delta_{ji}(x-z)G^{\mu}_{jl}\D^{\nu}_{x} 
+M_{jl}\Delta_{ji}(x-z)\D^{\nu}_{x}+  \\
&\qquad+\D_{\mu}\D_{\rho}\Delta_{ji}(x-z)K^{\mu\rho}_{jl}\D^{\nu}_{x}+ 
\D^{\nu}\Delta_{ji}(x-z)M_{jl}+ \\
&\qquad+\D^{\nu}\Delta_{ji}(x-z)G^{\mu}_{jl}\D_{\mu}^{x} 
+\D^{\nu}\Delta_{ji}(x-z)K^{\mu\sigma}_{jl}\D_{\mu}^{x}\D_{\sigma}^{x} 
\Bigr)T(\vp_{l},N)\K{x,y_{N}} \\
&=\left(
D_{lj}\Delta_{ji}(x-z)\D^{\nu}_{x}+ 
\D^{\nu}\Delta_{ji}(x-z)D_{jl}^{x}\right) 
T(\vp_{l},N)\K{x,y_{N}} \\
&=\D^{\nu}\Delta_{ji}(x-z)D_{jl}^{x} T(\vp_{l},N)\K{x,y_{N}},
\end{split}
\end{multline} 
since $\Delta$ is a solution of the free field equation. 
Now we commute the anomaly with the free basic fields and use \Ndrei:
\begin{multline}
\label{eq:acommutator}
[a^{\nu}\K{x,y_{N}},\vp_{i}(z)]=\\
\begin{split}
&=\D^{x}_{\mu} 
\left[ T(\Tmnnc,N)\K {x,y_{N}},\vp_{i}(z)\right]+\\
&\quad-i\sum_{k=1}^{n}\delta(y_{k}-x) \D^{\nu}_{k}\left[ 
T(N)\K{y_{N}},\vp_{i}(z)\right]\\
&=i\D_{\mu}^{x}\biggl\{ T\left(\frac{\D\Tmnnc}{\D\vp_{j}},N 
\right)\K{x,y_{N}}\Delta_{ji}(x-z)+\\
&\qquad+ T\K{\parz{\Tmnnc}{\vp_{j,\rho}},N}\K{x,y_{N}} 
\D_{\rho}\Delta_{ji}(x-z) \biggr\}+ \\
&\qquad+i\sum_{l=1}^{n}\D_{\mu}^{x}     
T\left(\Tmnnc,W_{1}\dots,\frac{\D W_{l}}{\D\vp_{j}},\dots,W_{n}\right)
\K{x,y_{N}}\Delta_{ji}(y_{l}-z)+\\
&\qquad+\sum_{k=1}^{n}\delta(y_{k}-x)\D_{k}^{\nu}
\left\{\sum_{l=1}^{n} T\left(W_{1},\dots,\frac{\D W_{l}}{\D\vp_{j}},
\dots,W_{n}\right)\K{y_{N}}\Delta_{ji}(y_{l}-z)\right\}.
\end{split}
\end{multline}
If the derivative on the last line acts on $T$ it cancels the third 
line by the induction hypothesis. Only the term with $l=k$ and the 
derivative on $\Delta$ remains. The $\delta$ function allows to put 
$y_{k}=x$ and by inserting \eqref{eq:hilfseq1} we end up with
\begin{align}
\mbox{}&=\D^{\nu}\Delta_{ji}(x-z) 
\biggl[iD_{jl}^{x}T(\vp_{l},N)\K{x,y_{N}}+\\
&\qquad+\sum_{k=1}^{n}\delta(y_{k}-x) 
T\left(W_{1},\dots,\frac{\D W_{k}}{\D\vp_{j}},\dots, 
W_{n}\right)\biggr]\\
&=0,
\end{align}
because of (\Nvier). Since the Ward identity commutes with all free
fields the anomaly is a $\CC$-number. Because of the causal Wick
expansion \eqref{eq:causalwick} it can only appear in the vacuum
sector:
\begin{equation}
  \label{eq:avakuum}
  a^\nu(x,y_N)=\scp{\Omega}{a^\nu(x,y_N)\Omega}.
\end{equation}

\subsection*{Step 2}
We prove that the Ward identity is compatible with the normalization
condition (\Nvier). Using the equivalent formulation 
\eqref{eq:nvierintegrated}
\begin{multline}
\label{eq:vaknvier} 
\scp{\Omega}{T(N,\vp_i)\K{y_{N},z}\Omega}= \\
\begin{split}
        &=i\sum_{k=1}^n \Delta^F_{ij}(z-y_k) 
        \scp{\Omega}{T\left(W_1,\dots,\parz{W_k}{\vp_j},\dots, 
        W_n\right)\K{y_{N}}\Omega}+ \\
&\quad-i\sum_{k=1}^n \D_{\rho}\Delta^F_{ij}(z-y_k) 
\scp{\Omega}{T\left(W_1,\dots,\parz{W_k}{\vp_{j,\rho}},\dots, 
W_n\right)\K{y_{N}}\Omega}. 
\end{split}
\end{multline}
with $\Delta^F_{ij}(y-x)=\scp{\Omega}{\vp_i(y)\vp_j(x)\Omega}$. Then a 
calculation along the lines of \eqref{eq:hilfseq1} shows
\begin{multline}
\label{eq:hilfseq2}
\D_{\mu}^{x}\biggl\{\Delta^F_{ij}(z-x) \scp{\Omega}{ 
T\left(\frac{\D\Tmnnc}{\D\vp_{j}},N\right)\K{x,y_{N}}\Omega}+\\
\shoveleft{\quad- 
\D_{\rho}\Delta^F_{ij}(z-x)\scp{\Omega}{ 
T\left(\parz{\Tmnnc}{\vp_{j,\rho}},N\right)\K{x,y_{N}}\Omega} 
\biggr\}=}\\
\begin{split} 
&=\left( D_{jl}\Delta^F_{ij}(z-x)\D^{\nu}_{x}- 
\D^{\nu}\Delta^F_{ij}(z-x)D_{jl}^{x}\right) 
\scp{\Omega}{T(\vp_{l},N)\K{x,y_{N}}\Omega} \\
&=\left(
\delta_{li}\delta(z-x)\D^{\nu}_{x}- 
\D^{\nu}\Delta^F_{ij}(z-x)D_{jl}^{x}\right) 
\scp{\Omega}{T(\vp_{l},N)\K{x,y_{N}}\Omega},
\end{split}
\end{multline}
since $\Delta^F$ is a Green's function of the equation of motion: 
$D_{jl}\Delta^F_{ij} = D_{lj}\Delta^F_{ji} = \delta_{li}\delta$. We 
set $a^\nu\left(x,y_N,z\right)$ like before with $W_{n+1}\doteq\vp_i$ 
and compute its vacuum expectation value. We obtain
\begin{multline}
a^\nu\left(x,y_N,z\right)=\\
\begin{split} 
&=i\D_{\mu}^{x}\biggl\{\Delta^F_{ij}(z-x) \scp{\Omega}{%
T\left(\frac{\D\Tmnnc}{\D\vp_{j}},N\right)\K{x,y_{N}}\Omega}+\\
&\qquad-\D_{\rho}\Delta^F_{ij}(z-x)\scp{\Omega}{%
T\left(\parz{\Tmnnc}{\vp_{j,\rho}},N\right)\K{x,y_{N}}\Omega} 
\biggr\}+\\ 
&\qquad+i\sum_{l=1}^n\Delta^F_{ij}(z-y_l)\D_\mu^x 
\scp{\Omega}{%
T\left(\Tmnnc,W_1,\dots,\frac{\D W_l}{\D\vp_j},\dots,W_n\right)
\K{x,y_{N}}\Omega}
+\\
&\qquad-i\delta(z-x)\D^\nu_z 
\scp{\Omega}{T(\vp_i,N)\K{z,y_{N}}\Omega}+\\
&\qquad+\sum_{k=0}^n\delta(y_k-x)\D^\nu_{k}\left\{ 
\sum_{l=1}^n\Delta^F_{ij}(z-y_l)\scp{\Omega}%
{T\left(W_1,\dots,\parz{W_l}{\vp_j},\dots,W_n\right)\K{y_{N}}\Omega} 
\right\}.
\end{split}
\end{multline} 
Again, if in the last line the derivative acts 
on the $T$-products these terms cancel the third line by the 
induction hypothesis. The term $l=k$ with the derivative on $\Delta^F$ 
remains. Inserting \eqref{eq:hilfseq2} gives:
\begin{align} 
&=-i\D^\nu\Delta^F_{ij}(z-x)D^x_{jl} \scp{\Omega}
{ T(\vp_l,N)\K{x,y_{N}}\Omega}+\\
&\qquad-\sum_{k=0}^n\delta(y_k-x)\D^\nu\Delta^F_{ij}(z-y_k) 
\scp{\Omega}%
{T\left(W_1,\dots,\frac{\D W_k}{\D\vp_j},\dots,W_n\right)\K{y_{N}}\Omega}\notag\\
&=0,
\end{align}
because of (\Nvier). 

\subsection*{Step 3}
We show how the anomaly can be removed by an appropriate
normalization. The above steps have shown that it has the following
form:
\begin{align}
a^\nu(x,y_N) &=\D^x_\mu 
\scp{\Omega}{T(\Tmnnc,N)\K{x,y_N}\Omega} 
-i\sum_{k=1}^{n}\delta(y_k-x) \D_k^\nu 
\scp{\Omega}{T(N)\K{y_N}\Omega} \\
&=M^\nu(\D)\delta(y_1-x)\dots\delta(y_n-x).
\label{eq:adelta}
\end{align}
Here, $\D=(\D_1,\dots,\D_n)$ and $M^\nu$ is a Lorentz vector valued
polynomial of degree $\leq 5$, since $\singord
\scp{\Omega}{T(\Tmnnc,N),\Omega} = \dim \Tmnnc + \sum_{i=1}^n \dim W_i
-4n \leq 4$ according to \eqref{eq:scaledegT}. If $M^\nu(\D)$ has the form
\begin{equation}
  \label{eq:PvonD}
  M^\nu(\D)=\sum_{i=1}^n\D_\mu^i M_1^{\mu\nu}(\D),
\end{equation}
with $M_1$ again a polynomial, the normalization
$T(\Tmnnc,N) \rightarrow T(\Tmnnc,N)+
M_1^{\mu\nu}(\D)\delta$ removes the anomaly. 

We show that this is the case. We introduce the free momentum
operator:%
\footnote{One has to give some meaning to the formal integral in
  \eqref{def:Pm}. We refer to the method of Requardt
  \cite{pap:requardt}. This shows that a charge like $P^\mu$ can be
  defined for massive theories in general and for certain massless
  theories, if the infrared behaviour is not ``too bad''. We
  explicitly show the existence in section~\ref{sec:intcharge} for the
  massless $\vp^4$-model. But the same conclusion also holds if the
  mass dimension of $\Tmnnc$ is not less than four. This is the case
  here, if the fields in \eqref{def:Tmnn} contracted with $K$ are
  bosonic and the ones contracted with $G$ are fermionic, as usual.
}
\begin{equation}
  \label{def:Pm}
  P^\mu\doteq\int\dif^3\xvek\,\Tnnnc(x).
\end{equation}
It is a hermitian operator that annihilates the vacuum.

For every $(y_1,\dots,y_n)$ we take a 
double cone \Ocal\ with $y_i\in\Ocal$ for all $i=1,\dots,n$. Choosing 
a $g$, with $g\!\!\upharpoonright_{\overline{\Ocal}}\; =1$ we 
decompose $\D_\mu g = a_\mu-b_\mu$, such that $\supp a_\mu 
\cap(\blc+\Ocal) = \supp b_\mu \cap(\flc+\Ocal) = \emptyset$. We smear 
out the first term on the r.h.s of \eqref{eq:anomaly} with this 
$g$ and use the causal factorization of the $T$-products:
\begin{align}
  \int\dif x\, \D_\mu^x & T(\Tmnnc,N)\K{x,y_N}g(x)=\\
  &=-\Tmnnc(a_\mu)T(N)\K{y_N}+ T(N)\K{y_N}\Tmnnc(b_\mu) \\
  &=-\left[ \Tmnnc(a_\mu),T(N)\K{y_N}\right] 
  -T(N)\K{y_N}\Tmnnc(\D_\mu g).\notag\\
  \intertext{The second term vanishes because $\Tmnnc$ is a conserved
    current. Then, in the commutator $\Tmnnc(a_\mu)$ can be replaced
    by $-P^\nu$, since $T(N)$ is localized in \Ocal:} 
&=\left[P^\nu,T(N)\K{y_N}\right].
\label{eq:anomalyg}
\end{align}
Therefore the vacuum expectation value of \eqref{eq:anomalyg}
vanishes.  Smearing the second term of \eqref{eq:anomaly} with $g$ and
taking the vacuum expectation value we find:
\begin{equation}
i\int\dif x \sum_{j=1}^n\delta(y_j-x)\D^\nu_j 
\scp{\Omega}{T(N)\K{y_N}\Omega}g(x)
=i\sum_{j=1}^n\D^\nu_j \scp{\Omega}{T(N)\K{y_N}\Omega} =0,
\end{equation}
because of translation invariance. Hence we get%
\footnote{Note that the \rhs\ of \eqref{eq:adelta} is of compact
  support in the difference variables $y_i-x$.}
\begin{equation}
  \label{eq:intag}
\int\dif x\,a^\nu(x,y_N)=0.
\end{equation}
To prove \eqref{eq:PvonD} we work in Fourier space:
\begin{align}
\widehat{a^\nu}(x,p_1,\dots,p_n) 
&=\int\dif y_1\dots\dif y_n\,a^\nu(x,y_1,\dots,y_n) 
e^{i(p_1y_1+\dots+p_ny_n)}\\
&=M^\nu(-ip_1,\dots,-ip_n)e^{i(p_1+\dots+p_n)x}.
\end{align}
If we integrate over $x$ and use \eqref{eq:intag} we find:
\begin{gather}
\int\dif^4x\,\widehat{a^\nu}(x,p_1,\dots,p_n) =(2\pi)^4 
M^\nu(-ip_1,\dots,-ip_n)\delta(p_1+\dots+p_n)=0, \\
\Rightarrow 
M^\nu(-ip_1,\dots,-ip_n)\Bigr\vert_{p_1+\dots+p_n=0}=0.
\label{eq:Pnull} 
\end{gather}
We set $q=\sum_{i=1}^np_i$, and write $\widetilde{M^\nu}
(q,p_1,\dots,p_{n-1}) = M^\nu(-ip_1,\dots,-ip_n)$. Performing a Taylor
expansion at the origin we get:
\begin{equation}
  \label{eq:taylorP}
\widetilde{M^\nu}(q,p_1,\dots,p_{n-1}) = 
\sum_{k=1}^{\text{degree}\widetilde{M^\nu}} 
\sum_{|\alpha|+|\beta|=k}\frac{q^\alpha p^\beta}{\alpha!\beta!} 
\D^q_\alpha \D^p_\beta \widetilde{M^\nu}(0,\dots,0),
\end{equation}
$p=(p_1,\dots,p_{n-1})$.  Because of \eqref{eq:Pnull} there are only terms
with $|\alpha|\geq1$ in the second sum. If we Fourier transform back
into coordinate space this implies \eqref{eq:PvonD}.

The normalization term $M_1^{\mu\nu}$ cannot be added to the first
two terms of $T(\Tmnnc)$ only. The next two terms
(conf. \eqref{def:Tmnn}) are multiples of the traces of the first
ones. This symmetry has to be preserved since we demand linearity of
the $T$-products:
\begin{align}
  \eta_{\mu\nu}T(\D^\mu\vp_j\D^\nu\vp_l,N)
&=T(\D^\mu\vp_j\D_\mu\vp_l,N) \\
\Rightarrow
\eta_{\mu\nu}(\D^\mu\vp_j\D^\nu\vp_l)_{g\Ll}
&=(\D^\mu\vp_j\D_\mu\vp_l)_{g\Ll}.
\end{align}
Therefore we add the normalization terms according to
\begin{align}
  K^{\mu\rho}_{lk}T(\D_\rho\vp_l\D^\nu\vp_k,N)
&\rightarrow  K^{\mu\rho}_{lk}T(\D_\rho\vp_l\D^\nu\vp_k,N)
+\K{M_1^{\mu\nu}-\frac{1}{6}\eta^{\mu\nu}{M_1^\rho}_\rho}(\D)\delta, \\
\Rightarrow K^{\mu\rho}_{lk}T(\D_\rho\vp_l\D_\mu\vp_k,N)
&\rightarrow  K^{\mu\rho}_{lk}T(\D_\rho\vp_l\D_\mu\vp_k,N)
+\frac{1}{3}{M_1^\rho}_\rho(\D)\delta.
\end{align}
If the normalization has to be performed on the other terms we put:
\begin{align}
  G^{\mu}_{lk}T(\D^\nu\vp_l\vp_k,N)
&\rightarrow  G^{\mu}_{lk}T(\D^\nu\vp_l\vp_k,N)
+2\K{M_1^{\mu\nu}-\frac{1}{3}\eta^{\mu\nu}{M_1^\rho}_\rho}(\D)\delta, \\
\Rightarrow 
G^{\mu}_{lk}T(\D_\mu\vp_l\vp_k,N)
&\rightarrow  G^{\mu}_{lk}T(\D_\mu\vp_l\vp_k,N)
-\frac{2}{3}{M_1^\rho}_\rho(\D)\delta.
\end{align}
These normalizations remove the anomaly.


\section{The interacting momentum operator}
\label{sec:intcharge}
Now we investigate the interacting charge generated by the conserved
energy momentum tensor. It defines the interacting momentum operator
since its commutator implements the infinitesimal action on $ \Aag
(\Ocal)$ according to:
\begin{equation}
\label{eq:momentumcomm}
\left[\Tnncg(f),\Wg(x)\right]
=i\D^\nu \Wg(x)
\end{equation}
for all $W \in \Ba$ containing no derivated fields. The test function
$f\in\Dd(\MM)$ is supposed to be $f(y)=h(y^0)$ for all
$y=(y^0,\yvek)\in\MM$ in a neighbourhood of $x+(\flc\cup\blc)$ and
$\int\dif y^0\,h(y^0)=1$.

\subsection{Proof of \eqref{eq:momentumcomm}}

\begin{figure}[htbp]
  \begin{center}
    \input{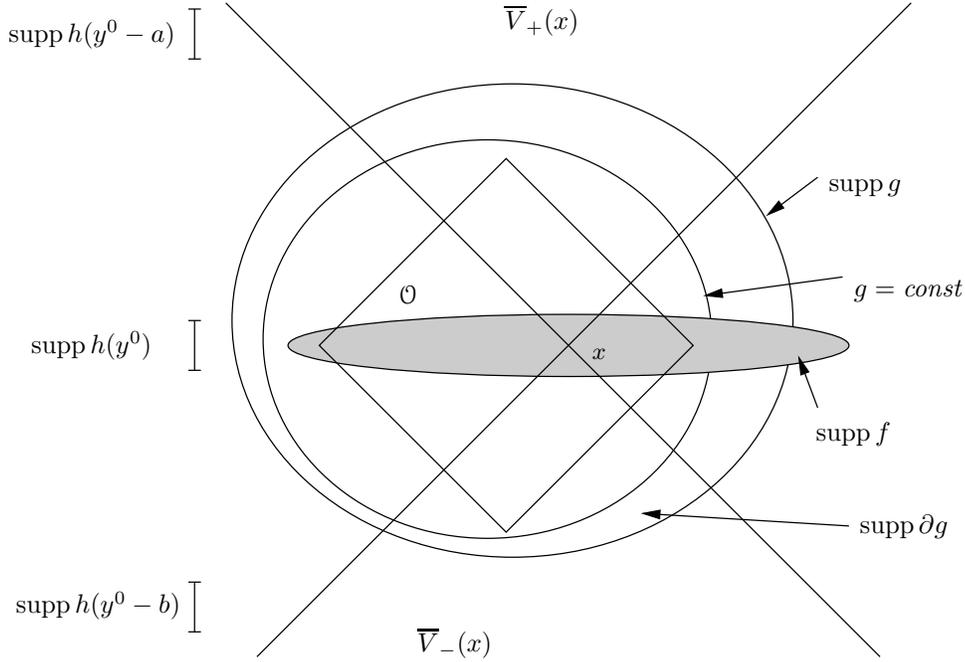}
    \caption{Supports of $f,h,g,\D g$ and $R$-products.}
    \label{fig:supp}
  \end{center}
\end{figure}

We follow the idea of \cite{pap:duet-fred1}. We use the abbreviation
$\dif y_N \, g(y_N) \doteq \prod_{i\in N} \dif y_i g(y_i)$. With the
definition of the commutator \eqref{eq:intcomm} the support properties
of the $R$-products and the choice of $f$ from above we have:
\begin{multline}
  \left[\Tnnncg(f),\Wg(x)\right]
=\int\dif y\, h(y^0)  \left[\Tnnncg(y),\Wg(x)\right]=\\
\begin{split}
&=-\sum_{n=0}^\infty\frac{i^n}{n!}\int
\dif y_N\,\dif y\, g(y_N)\times\\
&\quad\times\Bigl[(h(y^0)-h(y^0-a)+h(y^0-a))R\K{N,\Tnnnc;W}\K{y_N,y;x}+\\
&\quad-(h(y^0)-h(y^0-b)+h(y^0-b))R\K{N,W;\Tnnnc}\K{y_N,x;y}\Bigr],
\end{split}
\end{multline}
where $W_i=\Ll, i=1,\dots,n$. Since $h$ has compact support we can
choose $a>0$ and $b<0$ big enough that the contributions of $h(y^0-a)$
and $h(y^0-b)$ vanish due to the support properties of the
$R$-products, see figure~\ref{fig:supp}. 
If we define the following two functions
\begin{align}
  k(y)\doteq k(y^0)
&=\int_{-\infty}^{y^0}\dif z\, (h(z)-h(z-a)),\\
  \ktilde(y)\doteq \ktilde(y^0)
&=\int_{y^0}^\infty\dif z\, (h(z)-h(z-b)),
\end{align}
the commutator becomes
\begin{multline}
  \left[\Tnnncg(f),\Wg(x)\right]=\\
\begin{split}
&=\sum_{n=0}^\infty\frac{i^n}{n!}\int
\dif y_N\,\dif y\, g(y_N)\times\\
&\quad\times\Bigl[k(y)\D^y_\mu R\K{N,\Tmnnc;W}\K{y_N,y;x}
+\ktilde(y)\D^y_\mu R\K{N,W;\Tmnnc}\K{y_N,x;y}\Bigr]
\\
&=i(k(x)+\ktilde(x))\D^\nu\Wg(x)+\\
&\quad+i\sum_{n=1}^\infty\frac{i^n}{n!}\int
\dif y_N\, g(y_N)\times\\
&\quad\times\sum_{j=1}^n\Bigl[k(y_j)\D^\nu_j R\K{N;W}\K{y_N;x}
+\ktilde(y_j)\D^\nu_j R\K{N\setminus j,W;j}\K{y_{N\setminus j},x;y_j}
\Bigr]\\
&=i\D^\nu\Wg(x)+\\
&\quad-i\eta^{0\nu}\sum_{n=1}^\infty\frac{i^n}{n!}
\int\dif y_N\, g(y_N)\times\\
&\quad\times\sum_{j=1}^n\Bigl[(h(y_j^0)-h(y_j^0-a)) R\K{N;W}\K{y_N;x}
+\\
&\qquad-
(h(y_j^0)-h(y^0_j-b)) R\K{N\setminus j,W;j}\K{y_{N\setminus j},x;y_j}
\Bigr]+\\
&\quad-i\sum_{n=1}^\infty\frac{i^n}{n!}
\int\dif y_N\sum_{j=1}^n g(y_1)\dots\D^\nu g(y_j)\dots g(y_n)\times\\
&\quad\times\Bigl[k(y_j)R\K{N;W}\K{y_N;x}+
\ktilde(y_j)R\K{N\setminus j,W;j}\K{y_{N\setminus j},x;y_j},
\Bigr]
\end{split}
\label{eq:commcan2}
\end{multline}
where we have inserted the Ward identities \eqref{eq:RWI},
\eqref{eq:ward1R2} and used the fact that
\begin{equation}
  k(x)+\ktilde(x)
=\int\dif z\,h(z) 
-\int_{-\infty}^{x^0}\dif z\,h(z-a)
-\int_{x^0}^\infty\dif z\, h(z-b)=1.
\end{equation}
The support of $f$ in the time direction and therefore of $h$
can be made sufficiently small, see figure~\ref{fig:supp}. Then the
term $\D^\nu g(y_j) k(y_j) R(N;W)(y_N;x)$ vanishes due to the
support of properties. The same is true for the second term in the
last integrand of \eqref{eq:commcan2}.
In the first integrand the $h(y_j^0-a)$ and $h(y_j^0-b)$ can be
omitted, again. We obtain
\begin{multline}
  \left[\Tnnncg(f),\Wg(x)\right]=\\
\begin{split}
&=i\D^\nu\Wg(x)+\\
&\quad-i\eta^{0\nu}\sum_{n=1}^\infty\frac{i^n}{n!}
\int\dif y_N\, g(y_N)\times\\
&\quad\times\sum_{j=1}^n h(y_j^0)\Bigl[ R\K{N;W}\K{y_N;x}
- R\K{N\setminus j,W;j}\K{y_{N\setminus j},x;y_j}
\Bigr]\\
&=i\D^\nu\Wg(x)+\\
&\quad+\eta^{0\nu}\sum_{n=1}^\infty\frac{i^{n-1}}{(n-1)!}
\int\dif y_N\sum_{j=1}^n 
g(y_1)\dots g(y_j)h(y_j^0)\dots g(y_n) \times\\
&\quad\times\frac{1}{n} 
\Bigl[ R\K{N\setminus j,j;W}\K{y_{N\setminus j},y_j;x}
- R\K{N\setminus j,W;j}\K{y_{N\setminus j},x;y_j}
\Bigr]\\
&=i\D^\nu\Wg(x)
-\eta^{0\nu}\left[\Llg(gf),\Wg(x)\right],
\end{split}
\label{eq:commcan3}
\end{multline}
since all terms in the sum of the last integrand are equal because of
the symmetry. Hence we find:
\begin{align}
\left[\Tnncg(f),\Wg(x)\right]
&=\left[\Tnnncg(f),\Wg(x)\right]
+\eta^{0\nu}\left[\Llg(gf),\Wg(x)\right]\notag\\
&=i\D^\nu\Wg(x).
\end{align}

We show that one can define a momentum operator according to
\begin{equation}
  \label{def:intmomentum}
  \Png\doteq\int \dif^3\xvek\, \Tnncg(x^0,\xvek).
\end{equation}
Due to the definition of interacting fields \eqref{eq:Wg} we start
investigating the free contribution. Following the method of Requardt
\cite{pap:requardt} we consider the localized momentum operator:
\begin{equation}
  \label{eq:intmomentumscal}
 _\lambda P^\nu \doteq\int \dif^4 x\,k_\lambda(x^0)
h_\lambda(\xvek) \Tnnnc(x^0,\xvek),
\end{equation}
with test functions $h\in\Dd(\RR^3)$, $h(0)\equiv 1$ and $k\in\Dd(\RR)$
with $\int\dif x^0 k(x^0)=1$. We set $h_\lambda(\xvek)\doteq
h(\lambda\xvek)$ and $k_\lambda(x^0) \doteq \lambda k(\lambda x^0)$.
Calculation of the correlation function of two \emt s results in:
\begin{multline}
\scp{\Omega}{\Tnmnc(x)\Tnnnc(y)\Omega}=\\
\begin{split}
  &=\D^0\D^0 D_+(x-y)\D^\mu\D^\nu D_+(x-y)
  +\D^0\D^\mu D_+(x-y)\D^0\D^\nu D_+(x-y)+\\
  &\quad-2\eta^{0(\mu}\D^{\nu)}\D^\rho D_+(x-y)\D_\rho\D^0 D_+(x-y)+\\
  &\quad+\frac{1}{2}\eta^{0\mu}\eta^{0\nu}\D_\rho\D_\sigma D_+(x-y)
  \D^\rho\D^\sigma D_+(x-y)\\
  &=\lambda^8\scp{\Omega}{\Tnmnc(\lambda x)\Tnnnc(\lambda y)\Omega},
\end{split}
\end{multline}
since the massless two-point function $D_+$ is homogenous of degree
$-2$. We have
\begin{equation}
\scp{\Omega}{{_\lambda P}^\mu\,{_\lambda P}^\nu\Omega}
=\lambda^2\int \dif^4x\,\dif^4 y\, k(x^0)k(y^0)h(\xvek)h(\yvek)
\scp{\Omega}{\Tnmnc(x)\Tnnnc(y)\Omega},
\end{equation}
and this implies $\lim_{\lambda\rightarrow 0}||_\lambda P^\nu\Omega ||
=0$. For an arbitrary Wick  polynomial $W$ we additionally have
\begin{equation}
  \lim_{\lambda\rightarrow 0}\left[_\lambda P^\nu,W(x)\right]
=\D^\nu W(x).
\end{equation}
The domain $\Do$ is the linear hull of all $\Phi = W_1(f_1) \dots W_r(f_r)
\Omega$, with $W_i$ Wick monomials and $f_i$ test functions. With the
derivation property of the commutator, this defines the momentum $P^\nu
= \lim_{\lambda\rightarrow 0} {_\lambda P}^\nu$ in a
strong limit on $\Do$: $P\Phi\Omega = [P,\Phi]\Omega$.

In the interacting contribution of \eqref{def:intmomentum} the space
integral is restricted to the hypersurface of constant $x^0$
intersecting $\flc(\supp g)$ and hence compact because of the support
properties of the $R$-products. The interacting canonical tensor
further contains the interaction term (cf.\ \eqref{def:Tmng})
which is localized. Hence the integral in \eqref{def:intmomentum}
exists.

\section{The interacting improved tensor in massless $\vp^4$-theory}
\label{sec:impemtq}
This section treats the possibilities for defining an improved
\emt. It results in the unavoidable appearance of the well known trace
anomaly. The definition of a suitable improvement tensor requires the
validity of a further differential equation involving interacting
fields. This equation is proved by a corresponding Ward identity 
in the next section. 

We consider the free massless scalar field $\vp$. It satisfies the wave
equation and the commutation relation:
\begin{align}
  \square\vp&=0,&
[\vp(x),\vp(y)]&=iD(x-y),
\end{align}
for $\vp(x)=T(\vp)(x)$. $D$ is the Pauli-Jordan distribution. The
corresponding Feynman propagator is denoted by $D^F$. The free field
allows to define a conserved and traceless improved \emt\ by the 
expression from classical field theory in form of Wick products:
\begin{align}
\wick{\Tmnni}&=\wick{\Tmnnc}-\frac{1}{3}\wick{\Imn}, \\
\wick{\Imn}&=\wick{\D^\mu\vp\D^\nu\vp}
+\wick{\vp\D^\mu\D^\nu\vp}
-\eta^{\mu\nu}\wick{\D_\rho\vp\D^\rho\vp}\label{eq:Iwick1}\\
&=\D^\mu\wick{\vp\D^\nu\vp}-\eta^{\mu\nu}\D_\rho\wick{\vp\D^\rho\vp}
\label{eq:Iwick2}\\
&=\frac{1}{2}(\D^\mu\D^\nu-\eta^{\mu\nu}\square)\wick{\vp^2}.
\label{eq:Iwick3}
\end{align}
It is well known that the dilatations can be implemented as a unitary
symmetry on the invariant domain $\Do$ of Fock space $U: \Do \mapsto
\Do, U_\lambda \vp(x) U_\lambda^{-1} = \lambda\vp(\lambda x)$. The
infinitesimal transformation is given by the commutator of the
dilatation charge $Q_D^R=\int \dif x^0 \, \alpha (x^0) f_R(\xvek)
D^0(x)$ for sufficiently large $R$, where $\int \dif x^0 \alpha(x^0) =
1$ and $f_R$ is a smooth version of the characteristic function of the
ball of radius $R$. For large $R$ the commutator becomes independent
of $\alpha$ and $f_R$ \cite{pap:maison-reeh}. The dilatation current
is $D^\mu_0(x) = x_\nu \wick{\Tmnni(x)}$. Since the symmetry is
conserved we have $\lim_{R\rightarrow\infty} \omega_0 ([Q_D^R,W]) =
0$, for any observable $W = \int \dif x_1 \dots \dif x_n $
$\mbox{\wick{W_1(x_1)} \dots \wick{W_n(x_n)}} f(x_1) \dots f(x_n)$
with $W_i \in \Ba, f_i \in
\Dd(\MM)$.%
\footnote{Instead of writing $\lim_{R\rightarrow\infty} [Q_D^R,W]$ or
  using the better convergent definition by Requardt
  \cite{pap:requardt} we use the expression $[D^0(f),W]$ with a
  suitable test function $f$.}

Switching on the interaction, the field equation becomes
\begin{equation}
  \label{eq:intscalarfield}
  \square\vpg=-g\K{\parz{\Ll}{\vp}}_{g\Ll}.
\end{equation}
Since the model is just a special case of the general situation
discussed in the last sections this leads to a locally conserved
canonical \emt\ \eqref{def:Tmnn},\eqref{def:Tmng},\eqref{eq:consemtg}:
\begin{equation}
\Tmncg=\Kg{\D^\mu\vp\D^\nu\vp}
-\frac{1}{2}\eta^{\mu\nu}\Kg{\D_\rho\vp\D^\rho\vp}
+g\eta^{\mu\nu}\Llg.
\end{equation}

In order to define an interacting improvement tensor based on
\eqref{eq:Iwick2} we require the
following identity for some $c\in\RR, c\not=\frac{1}{4}$:
\begin{multline}
  \D^\mu\Kg{\vp\D^\nu\vp} =\\
  =\Kg{\D^\mu\vp\D^\nu\vp}
  +\Kg{\vp\D^\mu\D^\nu\vp}-c\eta^{\mu\nu}\Kg{\vp\square\vp}
  -c\eta^{\mu\nu}g\Kg{\vp\parz{\Ll}{\vp}}.
\label{eq:sternid}
\end{multline}
The equation is obviously symmetric in $\mu,\nu$.  We show this
identity to be satisfied by a corresponding Ward identity in the next
section. The exclusion of the case $c=\frac{1}{4}$ is due to the fact
that \eqref{eq:sternid} has to be satisfied with \Weins\ 
simultaneously. Since the latter one fixes the normalization of
$\Kg{\D^\mu\vp\D^\nu\vp}$ the new interacting field
$\Kg{\vp\D^\mu\D^\nu\vp}$ must not appear in a traceless combination.

Now the improvement tensor is defined by
\begin{equation}
  \label{def:Imng}
  \Imng\doteq\D^\mu\Kg{\vp\D^\nu\vp}
  -\eta^{\mu\nu}\D_\rho\Kg{\vp\D^\rho\vp}.
\end{equation}
It is conserved due to the $\mu,\nu$-symmetry of \eqref{eq:sternid}.
\begin{equation}
  \label{eq:consImng}
  \D_\mu\Imng=0.
\end{equation}
To discuss the consequences of the improvement we introduce
the dimension operator $d$ on monomials $W\in\Ba$ according to:
\begin{equation}
  \label{def:dimop}
  dW\doteq\sum_r (r+1) \vp_{,\mu_1\dots\mu_r}\parz{W}{\vp_{,\mu_1\dots\mu_r}}.
\end{equation}
The 1 in parenthesis refers to the dimension of the scalar field
$\vp$.  Obviously, $d$ has integer eigenvalues.  In case of a pure
$\Ll\propto\vp^4$-interaction we have:
\begin{equation}
  \label{eq:confcoupling}
  d\Ll=\vp\parz{\Ll}{\vp}=4\Ll.
\end{equation}
Now we define the interacting improved \emt\ according to
\eqref{def:Tmnicl3} by:
\begin{equation}
  \label{def:Tmnig}
  \Tmnig\doteq\Tmncg-\frac{1}{3}\Imng.
\end{equation}
But the trace of that tensor is not zero. We find:
\begin{align}
  \eta_{\mu\nu}\Tmnig
  &=-\Kg{\D_\rho\vp\D^\rho\vp}+4g\Llg+\D_\rho\Kg{\vp\D^\rho\vp}\\
  &=(1-4c)\left(\Kg{\vp\square\vp}+g\Kg{\vp\parz{\Ll}{\vp}}\right).
\end{align}
This is the well known trace anomaly of the \emt. We see that it is
undetermined up to a multiplicative real parameter. The anomaly is zero
in case that one of the factors vanishes. But this contradicts the non
existence of a scale invariant renormalization.%
\footnote{To show the contradiction we have to treat the two cases
  \begin{align}
  &(i) \quad c =\frac{1}{4} \text{ and } \\ 
  &(ii)\quad
  \Kg{\vp\square\vp} = -g\Kg{\vp\parz{\Ll}{\vp}}.
  \label{eq:innerintfieldeq}
\end{align}
Assume $(i)$ is true. With 
\begin{align}
\Imnn&=\D^\mu\vp\D^\nu\vp+\vp\D^\mu\D^\nu\vp
-\eta^{\mu\nu}\D_\rho\vp\D^\rho\vp
-\frac{1}{4}\eta^{\mu\nu}\vp\square\vp,
\label{def:Imnn}\\
\Imng&=\Imnng+\frac{3}{4}\eta^{\mu\nu}g\Kg{d\Ll},
\end{align}
we could define the locally conserved dilatation current by
\begin{align}
  \Dmg\doteq x_\nu\Tmnig
&=x_\nu\left(\Tmncg-\frac{1}{3}\Imng\right)\label{def:dilatg}\\
&=x_\nu\left(\Tmnncg-\frac{1}{3}\Imnng\right)\\
&=\Kg{x_\nu\left(\Tmnnc-\frac{1}{3}\Imnn\right)}.
\end{align}
The conservation is equivalent to the Ward identity
\begin{equation}
  \label{eq:wardD}
  \D_\mu^x T(D^\mu,N)(x,y_N)
  =i\sum_{k=0}^n\delta(x-y_k)(d_k+y_k\cdot\D_k)T(N)(y_N),
\end{equation}
where $d_k$ denotes the $k$'th dimension (see next footnote). Passing
to the integrated Ward identity by integration with a function $g$
chosen like in step 3 of section~\ref{sec:ward1} this leads to
\begin{equation}
  \label{eq:intwardD}
  [D^0(f),T(N)(y_N)]
  =\sum_{k=1}^n(d_k+y_k\cdot\D^k)T(N)(y_N),
\end{equation}
where $f$ is a test function like in section~\ref{sec:intcharge}. 
Since the dilatations are the infinitesimal symmetry transformations of
the unitarily implementable scale transformations on the free field
algebra, the \rhs\ vanishes in the vacuum state $\omega_0$. But
this implies a scale invariant renormalization of $\omega_0(T(N))$
which is not possible. Therefore $c\not=\frac{1}{4}$. 

If we assume $(ii)$ to be true, the last two terms of \eqref{eq:sternid}
vanish leading to the conserved improvement tensor $\Imng$ with
\begin{equation}
\Imn=\D^\mu\vp\D^\nu\vp+\vp\D^\mu\D^\nu\vp
-\eta^{\mu\nu}\D_\rho\vp\D^\rho\vp
-\eta^{\mu\nu}\vp\square\vp.
\end{equation}
The dilatations according to \eqref{def:dilatg} are conserved and both
Ward identities \eqref{eq:wardD},\eqref{eq:intwardD} hold with $D^\mu$
replaced by $D_0^\mu=x_\nu(\Tmnnc-\frac{1}{3}\Imn)$. Because of the same
argument this is a contradiction.
}%

The improved \emt\ defines the same momentum operator:
\begin{equation}
  \label{eq:momentumcomm2}
  \left[\Tnnig(f),\Wg(x)\right]
  =\left[\Tnncg(f),\Wg(x)\right]
  =\D^\nu\Wg(x),
\end{equation}
because the $I_{g\Ll}^{0\nu}$-component is either a derivative (or
divergence) w.r.t.\ to the space components and $\D_jf(y) = 0,
y\in\flc(x)\cup\blc(x)$:
\begin{align}
  I_{g\Ll}^{00}&=\D_j\Kg{\vp\D^j\vp},\\
  I_{g\Ll}^{0j}&=\D^j\Kg{\vp\D^0\vp}.  
\end{align}

The trace of the improved \emt\ expresses the breaking of scale
invariance. If the dilatations are defined by (cp.\
\eqref{def:dilcurrent}) 
\begin{equation}
  \label{def:intD}
  D^\mu_{g\Ll}=x_\nu\Tmnig,
\end{equation}
we obtain
\begin{equation}
  \label{eq:breakintD}
  \D_\mu D^\mu_{g\Ll}={{\Theta_{\mathrm{imp}\, g\Ll}}^\mu}_\mu
  +x^\mu\D_\mu g \Llg.
\end{equation}
If we define the dilatations alternatively by (cp.\ \eqref{eq:classdilat})
\begin{equation}
  \label{def:intDtilde}
  \widetilde{D}^\mu_{g\Ll}=x_\nu\Tmncg+\Kg{\vp\D^\mu\vp},
\end{equation}
we find the same breaking \eqref{eq:breakintD}. The (not time
independent) charge remains unchanged due to:
\begin{equation}
  \widetilde{D}^0_{g\Ll}-D^0_{g\Ll}
  =\frac{2}{3}\D_j\Bigl(x^{[j}\Kg{\vp\D^{0]}\vp}\Bigr).
\end{equation}
The next section gives a proof of \eqref{eq:sternid}.

\section{Proof of the Ward identity}
\label{sec:ward2}
We prove equation \eqref{eq:sternid} in analogy to the conservation
of the canonical \emt\ by the validity of the following Ward
identity:%
\footnote{$d_k$ is the dimension of the $k$'th monomial in the
  $T$-product:
$d_k T(N) = T (W_1,\dots,d W_k,\dots, W_n)$ and $d$ given by
\eqref{def:dimop}.}
\begin{multline}
  \D^\mu_xT(\vp\D^\nu\vp,N)(x,y_N)=
  T(\D^\mu\vp\D^\nu\vp,N)(x,y_N)+T(\vp\D^\mu\D^\nu\vp,N)(x,y_N)+\\
  -c\eta^{\mu\nu}T(\vp\square\vp,N)(x,y_N)
  +ic\eta^{\mu\nu}\sum_{k=1}^n\delta(y_k-x)d_kT(N)(y_N).
\tag{\Wzwei}
\end{multline}
The proof follows the procedure in section~\ref{sec:ward1}. For $N =
\emptyset$ \Wzwei\ is obviously fulfiled. Then we make a double
induction over $n=|N|$ and the degree of the Wick monomials.  Under
the assumption that \Wzwei\ is fulfiled in lower orders we denote the
possible anomaly by
\begin{multline}
\begin{split}
a^{\mu\nu}(N)(x,y_N)
  &\doteq\D^\mu_xT(\vp\D^\nu\vp,N)(x,y_N)
  -T(\D^\mu\vp\D^\nu\vp,N)(x,y_N)+\\
  &\quad-T(\vp\D^\mu\D^\nu\vp,N)(x,y_N)
  +c\eta^{\mu\nu}T(\vp\square\vp,N)(x,y_N)+\\
  &\quad-i\eta^{\mu\nu}c\sum_{k=1}^n\delta(y_k-x)d_kT(N)(y_N).
\end{split}
\label{def:anomly2}
\end{multline}
We show that a normalization of the $T(\vp\D^\mu\D^\nu,N)$ exists such
that the anomaly vanishes if we require the following normalization
condition for the twice derivated basic field:
\begin{equation}
  \label{eq:twicederiv}
  T(\D_\mu\D_\nu\vp,N)(x,y_N)\doteq\D_\mu^x\D_\nu^x T(\vp,N)(x,y_N).
\end{equation}
By comparison to \eqref{eq:nvierintegrated}, the integrated form of
\Nvier, this is achieved by fixing the two point function: $\omega_0
(T(\D^\mu\D^\nu\vp,\vp)(x,y)) = i \D^\mu \D^\nu D^F (x-y)$. For the
corresponding interacting fields we have $\Kg{\D^\mu\D^\nu\vp} =
\D^\mu \D^\nu \vpg$. In order to condense the notation we introduce
the following abbreviation:
\begin{equation}
  \label{eq:defNk}
  \Nk\doteq\left\{W_1,\dots,\parz{W_k}{\vp},\dots,W_n\right\}.
\end{equation}
\subsection*{Step 1}
We commute the anomaly with the free field $\vp(z)$:
\begin{multline}
  [a^{\mu\nu}(N)(x,y_N),\vp(z)]=\\
  \begin{split}
    &=\D^\mu_x[T(\vp\D^\nu\vp,N)(x,y_N),\vp(z)]
    -[T(\D^\mu\vp\D^\nu\vp,N)(x,y_N),\vp(z)]+\\
    &\quad-[T(\vp\D^\mu\D^\nu\vp,N)(x,y_N),\vp(z)]
    +c\eta^{\mu\nu}[T(\vp\square\vp,N)(x,y_N),\vp(z)]+\\
    &\quad-ic\eta^{\mu\nu}
    \sum_{l=1}^n\delta(y_l-x)d_k[T(N)(y_N),\vp(z)]\\
    &=\D^\mu_x\D^\nu_x T(\vp,N)(x,y_N)iD(x-z)
    +\D^\nu_xT(\vp,N)(x,y_N)\D^\mu D(x-z)+\\
    &\quad+\D^\mu_x T(\vp,N)(x,y_N)i\D^\nu D(x-z)
    +T(\vp,N)(x,y_N)i\D^\mu\D^\nu D(x-z)+\\
    &\quad-\D^\nu_x T(\vp,N)(x,y_N)i\D^\mu D(x-z)
    -\D^\mu_x T(\vp,N)(x,y_N)i\D^\nu D(x-z)+\\
    &\quad-T(\D^\mu\D^\nu\vp,N)(x,y_N)iD(x-z)
    -T(\vp,N)(x,y_N)i\D^\mu\D^\nu D(x-z)+\\
    &\quad+c\eta^{\mu\nu}T(\square\vp,N)(x,y_N)iD(x-z)+\\
    &\quad-ic\eta^{\mu\nu}
    \sum_{k=1}^n\delta(y_k-x)T(\Nk)(y_N)iD(y_k-z)+\\
    &\quad+\sum_{k=1}^n a^{\mu\nu}(\Nk)(x,y_N)iD(y_k-z)\\
    &=0,
\end{split}
\end{multline}
if we apply the induction assumption ($a^{\mu\nu}(\Nk)=0$) to the last
line and and the normalization \eqref{eq:twicederiv} and normalization
condition \Nvier\ to the previous lines. Therefore, the anomaly has to
be a vacuum expectation value:
\begin{equation}
  \label{eq:anomalyvac}
  a^{\mu\nu}(N)(x,y_N)=\omega_0\K{a^{\mu\nu}(N)(x,y_N)}.
\end{equation}

\subsection*{Step 2}
We show that the anomaly vanishes if one Wick monomial is a basic
generator $\vp$. Because of Step 1 we only consider vacuum expectation
values and use \eqref{eq:nvierintegrated}:
\begin{multline}
  \omega_0\K{a^{\mu\nu}(N,\vp)(x,y_N,z)}=\\
  \begin{split}
    &=\D^\mu_x\omega_0\K{T(\vp\D^\nu\vp,N,\vp)(x,y_N,z)}+\\
    &\quad-\omega_0\K{T(\D^\mu\vp\D^\nu\vp,N,\vp)(x,y_N,z)}+\\
    &\quad-\omega_0\K{T(\vp\D^\mu\D^\nu\vp,N,\vp)(x,y_N,z)}+\\
    &\quad+c\eta^{\mu\nu}\omega_0\K{T(\vp\square\vp,N,\vp)(x,y_N,z)}+\\
    &\quad-ic\eta^{\mu\nu}
    \sum_{l=1}^n\delta(y_l-x)\omega_0(d_lT(N,\vp)(y_N,z))+\\
    &\quad-ic\eta^{\mu\nu}\delta(z-x)\omega_0(T(N,\vp)(y_N,z))\\
    &=-i\D^\mu D^F(z-x)\omega_0\K{T(\D^\nu\vp,N)(x,y_N)}+\\
    &\quad+iD^F(z-x)\D^\mu_x\omega_0\K{T(\D^\nu\vp,N)(x,y_N)}+\\
    &\quad+i\D^\mu\D^\nu D^F(z-x)\omega_0\K{T(\vp,N)(x,y_N)}+\\
    &\quad-i\D^\nu D^F(z-x)\D^\mu_x\omega_0\K{T(\vp,N)(x,y_N)}+\\
    &\quad+i\D^\mu D^F(z-x)\omega_0\K{T(\D^\nu\vp,N)(x,y_N)}+\\
    &\quad+i\D^\nu D^F(z-x)\omega_0\K{T(\D^\mu\vp,N)(x,y_N)}+\\
    &\quad-iD^F(z-x)\omega_0\K{T(\D^\mu\D^\nu\vp,N)(x,y_N)}+\\
    &\quad-i\D^\mu\D^\nu D^F(z-x)\omega_0\K{T(\vp,N)(x,y_N)}+\\
    &\quad+ic\eta^{\mu\nu}D^F(z-x)\omega_0\K{T(\square\vp,N)(x,y_N)}+\\
    &\quad+ic\eta^{\mu\nu}\square D^F(z-x)\omega_0\K{T(\vp,N)(x,y_N)}+\\
    &\quad-ic\eta^{\mu\nu}\sum_{k=1}^n
    \delta(x-y_k)\omega_0(T(\Nk)(y_N))iD^F(z-y_k)+\\
    &\quad-ic\eta^{\mu\nu}\delta(z-x)\sum_{k=1}^n
    iD^F(z-y_k)\omega_0(T(\Nk)(y_N))+\\
    &\quad+\sum_{k=1}^n\omega_0\K{a^{\mu\nu}(\Nk)(x,y_N)}iD^F(z-y_k)\\
    &=0.
\end{split}
\end{multline}
We have used $\square D^F=\delta$, the normalization \eqref{eq:twicederiv},
\Nvier\ and the induction assumption.

\subsection*{Step 3}
Because of the induction assumption the anomaly is a local term:
\begin{equation}
  \label{eq:anomalylocal}
  \omega_0(a^{\mu\nu}(N)(x,y))=M^{\mu\nu}(\D)\delta(x,y_N),
\end{equation}
where $\delta(x,y_N)=\delta(x-y_1)\dots\delta(x-y_n)$ and $M^{\mu\nu}$
is a polynomial in $\D=(\D_1,\dots,\D_n)$ of degree $\leq 4$.
Therefore we can absorb the anomaly by the following normalization:
\begin{align}
  \omega_0\K{T(\vp\D^\mu\D^\nu\vp,N)(x,y_N)}&\rightarrow
  \omega_0\K{T(\vp\D^\mu\D^\nu\vp,N)(x,y_N)}+M^{\mu\nu}(\D)\delta(x,y_N)+
  \notag\\
  &\quad-\frac{c}{4c-1}\eta^{\mu\nu}{M^\rho}_\rho(\D)\delta(x,y_N),\\
\Rightarrow\omega_0\K{T(\vp\square\vp,N)(x,y_N)}&\rightarrow
  \omega_0\K{T(\vp\square\vp,N)(x,y_N)}
  -\frac{1}{4c-1}{M^\rho}_\rho(\D)\delta(x,y_N),\\
\Rightarrow\omega_0(a^{\mu\nu}(N)(x,y_N))&\rightarrow 0.
\end{align}
As long as the trace of the anomaly ${M^\rho}_\rho$ is non vanishing
the normalization can only be done for $c\not=\frac{1}{4}$.


\section{The anomalous dimension}
\label{sec:anomal}
Although the dilatation current is not conserved in the interacting
theory we calculate the commutator of the corresponding charge in
order to obtain a term that measures the anomalous dimension. From
\Wzwei\ we derive the following Ward identities for the
$R$-products:
\begin{multline}
\D_\mu^y R(N,\vp\D^\mu\vp;W)(y_N,y;x)=\\
\begin{split}
    &=R(N,\D_\mu\vp\D^\mu\vp;W)(y_N,y;x)
  +(1-4c)R(N,\vp\square\vp;W)(y_N,y;x)+  \\
  &\quad+4ic\sum_{j\in N}\delta(y_j-y)d_jR(N;W)(y_N;x)
  +4ic\delta(x-y)R(N;dW)(y_N;x).
\end{split}
\label{eq:ward2R1}
\end{multline}
If we set $M=\{V_1,\dots,V_m\}, V_i\in\Ba$ we find:
\begin{multline}
\D_\mu^x R(N;\vp\D^\mu\vp,M)(y_N;x,x_M)=\\
\begin{split}
  &=R(N,\D_\mu\vp\D^\mu\vp,M)(y_N;x,x_M)
  +(1-4c)R(N;\vp\square\vp,M)(y_N;x,x_M)+ \\
  &\quad+4ic\sum_{j\in N}\delta(y_j-x)d_j
  R(N\setminus j;j,M)\K{y_{N\setminus j};y_j,x_M}+\\
  &\quad+4ic\sum_{j\in M}\delta(x_j-x)d_j
  R(N;M)(y_N;x_M).
\end{split}
\label{eq:ward2R2}
\end{multline}
A calculation similar to the one given in section~\ref{sec:intcharge}
shows that the following commutation relation holds:
\begin{multline}
  \left[D_{g\Ll}^0(f),\Wg(x)\right]=\\
  \begin{split}
    &=i\Kg{dW}(x)+ix^\mu\D_\mu\Wg(x)+\\
    &\quad+(1-4c)\biggl\{
    \sum_{n=0}^\infty\frac{i^n}{n!}
    \int\dif y_N\dif y \, g(y_N)\times\\
    &\quad\times\Bigl[
    \Bigl(R(N,gd\Ll;W)(y_N,y;x)+R(N,\vp\square\vp;W)(y_N,y;x)\Bigr)k(y)+\\
    &\quad+
    \Bigl(R(N,W;gd\Ll)(y_N,x;y)+R(N,W;\vp\square\vp)(y_N,x;y)\Bigr)
    \ktilde(y)\Bigr]+\\
    &\quad-i\Kg{dW}(x)
    \biggr\}.
  \end{split}
\end{multline}
The terms in braces are the anomalous contributions. They are
necessarily non vanishing (because of the normalization that excludes
case $(ii)$ in \eqref{eq:innerintfieldeq}). Moreover they are operator
valued.  We compute them for the case $W=\vp$, where the normalization
of the $R$-products is known due to \Nvier\ \cite{pap:duet-fred1}.
\begin{multline}
  \left[D_{g\Ll}^0(f),\vpg(x)\right]=\\
  \begin{split}
    &=i\vpg(x)+ix^\mu\D_\mu\vpg(x)+\\
    &\quad+i(1-4c)\Biggl\{
    \sum_{n=0}^\infty\frac{i^n}{n!}
    \int\dif y_N\dif y \, g(y_N)\times\\
    &\quad\times\biggl\{
    \sum_{l=1}^n
    \biggl[ 
      \Dret(x-y_l) \biggl( g(y)
      R\K{N\setminus l,d\Ll;\parz{\Ll}{\vp}}\K{y_{N\setminus l},y;y_l}+\\
      &\qquad\qquad\qquad+ R\K{N\setminus l,\vp\square\vp;\parz{\Ll}{\vp}}
      \K{y_{N\setminus l},y;y_l}
      \biggr)k(y)+\\
      &\qquad\qquad\qquad+ \Dav(x-y_l) \biggl( g(y)
      R\K{N^{(e_l)};d\Ll}\K{y_N;y}+ \\
      &\qquad\qquad\qquad+R\K{N^{(e_l)};\vp\square\vp}\K{y_N;y}
      \biggr)\ktilde(y)
    \biggr]+ \\
    &\quad+\bigl(\Dret(x-y)k(y)+\Dav(x-y)\ktilde(y)\bigl)
    R\K{N;\vp\parz{^2\Ll}{\vp^2}}(y_N;y)k(y)
    \biggr\}\Biggr\}.
  \end{split}
\end{multline}
On the other hand we can study the interacting Ward identities of the
dilatations. Since time ordered products of interacting fields are
already determined by time ordered products of free fields with an
arbitrary number of interactions according to \eqref{eq:Tint}, we find with
\eqref{eq:ward2R1}, \eqref{eq:ward2R2}:
\begin{multline}
  \D_\mu^x T\Kg{D^\mu_{g\Ll},\vp,\dots,\vp}(x,x_1,\dots,x_m)=\\
  \begin{split}
    &=T\Kg{ {{\Theta_{\mathrm{imp}}}^\mu}_\mu,\vp,\dots,\vp}
    (x,x_1,\dots,x_m)+\\
    &\quad+x^\mu \D_\mu g(x)
    T\Kg{\Ll,\vp,\dots,\vp}(x,x_1,\dots,x_m)+\\
    &\quad+i\sum_{l=1}^m
    \delta(x_l-x)(4c+x^\mu_l\D^l_\mu)T\Kg{\vp,\dots,\vp}(x_1,\dots,x_m).
  \end{split}
\end{multline}
The anomalous terms defined above are no anomalous dimensions in the
form of a formal (local) power series in the coupling that multiply
the interacting fields. This is no surprise since already for the free
massive field the dilatations are no symmetry and the above method of
commuting with the (non time invariant) dilatation charge does not
produce a number in that case. Nevertheless also the massive scalar
field is given the canonical dimension one. In \cite{pap:coleman} the
authors suggest to define the dimension by equal time commutators.
For the free field this method is well defined and produces the right
result. But it is not clear that this carries over to the interacting
fields since they may become more singular objects in the time
coordinate as their free counterparts.


\chapter{Operator product expansions}
\label{chap:ope}

In \cite{pap:wil1} Wilson suggested that a product of interacting
field operators on separated points could be expanded into a sum of
local operators if the separation goes to zero. Such an expansion is
called \emph{operator product expansion} (\ope).  

Zimmermann has introduced his notion of perturbative normal products in
\cite{lect:zimmermann,pap:zimmermann1}. In his approach (interacting)
operators are always defined via their Green's functions, namely vacuum
expectation values of time ordered products with an arbitrary number
of fields. The Green's functions are renormalized by BPHZ- or in the
massless case by BPHZL subtraction. 

In \cite{lect:zimmermann,pap:zimmermann2} he gave a generalization of
these local products to multi local ones. They admit a restriction of
all coordinates to one yielding the local normal product. By
relating a bilocal product to a time ordered one he derived an
\ope\ verifying Wilson's hypothesis perturbatively. He found explicit
formulas for the expansion coefficients in the form of
Green's functions.

Since in the framework of Bogoliubov and Epstein-Glaser the operators
are defined directly we try to mimick Zimmermann's procedure. We 
define a new time ordered product containing a bilocal expression
which allows for setting its coordinates to the same value (in the
sense of a restriction of a distribution). We are concerned with
scalar fields only and our bilocal $T$-product has only the basic
generators in the bilocal insertion. 

The definition of a bilocal $T$-product gives rise to a corresponding
bilocal interacting field. Following Zimmermann's notation we also
call this object a normal product. The transition from the
$T$-products to the interacting fields automatically generates an
\ope\ for the time ordered product of two interacting fields. The
coefficients depend on the coupling only locally. In $\vp^4$-theory
two coefficients appear.  One consist of graphs that contribute
to wave function and mass renormalization only. The other collects
graphs contributing only to coupling constant renormalization.
 
With the normal product defined we investigate the first step towards
the definition of a state on the local algebra. We find the
corresponding two point function to be positive in an appropriate
sense as a formal power series.


\section{Bilocal time ordered products}
\label{sec:bilocal}
The word \emph{bilocal time ordered product} means a usual time
ordered product where only \emph{one} entry is a bilocal expression.
We derive an explicit formula that defines these products for two
scalar fields. Let us mention that our expression is only explicit up
to normalization terms which restore broken Lorentz covariance (cf.
chapter~\ref{chap:timeordered}, section~\ref{sec:poincare}). We state
the problem first:

Consider the case of the interacting scalar fields $\vpg$. We aim at
the definition of a \emph{normal product} $\wpvg(x_1,x_2)$ with the
property 
\begin{equation}
\lim_{\xi\rightarrow 0} \wpvg(x+\xi,x-\xi) =
\Kg{\vp^2}(x),
\label{eq:xinull}
\end{equation}
where $x=\frac{x_1+x_2}{2}$ denotes the central coordinate and
$\xi=\frac{x_1-x_2}{2}$ the difference coordinate. Taking the
definition of interacting fields into account
(chapter~\ref{chap:intfields}), we notice that it suffices to define
the corresponding $T$-products. This is illustrated in

\subsection{The easiest example.} 
\begin{figure}[htbp]
\[
\parbox{30mm}{
\centering
\begin{fmfgraph*}(50,50)
\fmfleft{i1,i2}\fmfright{o1,o2}
\fmf{plain}{i1,v,o1}
\fmf{plain}{i2,v,o2}
\fmfdot{i1,i2,v}
\fmflabel{$x_1$}{i1}
\fmflabel{$x_2$}{i2}
\fmflabel{$y$}{v}
\end{fmfgraph*}
}
\quad
\stackrel{\xi\rightarrow0}\longrightarrow
\quad
\parbox{30mm}{
\centering
\begin{fmfgraph*}(60,60)
\fmfleft{i1}\fmfright{o1,o2}
\fmf{plain,left,tension=.3}{i1,v,i1}
\fmf{plain}{v,o1}
\fmf{plain}{v,o2}
\fmfdot{i1,v}
\fmflabel{$x$}{i1}
\fmflabel{$y$}{v}
\end{fmfgraph*}
}
\]
\caption{\slshape A loop is generated if $\xi\rightarrow0$.}
\label{fig:loop}
\end{figure}
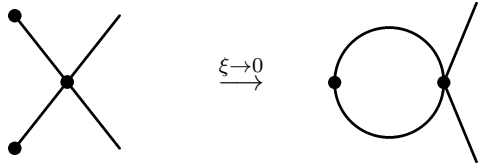

We consider scalar $\frac{\vp^4}{4!}$-theory. The task is to determine
$T(\Ll,\wick{\vp,\vp})(y,x_1,x_2)$.  We proceed in the following way:
First we define ${^0T}(\Ll,\wick{\vp,\vp})(y,x_1,x_2)$ which consists
of the usual $T$-product with $x_1,x_2$-contractions omitted. Then we
subtract a suitable term with support on $y=x$ that allows for the
restriction. Obviously we have
\begin{align}
  \label{eq:Twick1}
^0T(\Ll,\wick{\vp,\vp})(y,x_1,x_2)
&=\wick{\frac{\vp(y)^4}{4!}\vp(x_1)\vp(x_2)}+ \\
&\quad+i\Delta_F(y-x_1)\wick{\frac{\vp(y)^3}{3!}\vp(x_2)}\\
&\quad+i\Delta_F(y-x_2)\wick{\frac{\vp(y)^3}{3!}\vp(x_1)}+\\
&\quad+i\Delta_F(y-x_1)i\Delta_F(y-x_2)\wick{\frac{\vp(y)^2}{2!}}.
\end{align}
The problem for the definition of the restriction of $^0T$ emerges
on the last line, where after setting $x_1=x_2\ (=x)$ we obtain a
$\Delta_F(y-x)^2$-term that is not well defined. Graphically this
procedure produces a loop, see figure~\ref{fig:loop}.
But we can find a term which subtraction allows for putting
$x_1=x_2=x$, yielding $T(\Ll,\vp^2)(y,x)$ with its
respective normalization. We claim that with
\begin{equation}
  \label{eq:defezwei}
  e^{(2)}(\xi)\doteq\int\dif z\,i\Delta^F(z-\xi)i\Delta^F(z+\xi)w(z)
\end{equation}
the subtracted distribution
\begin{equation}
  \label{def:DeltaFren}
\Bigl. i\Delta^F(y-x_1)i\Delta^F(y-x_2)\Bigr\vert_R
=i\Delta^F(y-x_1)i\Delta^F(y-x_2)-\delta(x-y)e^{(2)}(\xi)
\end{equation}
allows for the coincidence $x_1=x_2\Leftrightarrow \xi=0$ after
smearing with a test function. We calculate
\begin{multline}
\label{eq:calc1}
  \int\dif y\,\Bigl.i\Delta_F(y-x_1)i\Delta_F(y-x_2)\Bigr\vert_R
  f(y)=\\
  =\int\dif y\,i\Delta_F(y-x_1)i\Delta_F(y-x_2)(f(y)-w(y-x)f(x)),
\end{multline}
and this implies
\begin{equation}
  \label{eq:limesxi}
  \lim_{\xi\rightarrow 0}
\scp{\Bigl.i\Delta^F(\cdot+\xi)i\Delta^F(\cdot-\xi)\Bigr\vert_R}{f_x}
=\scp{i^2(\Delta^F)^2}{\W{0}{w}f_x},
\end{equation}
with $f_x(y)=f(y+x)$, which proves our claim. Note, that the
coefficient in front of the $\delta$ term is given by
\begin{equation}
  \label{eq:ezwei}
  e^{(2)}(\xi)=\int\dif z\,\omega_0
\K{^0T(\Ll^{(2)},\wick{\vp,\vp})(z;\xi,-\xi)}w(z),
\end{equation}
where the $\mbox{}^{(2)}$ again denotes twice differentiation with
respect to $\vp$. Now we return to the complete
$T$-product. Collecting everything into one expression we find
\begin{equation}
  \label{eq:Tfirstorder}
  T(\Ll,\wick{\vp,\vp})(y,x_1,x_2)
  ={^0T}(\Ll,\wick{\vp,\vp})(y,x_1,x_2)
  -\delta(y-x)e^{(2)}(\xi)\frac{\wick{\vp(x)^2}}{2}.
\end{equation}
Because of the $\delta$ distribution we have set the coordinate of the
last Wick monomial to $x$.
\begin{rem}
  As well as the center coordinate $x$ we could have chosen any other
  point on the straight line between $x_1$ and $x_2$ (or also in the
  causally completed region spanned by these two points) for
  subtraction. But our choice is inspired by Zimmermann's work,
  moreover yielding a symmetrical solution.
\end{rem}

Now we generalize the idea of subtracting a local term that
compensates the overall divergence. Thereby we make use of the method
of Epstein-Glaser where all lower order divergencies are appropriately
handled by an

\subsection{Inductive causal construction.}
We begin with a brief overview of the construction. Motivated by our
example above we define the bilocal $T$-pro\-ducts in the following way:
Denoting the sub manifold 
\begin{equation}
\Diag^x_n \doteq \left\{(y_1,\dots,y_n,x_1,x_2)
\in \MM^{n+2} | y_1 = \dots = y_n = \frac{x_1+x_2}{2} \right\},
\label{eq:def:diagxn}
\end{equation}
we require the bilocal $T$-product of order $n$ (where $n$ is the
number of the coordinates not including the two bilocal points) to be
given by all bilocal $T$-products of lower and all local $T$-products
of lower and same order on $\MM^{n+2}\setminus\Diag^x_n$. This
provides for a $^0T$-product which yields the local $^0T$-product in
the limit $\xi \rightarrow 0$. Then we subtract a term with support on
$\Diag^x_n$ such that the limit $\xi \rightarrow 0$ exists and yields
the corresponding $T$-product. Hence in every order the difference
between a local $T$-product (with $x_1,x_2$-contractions omitted) and
a bilocal one only consists of these local terms.

In zero'th order we define: $ T(\wpv)(x_1,x_2) \doteq
\wick{\vp(x_1)\vp(x_2)} $. Using our shorthand
notation for the arguments ($N$ can be any set of Wick monomials) we
require the following causal factorization properties:
\begin{multline}
T(N\wpv)(y_N,x_1,x_2)=\\
=\begin{cases}
T(I)(y_I)T\K{N\setminus I,\wpv}\K{y_{N\setminus I},x_1,x_2},
&\text{if }I\gtrsim N\setminus I,x_{1},x_{2},
I \not=\emptyset, \\[1ex]
T\K{I,\wpv}(y_I,x_1,x_2)T(N\setminus I)\K{y_{N\setminus I}},
&\text{if }I,x_{1}, x_{2},\gtrsim N\setminus I,
I\not=N, \\[1ex]
T(N,\vp,\vp)\K{y_N,x_1,x_2}+\\
+T(I)(y_I)\bigl[T\K{N\setminus I,\wpv}\K{y_{N\setminus I},x_1,x_2}+\\
-T(N\setminus I,\vp,\vp)\K{y_{N\setminus I},x_1,x_2}\bigr],
&\text{if }I, x_{1}\gtrsim N\setminus I, x, x_{2},I\not=\emptyset,\\[1ex]
T(N,\vp,\vp)\K{y_N,x_1,x_2}+\\
+\bigl[T(I,\wpv)(y_I,x_1,x_2)+\\
-T(I,\vp,\vp)(y_I,x_1,x_2)\bigr]
T\K{N\setminus I}\K{y_{N\setminus I}},
&\text{if }I, x_{1}, x,\gtrsim N\setminus I, x_{2},I\not=N,\\[1ex]
\text{the last two expressions with $x_1 \leftrightarrow x_2$.}
\end{cases}
\label{def:zeroT}
\end{multline}
We convince ourselves that this a reasonable causal factorization. If
$x_1, x_2$ are contracted to a point, then the first two equations
obviously give the right causal decompositions. We investigate the last
term on the third line, where $I, x_{1}\gtrsim N\setminus I, x,
x_{2}$. If $\xi\rightarrow 0$ we find
\begin{multline}
T(I)(y_I)T(N\setminus I,\vp,\vp)\K{y_{N\setminus I},x_1,x_2}=\\
\begin{split}
  &=T(I)(y_I)\vp(x_1)T(N\setminus I,\vp)\K{y_{N\setminus I},x_2}\\
  &\stackrel{\xi\rightarrow 0}{=}
  T(I,\vp)(y_I,x_1)T(N\setminus I,\vp)\K{y_{N\setminus I},x_2}\\
  &=T(N,\vp,\vp)\K{y_N,x_1,x_2},
\end{split}
\end{multline}
since $x_1$ becomes earlier than all $y_I$. This term cancels
the first term of the third line in \eqref{def:zeroT} leaving the
first line of \eqref{def:zeroT}. A similar
consideration leads to the same conclusion also for the last line of
\eqref{def:zeroT}.

This shows that the bilocal $T$-product is completely determined by
\eqref{def:zeroT} up to the sub manifold $\Diag_n^x$. In contrast to
the definition of a local $T$-product where one has to perform an
extension of the numerical distributions involved, the bilocal product
can be defined by a suitable subtraction. This is due to the fact that
all terms are well defined distributions in $n+2$ variables. We state
the solution:
\begin{multline}
T\K{N,\wpv}(y_N,x_1,x_2)=\\
\begin{split}
  &=T(N,\vp,\vp)(y_N,x_1,x_2)
  -\omega_0\K{T(\vp,\vp)(x_1,x_2)}T(N)(y_N)+\\
  &\quad- \sum_{\substack{I\subset N\\ I\not=\emptyset}}
  \sum_{|\gamma|\leq \omega_I}\sum_{|\alpha|\leq\omega_I-|\gamma|}
  \sum_{\mu+\nu=\alpha}\frac{(-)^{|\mu|}}{\mu!\nu!\gamma!}
  \D^\mu\delta(I-x)\times\\
  &\quad\times
  \bigl(e^{\alpha(\gamma)}_I(\xi)-a^{\alpha(\gamma)}_I(\xi)\bigr)
  T\K{N\setminus I, \D^\nu \vp^\gamma}\K{y_{N\setminus I},x},
\end{split}
\label{def:bilocalT}
\end{multline}
where $\gamma\in\NN^{|I|}$ is a multi index and $\mu,\nu\in\NN^{4|I|}$
are multi quadri indices.  Therefore the term $\D^\mu\delta(I-x) =
\prod_{i\in I} \D^{\mu_i}\delta(y_i-x)$ (and $\mu_1$ can be
$\mu\nu\rho$ for example, with $\mu,\nu,\rho$ usual Lorentz indices).
The number $\omega_I$ refers to the singular order of the numerical
distribution $\omega_0\K{T(I,\vp^2)(y_I,x)}$ and is given by $\omega_I
= \sum_{i\in I} \dim W_i + 2 \dim \vp - 4|I|$. Hence the sum over $I$
only runs over subsets for which the coincidence of $x_1,x_2$ produces
distributions with non negative singular order. These are ordered into
subgraphs by the sum over $\gamma$. Only graphs for which their
corresponding order, namely $\omega_I - |\gamma|$, is non negative
($\gamma$ is a derivative w.r.t.\ $\vp$ and therefore the singular
order decreases with increasing $|\gamma|$) are taken into account.
The sum over $\alpha$ refers to the usual subtraction procedure
(running from 0 up to the order of singularity of the corresponding
distribution). As a matter of fact it has to be split into an action
on $\delta$ and on $T(\dots,\vp^\gamma)$ since we changed the
coordinate of $\vp^{\gamma_i}(y_i)$ into $\vp^{\gamma_i}(x)$ according
to the $\delta$ function (see also the example at the beginning).

To explain the coefficients, we have to introduce the corresponding
$^0T$-product which is the same expression like \eqref{def:bilocalT}
up to the last sum which does not contain the term $I=N$. So
\begin{multline}
{^0T}\K{N,\wpv}(y_N,x_1,x_2)=\\
\begin{split}
  &=T(N,\vp,\vp)(y_N,x_1,x_2)
  -\omega_0\K{T(\vp,\vp)(x_1,x_2)}T(N)(y_N)+\\
  &\quad-\sum_{\substack{I\subset N\\ I\not=\emptyset\\I\not=N}}
  \sum_{|\gamma|\leq \omega_I}\sum_{|\alpha|\leq\omega_I-|\gamma|}
  \sum_{\mu+\nu=\alpha}\frac{(-)^{|\mu|}}{\mu!\nu!\gamma!}
  \D^\mu\delta(I-x)\times\\
  &\quad\times
  \bigl(e^{\alpha(\gamma)}_I(\xi)-a^{\alpha(\gamma)}_I(\xi)\bigr)
  T\K{N\setminus I, \D^\nu \vp^\gamma}\K{y_{N\setminus I},x}.
\end{split}
\label{def:bilocalnullT}
\end{multline}
Hence $T$ and $^0T$ only differ by a term with support on $\Diag^x_n$.
The coefficients $e^{\alpha(\gamma)}_I$ are given by the expression
\begin{equation}
  \label{def:aI}
  e^{\alpha(\gamma)}_I(\xi)
=\int\dif z_I\, z^\alpha
\omega_0\K{{^0T}\K{I^{(\gamma)},\wpv}(y_I,\xi,-\xi)}w_{I^{(\gamma)}}(z_I),
\end{equation}
where again $I^{(\gamma)} = \{W_i^{(\gamma_i)},i\in I\}$ and the
exponent $\mbox{}^{(\gamma_i)}$ means $\gamma_i$ fold differentiation
with respect to $\vp$ in $\Ba$. The coefficients
$a_I^{\alpha(\gamma)}$ are chosen in such a way, that Lorentz
covariance is conserved. The function $w_{I^{(\gamma)}}$ is the
auxiliary function used in the extension process of the distribution $
\omega_0\K{T\K{I^{(\gamma)},\vp^2}(y_I,x)}$ of the $|I|$ difference
coordinates $y_1-x,\dots,y_{|I|}-x$.%
\footnote{Without loss of generality $I=\{1,\dots,|I|\}$.}

We require our bilocal products to fulfil normalization condition
\Ndrei\ in the corresponding form, namely
\begin{multline}
  \bigl[
T\K{N,\wpv}(y_N,x_1,x_2),\vp(z)\bigr]=\\
\begin{split}
  &=i\sum_{k=1}^n T\K{N^{(e_k)},\wpv}(y_N,x_1,x_2)\Delta(y_k-z)+\\
  &\quad+iT\K{N,\vp}(y_N,x_1)\Delta(x_2-z)
  +iT\K{N,\vp}(y_N,x_2)\Delta(x_1-z),
\end{split}
\tag{\Ndreistrich}
\end{multline}
and $e_k = (0,\dots,1,\dots,0) \in \NN^n$ with the $1$ in the $k$'th
position.


\subsection{Proof of the restriction property.}
We show that \eqref{def:bilocalT} yields the right $T$-product
by restricting $x_1=x_2$. Our proof requires the validity of
\Ndreistrich\ which is proven in the next subsection. Unfortunately
our proof still lacks an existence statement for the distributions
$a_I(\xi)$, necessary for the conservation of Poincar\'e
covariance. Hence we have to assume that they exist.

Assume that up to order $n-1$ the bilocal $T(N,\wpv)(y_N,x_1,x_2)$
exists and its restriction $x_1=x_2$ is given by $T(N,\vp^2)(y_N,x)$.
We show that $T$ has the right causal factorization \eqref{def:zeroT}.
We write \eqref{def:bilocalT} as
\begin{multline}
  T(N,\wpv)(y_N,x_1,x_2)= T(N,\vp,\vp)(y_N,x_1,x_2)+\\
-\sum_{I\subset N}\sum_{|\gamma|\leq \omega_I}
\sum_{|\mu|+|\nu|\leq\omega_I-|\gamma|}
E^{\mu\nu(\gamma)}(I,\vp,\vp)(y_I,x,\xi)
T(N\setminus I,\D^\nu\vp^\gamma)\K{y_{N\setminus I},x}.
\label{eq:bilocalT}
\end{multline}
The $E^{\mu\nu(\gamma)}$ are given by 
\begin{align}
  \label{def:E}
E^{\mu\nu(\gamma)}(I,\vp,\vp)(y_I,x,\xi)
&=\frac{(-)^\mu}{\mu!\nu!\gamma!}\D^\mu\delta(I-x)
\bigl(e_I^{\mu+\nu(\gamma)}(\xi)-a_I^{\mu+\nu(\gamma)}(\xi)\bigr),\\ 
\text{with }E^{\mu\nu(\gamma)}(\emptyset,\vp,\vp)(x,\xi) 
&=\delta^\mu_0\delta^\nu_0 \delta^\gamma_0 
\omega_0\K{T(\vp,\vp)(x_1,x_2)}\big\vert_{x_1-x_2=2\xi}\\
\text{and }
\supp E^{\mu\nu(\gamma)}(N,\vp,\vp)(y_N,x,\xi) 
&\subset \Diag_n^x, N\not=\emptyset.
\end{align}
 
If $L\gtrsim N\setminus L, x, x_1, x_2$, we have from
\eqref{eq:bilocalT}: 
\begin{multline}
  T(N,\wpv)(y_N,x_1,x_2)=\\
  \begin{split}
    &=T(L)(y_L)T(N\setminus L,\vp,\vp)\K{y_{N\setminus L},x_1,x_2}+\\
    &\quad-T(L)(y_L)E(\emptyset,\vp,\vp)(x,\xi)T(N\setminus L)
    \K{y_{N\setminus L}} +\\
    &\quad-\sum_{\substack{I\subset N\setminus L\\I\not=\emptyset}}
    \sum_{\gamma,\mu,\nu} E^{\mu\nu(\gamma)}(I,\vp,\vp)(y_I,x,\xi)
    T(N\setminus I,\D^\nu\vp^\gamma)\K{y_{N\setminus I},x}\\
    &=T(L)(y_L)T(N\setminus L,\vp,\vp)(y_N,x_1,x_2)+\\
    &\quad-T(L)(y_L)\sum_{I\subset N\setminus L} \sum_{\gamma,\mu,\nu}
    E^{\mu\nu(\gamma)}(I,\vp,\vp)(y_I,x,\xi)\times\\
    &\qquad\times T((N\setminus L)\setminus I,\D^\nu\vp^\gamma)
    \K{y_{(N\setminus L)\setminus I},x}\\
    &=T(L)(y_L)T(N\setminus L,\wpv)\K{y_{N\setminus L},x_1,x_2}.
\end{split}
\end{multline}
If $L,x_1\gtrsim N\setminus L, x_2, x$, with $x_1\gtrsim x_2$ we find:
\begin{multline}
  T(N,\wpv)(y_N,x_1,x_2)=\\
  \begin{split}
    &=T(L,\vp)(y_L,x_1)T(N\setminus L,\vp)\K{y_{N\setminus L},x_2}+\\
    &\quad-T(L)(y_L)E(\emptyset,\vp,\vp)(x,\xi)T(N\setminus L)
    \K{y_{N\setminus L}} +\\
    &\quad-\sum_{\substack{I\subset N\setminus L\\I\not=\emptyset}}
    \sum_{\gamma,\mu,\nu} E^{\mu\nu(\gamma)}(I,\vp,\vp)(y_I,x,\xi)
    T(N\setminus I,\D^\nu\vp^\gamma)\K{y_{N\setminus I},x}+\\
    &=T(L,\vp)(y_L,x_1)T(N\setminus L,\vp)\K{y_{N\setminus L},x_2}+\\
    &\quad-T(L)(y_L)\sum_{I\subset N\setminus L} \sum_{\gamma,\mu,\nu}
    E^{\mu\nu(\gamma)}(I,\vp,\vp)(y_I,x,\xi)\times\\ 
    &\qquad\times T((N\setminus L)\setminus I,\D^\nu\vp^\gamma)
    \K{y_{(N\setminus L)\setminus I},x}\\
    &=T(L,\vp)(y_L,x_1)T(N\setminus L,\vp)\K{y_{N\setminus L},x_2}
    -T(L)(y_L)\times\\
    &\quad\times\bigl(T(N\setminus L,\vp,\vp)\K{y_{N\setminus L},x_1,x_2}
    -T(N\setminus L,\wpv)\K{y_{N\setminus L},x_1,x_2}\bigr).
\end{split}
\end{multline}
A similar calculation also shows the right causality decomposition, if
$x$ is in the later set. 

Because of this causal factorization property and the inductive
assumption we immediately have
\begin{equation}
  \label{eq:zeroT}
  {^0T}(N,\wpv)(y_N,x,x)={^0T}(N,\vp^2)(y_N,x), 
\quad(y_N,x)\in\MM^{n+1}\setminus\Diag_{n+1}.
\end{equation}
We show that the same equation also holds for the
$T$-products. Because of \Ndreistrich\ it is sufficient to consider
vacuum expectation values only. Inserting the definition
\eqref{def:bilocalT} we find:
\begin{multline}
  \label{eq:vacdiff}
  \omega_0\K{T(N,\wpv)(y_N,x_1,x_2)}
  -\omega_0\K{{^0T}(N,\wpv)(y_N,x_1,x_2)}=\\
  =-\sum_{|\alpha|\leq\omega_N}\frac{(-)^{|\alpha|}}{\alpha!}
  \D^\alpha\delta(y_N-x)
  \bigl(e_N^{\alpha(0)}(\xi)-a_N^{\alpha(0)}(\xi)\bigr).
\end{multline}
Since the vacuum expectation values are translation invariant we use
the coordinates $z_i=y_i-x$ and $z=(z_1,\dots,z_n)$. We set 
\begin{align}
  t(z,\xi)&\doteq\omega_0\K{T(N,\wpv)(z,\xi,-\xi)},
  \label{def:tz}\\
  {^0t}(z,\xi)&\doteq\omega_0\K{{^0T}(N,\wpv)(z,\xi,-\xi)}.
\end{align}
It follows that
\begin{equation}
  \label{eq:abytz}
  e^{\alpha(0)}_N(\xi)=\int\dif z_N\,
  {^0t}(z,\xi)z^\alpha w_{N^{(0)}}(z)
\end{equation}
in this notation. Now, by smearing with $f\in\Dd(\MM^n)$ in the $z$
coordinates we find:
\begin{equation}
  \label{eq:tzf}
  \scp{t(\cdot,\xi)}{f}=\scp{{^0t}(\cdot,\xi)}{\W{\omega_N}{w_{N^{(0)}}}f}
  +\sum_{|\alpha|\leq\omega_N}
  \frac{a^{\alpha(0)}_N(\xi)}{\alpha!}\D^\alpha f(0). 
\end{equation}
Because of the sufficient subtraction on the test function we can put
$\xi=0$. Then we have
\begin{equation}
  \label{eq:tzfnull}
  t(z,0)=\omega_0\K{T(N,\vp^2)(z_N,0)},
\end{equation}
where the constants $a^{\alpha(0)}_N(0)$ can be chosen to produce any
normalization of the \rhs.

\subsection{Lorentz covariance}
\sloppy
Due to the definition the bilocal $T$-products are translation
covariant. Namely, they are products of translation invariant numerical
distributions and operator valued Wick products because of
\Ndreistrich. Therefore we have to consider Lorentz covariance for the
the numerical distributions only. 

\fussy
If ${^0t}$ transforms covariantly under the Lorentz group
\begin{equation}
  \Lambda{^0t}(z,\xi)
=D(\Lambda^{-1}){^0t}(z,\xi),
\end{equation}
w.r.t.\ both variables we find that $t$ transforms the same way, if%
\footnote{We suppress the indices $I$ and $\gamma$ which are fixed in
  this problem.}
\begin{equation}
\left({\Lambda^\alpha}_\beta D(\Lambda)\Lambda
-\delta^\alpha_\beta\right)e^\beta(\xi)
=\left({\Lambda^\alpha}_\beta D(\Lambda)\Lambda
-\delta^\alpha_\beta\right)a^\beta(\xi)
\label{eq:cocyc0}
\end{equation}
w.r.t.\ $\xi$. With $e^\beta$ from \eqref{def:aI} this leads to
\begin{equation}
\int\dif z\,{^0t}(z,\xi)z^\alpha(\Lambda w-w)(z)
=\left({\Lambda^\alpha}_\beta D(\Lambda)\Lambda
-\delta^\alpha_\beta\right)a^\beta(\xi).
\label{eq:cocyc}
\end{equation}
The \rhs\ is a one coboundary, which we have to solve for $a^\beta$.
We assume that there are solutions $\Dd'(\MM) \ni a^\beta\not =
e^\beta$ with the property that $a(0)$ exists. Unlike in the case of
usual $T$-products (cf. chapter~\ref{chap:timeordered},
section~\ref{sec:poincare}) we have no existence proof, so we have to
impose it as an assumption.  Moreover we see that $a^\beta(\xi)$ is
determined by the \rhs\ of \eqref{eq:cocyc} only up to terms
$h^\beta(\xi)$, with ${\Lambda^\alpha}_\beta D(\Lambda) \Lambda
h^\beta(\xi) = h^\alpha(\xi)$.

If the central solution ($w=1$) exists $a$ can be chosen as
\begin{equation}
a^\alpha(\xi) = \int\dif z\,{^0t} (z,\xi) z^\alpha
(w-1)(z),
\end{equation}
which fulfils all properties. It simply replaces the subtraction with
auxiliary function $w$ by the central subtraction.

\subsection{Proof of \Ndreistrich.}
We show that \Ndreistrich\ holds by evaluating both sides
independently. In the calculation we use the following formula,
taken from \cite{prep:duet-fred3}:
\begin{equation}
  \label{eq:michael}
  \frac{1}{\alpha!}\parz{(\D^\alpha V)}{\vp_r}
  =\sum_{\mu+\nu=\alpha}\frac{1}{\mu!\nu!}\D^\mu
  \left(\parz{V}{\vp}\right)\delta^\nu_r,
\end{equation}
where $V\in\Ba$ is supposed to contain no derivated fields. We first
investigate the contribution to the commutator of the \rhs\ of
\Ndreistrich\ arising from the sum in \eqref{def:bilocalT}. The
following term appears:
\begin{multline}
\frac{1}{\nu_1!\dots\nu_{|I|}!\gamma_1!\dots\gamma_{|I|}!}  
\left[
T\K{N\setminus I, 
\D^{\nu_1}\vp^{\gamma_1}\cdots\D^{\nu_{|I|}}\vp^{\gamma_{|I|}}}(y_N,x),
\vp(z)\right]=\\
\begin{split}
&=\sum_{k\in N\setminus I}\frac{1}{\nu!\gamma!}
T\K{(N\setminus I)^{(e_k)},\D^\nu\vp^\gamma}\K{y_{N\setminus I},x}
i\Delta(y_k-x)+\\
&+\quad\sum_{k\in I}\sum_{\sigma_k+\rho_k=\nu_k}
\frac{1}{\nu_1!\gamma_1!\dots\not k\dots\nu_{|I|}!\gamma_{|I|}!}
\frac{1}{\sigma_k!\rho_k!(\gamma_k-1)!}\times\\
&\quad\times
T\K{N\setminus I,  
\D^{\nu_1}\vp^{\gamma_1}\cdots
\D^{\sigma_k}\vp^{\gamma_k-1}\cdots\D^{\nu_{|I|}}\vp^{\gamma_{|I|}}}(y_N,x)
i\D^{\rho_k}\Delta(x-z).
\end{split}
\end{multline}
Without loss of generality we have put $I=\{1,\dots,|I|\}$. Denote the
$T$-product on the last line symbolically by $T^{\sigma_k}$. Note that
it only appears, if $|\gamma|>0$. With the equality
\begin{equation}
  \label{eq:Ddelta}
  \frac{(-)^{|\beta|}}{\beta!}\D^\beta\delta(y-x)f(y)
    =\sum_{\rho+\sigma=\beta}\frac{(-)^{|\rho|}}{\rho!\sigma!}
      \D^{\rho}\delta(y-x)\D^\sigma f(x),
\end{equation}
we have for every $k\in I$ (now $I$ can be any subset of $N$) the
following contribution:
\begin{multline}
  \sum_{\mu_k+\nu_k=\alpha_k}\frac{(-)^{|\mu_k|}}{\mu_k!}
  \D^{\mu_k}\delta(y_k-x)
  \sum_{\sigma_k+\rho_k=\nu_k}\frac{1}{\sigma_k!\rho_k!}
  T^{\sigma_k}\D^{\rho_k}\Delta(x-z)=\\
  \sum_{\mu_k+\nu_k=\alpha_k}\frac{(-)^{|\mu_k|}}{\mu_k!\nu_k!}
  \D^{\mu_k}\delta(y_k-x)T^{\nu_k}\Delta(y_k-z).
\end{multline}
Now we insert this result into the last term of \eqref{def:bilocalT},
commuted with $\vp(z)$:
\begin{multline}
  \left[\text{last term of }\eqref{def:bilocalT},\vp(z)\right]=\\
  \begin{split}
  &=\sum_{\substack{I\subset N\\ I\not=\emptyset}}
  \sum_{|\gamma|\leq \omega_I}\sum_{|\alpha|\leq\omega_I-|\gamma|}
  \sum_{\mu+\nu=\alpha}\frac{(-)^{|\mu|}}{\mu!\nu!\gamma!}
  \D^\mu\delta(I-x)\times\\
  &\quad\times
  \bigl(e^{\alpha(\gamma)}_I(\xi)-a^{\alpha(\gamma)}_I(\xi)\bigr)
  \sum_{k\in N\setminus I}
  T\K{N^{(e_k)}\setminus I, \D^\nu \vp^\gamma}\K{y_{N\setminus I},x}  
  i\Delta(y_k-x)+\\
  &\quad+\sum_{\substack{I\subset N\\ I\not=\emptyset}}
  \sum_{1\leq|\gamma|\leq \omega_I}\sum_{k\in I}
  \sum_{|\alpha|\leq\omega_I-|\gamma|}
  \sum_{\mu+\nu=\alpha}\frac{(-)^{|\mu|}}{\mu!\nu!(\gamma-e_k)!}
  \D^\mu\delta(I-x)\times\\
  &\quad\times
  \bigl(e^{\alpha(\gamma)}_I(\xi)-a^{\alpha(\gamma)}_I(\xi)\bigr)
  T\K{N\setminus I, \D^\nu \vp^{\gamma-e_k}}\K{y_{N\setminus I},x}  
  i\Delta(y_k-x).
\end{split}
\label{eq:ndreis3}
\end{multline}
If we shift the multi index $\gamma \rightarrow \gamma + e_k$ in the
second term and note that $e^{\alpha(\gamma+e_k)}_I =
e^{\alpha(\gamma)}_{I^{(e_k)}}$ by definition of $a^{\alpha(\gamma)}_I$,
\eqref{def:aI}, we have:
\begin{multline}
  \text{last term of }\eqref{eq:ndreis3}=\\
  \begin{split}
  &=\sum_{\substack{I\subset N\\ I\not=\emptyset}}
  \sum_{0\leq|\gamma|\leq \omega_I-1}\sum_{k\in I}
  \sum_{|\alpha|\leq\omega_I-|\gamma|-1}
  \sum_{\mu+\nu=\alpha}\frac{(-)^{|\mu|}}{\mu!\nu!\gamma!}
  \D^\mu\delta(I-x)\times\\
  &\quad\times
  \bigl(e^{\alpha(\gamma)}_{I^{(e_k)}}(\xi)
  -a^{\alpha(\gamma)}_{I^{(e_k)}}(\xi)\bigr)
  T\K{N\setminus I, \D^\nu \vp^{\gamma}}\K{y_{N\setminus I},x}  
  i\Delta(y_k-x).
\end{split}
\label{eq:ndreis4}
\end{multline}
Then we can sum up both terms and obtain:
\begin{multline}
  \left[\text{last term of }\eqref{def:bilocalT},\vp(z)\right]=\\
  \begin{split}
  &=\sum_{k=1}^n
  \sum_{\substack{I\subset N^{(e_k)}\\ I\not=\emptyset}}
  \sum_{|\gamma|\leq \omega_I}\sum_{|\alpha|\leq\omega_I-|\gamma|}
  \sum_{\mu+\nu=\alpha}\frac{(-)^{|\mu|}}{\mu!\nu!\gamma!}
  \D^\mu\delta(I-x)\times\\
  &\quad\times
  \bigl(e^{\alpha(\gamma)}_I(\xi)-a^{\alpha(\gamma)}_I(\xi)\bigr)
  T\K{N^{(e_k)}\setminus I, \D^\nu \vp^\gamma}\K{y_{N\setminus I},x}  
  i\Delta(y_k-x).
\end{split}
\label{eq:ndreis5}
\end{multline}
Note, that if $k\in I$ the singular order $\omega_I$ in
\eqref{eq:ndreis5} is automatically lowered by one compared to
\eqref{eq:ndreis4} since now $k \in N^{(e_k)}$. In \eqref{eq:ndreis4}
the singular order is measured with respect to the set $I \subset N$
without differentiation of the $k$'th symbol.

This shows the equality of the last term, if we insert
\eqref{def:bilocalT} into \Ndreistrich. The remaining two terms of
\eqref{def:bilocalT} on the \lhs\ of \Ndreistrich\ commute with $\vp$
according to
\begin{multline}
\left[
T(N,\vp,\vp)(y_N,x_1,x_2)
  -\omega_0\K{T(\vp,\vp)(x_1,x_2)}T(N)(y_N),\vp(z)\right]=\\
\begin{split}
&=\sum_{k=1}^n \Bigl[
T\K{N^{(e_k)},\vp,\vp}(y_N,x_1,x_2)+\\
&\quad-\omega_0\K{T(\vp,\vp)(x_1,x_2)}T\K{N^{(e_k)}}(y_N)
\Bigr]i\Delta(y_k-z)+\\
&\quad+T(N,\vp)(y_N,x_1)i\Delta(x_2-z)
+T(N,\vp)(y_N,x_2)i\Delta(x_1-z),
\end{split}
\end{multline}
such that they match the missing terms in the \rhs\ of
\Ndreistrich. This finishes the proof.

\section{The operator product expansion}
In order to derive the interacting normal product $\Kg{\wpv}$ we only
have to evaluate the $R$ products which are given in terms of
$T$-products according to \eqref{def:R}.
\begin{multline}
  R(N;\vp,\vp)(y_N,x_1,x_2)-R\K{N;\wpv}(y_N,x_1,x_2)=\\
  \begin{split}
  &=\sum_{I\subset N}(-)^{|I|}\Tbar(I)(y_I)
  \bigl[T(N\setminus I,\vp,\vp)\K{y_{N\setminus I},x_1,x_2}+\\
  &\quad-T(N\setminus I,\wpv)\K{y_{N\setminus I},x_1,x_2}\bigr]\\
  &=\sum_{I\sqcup J\sqcup K=N}\sum_{\mu,\nu,\gamma}
  E^{\mu\nu(\gamma)}(I,\vp,\vp)(y_I,x,\xi)
  (-)^{|J|}\Tbar(J)(y_J)T(K,\D^\nu\vp^\gamma)(y_K,x)\\
  &=\sum_{I\subset N}\sum_{\mu,\nu,\gamma}
  E^{\mu\nu(\gamma)}(I,\vp,\vp)(y_I,x,\xi)
  R\K{N\setminus I,\D^\nu\vp^\gamma}\K{y_{N\setminus I},x}.
\end{split}
\label{eq:bilocalR}
\end{multline}
If we now insert into the power series for the interacting fields
\eqref{eq:Wg} only the $\mu=0$ coefficients contribute, since
$g\restriction_\Ocal=\const$. We omit this index on the
$E^{\mu\nu(\gamma)}$-terms. The \rhs\ of \eqref{eq:bilocalR} is just
the $n$'th order contribution of the product of two power series. So
we find the expansion:
\begin{equation}
  \label{eq:ope1}
  T\Kg{\vp,\vp}(x_1,x_2)
  =\Kg{\wpv}(x_1,x_2)+\sum_{|\gamma|\leq 2}
  \sum_{\alpha\leq 2-|\gamma|}
  E^{\alpha(\gamma)}_{g\Ll}(\xi)\Kg{\D^\alpha\vp^\gamma}(x).
\end{equation}
This is the \emph{operator product expansion}. We used the fact, that
the maximal singular order $\omega_0\K{T(N,\vp^2)}$ is two in a
renormalizable field theory. If the interaction is a sum of Wick
monomials $g\cdot\Ll=\sum_{r=1}^s g_r \Ll_r$, the expansion
coefficients read:
\begin{multline}
  E^{\alpha(\gamma)}_{g\Ll}(\xi)=
  \sum_{n=0}^\infty\frac{i^n}{n!}\sum_{r_1,\dots,r_n=1}^s
  g_{r_1}(x)\dots g_{r_n}(x) \times \\
  \times\sum_{|\gamma|\leq\omega_{r_1,\dots,r_n}}
  \sum_{|\alpha|\leq\omega_{r_1,\dots,r_n}-|\gamma|}
  \biggl[-a^{\alpha(\gamma)}_{r_1,\dots,r_n}(\xi)+\\
  +\int\dif z_N\, z^\alpha
  \omega_0\K{{^0T}
  \K{\Ll_{r_1}^{(\gamma_1)},\dots,\Ll_{r_n}^{(\gamma_n)},\wpv}
  (z_N,\xi,-\xi)}w_{r_1\dots r_n}^{\gamma_1\dots\gamma_n}(z_N)
  \biggr].
\end{multline}
As was expected from the general theorem of the unitary equivalence of
the local algebras \cite{proc:brun-fred}, the expansion coefficients
depend only locally on $g$.

The operator product expansion \eqref{eq:ope1} has the general
form:
\begin{align}
  \label{eq:ope2}
  T\Kg{\vp,\vp}(x_1,x_2)
  &=\Kg{\wpv}(x_1,x_2)+E^{0(0)}_{g\Ll}(\xi)\eins_{g\Ll}
  +E^{0(1)}_{g\Ll}(\xi)\vpg(x)+\notag \\
  &\quad+E^{\mu(1)}_{g\Ll}(\xi)\D_\mu\vpg(x)
  +E^{0(2)}_{g\Ll}(\xi)\Kg{\vp^2}(x).
\end{align}
If we consider a pure $\vp^4$-coupling we have $E^{0(1)}_{g\Ll} =
E^{\mu(1)}_{g\Ll} = 0$, since the vacuum expectation value of an odd number
of fields is zero. In that case the coefficients read
\begin{align}
E^{0(0)}_{g\Ll}(\xi)
&=i\Delta_F(2\xi)+
\sum_{n=2}^\infty\frac{i^n}{n!}g(x)^n
\biggl[-a^{0(0)}_{N^{(0)}}(\xi)+ \notag\\
&\quad+\int\dif z_N\,
\omega_0\K{{^0T}\K{\Ll,\dots,\Ll,\wpv}\K{z_N,\xi,-\xi}}
w_{N^{(0)}}(z_N)\biggr] \\
E^{0(2)}_{g\Ll}(\xi)&=
\sum_{n=1}^\infty\frac{i^n}{n!}g(x)^n
\sum_{|\gamma|=2}\frac{1}{\gamma_1!\dots\gamma_n!}
\biggl[-a^0_{N^{(\gamma)}}(\xi)+\notag\\
&\quad+\int\dif z_N\,
\omega_0\K{{^0T}
\K{\Ll^{(\gamma_1)},\dots,\Ll^{(\gamma_n)},\wpv}(z_N,\xi,-\xi)}
w_{N^{(\gamma)}}(z_N)
\biggr].
\end{align}
A closer inspection of the two terms reveals that $E^{0(0)}_{g\Ll}$
contains the terms which appear in the mass and wave function
renormalization. If we put $w=1$ (which is allowed if $m>0$) and do a
resummation over one particle irreducible graphs, we would end up with
the usual geometric series found in the literature for the interacting
propagator (cf. \cite{bk:itzykson}). In that case all disconnected
graphs disappear. But this is only due to that special choice of 
normalization. The series $E^{0(2)}_{g\Ll}$ contain the contributions
to the renormalization of the coupling constant.

\section{Towards the definition of a state}
In this section we introduce the idea to define a state on the algebra
of local observables with the help of the \ope.  We remind the
definition of a state $\omega$ as a linear normed positive functional
in the free field theory. In that case there is also an \ope, namely
Wick's theorem (cf.  \eqref{def:wick1} -- \eqref{def:wick3}). The
vacuum state $\omega_0$ on the algebra of observables (of the free
field) was defined by $\omega_0(\wick{A})=0$ and $\omega_0(\eins)=1$.
Since Wick's theorem allows to expand any observable into a series
of Wick polynomials the state is uniquely defined.

In the interacting field theory the fields are (operator valued
distributional) formal power series in $g$. Denote by $\CC_g$ the
formal power series in $g$ with complex coefficients. Following
\cite{pap:duet-fred1} a state $\omega_{g\Ll}$ on $\Aag(\Ocal)$ is a
mapping:
\begin{align}
  \omega_{g\Ll}: \Aag(\Ocal) &\mapsto \CC_g, \\
  \omega_{g\Ll}\K{a_g A_{g\Ll} + B_{g\Ll}}
  &=a_g\omega_{g\Ll}(A_{g\Ll}) + \omega_{g\Ll}(B_{g\Ll}),
  \quad a_g\in\CC_g \\
  \omega_{g\Ll}\K{A_{g\Ll}^*}
  &=\overline{\omega_{g\Ll}\K{A_{g\Ll}}}\\
  \omega_{g\Ll}(\eins_{g\Ll})& = 1,\\
  \omega_{g\Ll}\K{A_{g\Ll}^*A_{g\Ll}}& \geq 0,
\end{align}
$A_{g\Ll},B_{g\Ll}\in\Aag(\Ocal)$.  Inspired by the definition of the
vacuum state for the free field algebra we set
\begin{align}
  \omega_{g\Ll}(\Kg{\wick{A}})&=0,\\
  \omega_{g\Ll}(\eins_{g\Ll})&=1.
\end{align}
Unfortunately we do not have an \ope\ for the general product of
interacting fields and time ordered products of them. But we can
already check, if the above criteria hold in our case. 

We consider pure $\vp^4$-interaction. Since $\Kg{\wpv}^*=\Kg{\wpv}$ on
$\Do$ the adjoint of \eqref{eq:ope2} is:
\begin{equation}
  \label{eq:opead}
  \Tbar\Kg{\vp,\vp}(x_1,x_2)
  =\Kg{\wpv}(x_1,x_2)+\overline{E^{0(0)}_{g\Ll}(\xi)}\eins_{g\Ll}
  +\overline{E^{0(2)}_{g\Ll}(\xi)}\Kg{\vp^2}(x).
\end{equation}
The singular order of $E^{0(0)}$ is 2 and of $E^{0(2)}$ is 0.
Therefore we can form
\begin{multline}
  \label{eq:opprod}
  \vpg(x_1)\vpg(x_2)=\\
  =\theta(x_1^0-x_2^0)T\Kg{\vp,\vp}(x_1,x_2)+
  \theta(x_2^0-x_1^0)\Tbar\Kg{\vp,\vp}(x_2,x_1).
\end{multline}
Then the interacting two point function is given by
\begin{equation}
  \label{def:inttowpt}
  \omega_{2g\Ll}(x_1,x_2)\doteq\omega_{g\Ll}(\vpg(x_1)\vpg(x_2))
  =\theta(\xi^0)E^{0(0)}_{g\Ll}(\xi)
  +\theta(-\xi^0)\overline{E^{0(0)}_{g\Ll}(\xi)}.
\end{equation}
For the notion of positivity we refer to the work
\cite{pap:duet-fred1}. A formal power series $\CC_g \ni b_g = \sum_n
b_n g^n$ is defined to be positive if it can be written as the square
of another power series $b_g = \overline{c_g} c_g$. This is equivalent
to the conditions: $b_n \in \RR, \forall n \in \NN_0$ and for the
first non vanishing $b_l$ it is required that $b_l > 0$ and $l$ is
even. So for $f \in \Dd(\MM)$ we have:
\begin{multline}
  \omega_{2,g\Ll}(\overline{f},f)=\\
  \begin{split}
    &=\int\dif x\,\dif y \left(\theta(x^0-y^0)E^{0(0)}_{g\Ll}(x-y)
      +\theta(y^0-x^0)\overline{E^{0(0)}_{g\Ll}(x-y)}\right)
    \overline{f(x)}f(y)\\
    &=2\int\dif x\,\dif y \text{ Re }
    \theta(x^0-y^0)E^{0(0)}_{g\Ll}(x-y) \overline{f(x)}f(y),
\end{split}
\end{multline}
where we exchanged $x \leftrightarrow y$ in the second integral and
used the fact that $E^{0(0)}_{g\Ll}$ is even. The first non vanishing
contribution comes from the free two point function which is positive,
cf.\ \eqref{eq:positivcheck1}. Under the above criterion our two point
function is positive.


\section{Remarks}
Our \ope\ is a consequence of the definition of our normal product of
two fields. The definition for other but scalar fields can be derived
straight forward by the applied methods. A generalization to composed
fields requires a modification of the terms arising by
$x_1,x_2$-contractions. Unfortunately, we have not succeeded in finding a
definition for higher order ($>$ bi) normal products.

On the other hand the bi-\ope\ already contains all contributions
which are necessary for the definition of a mass-, wavefunction- and
coupling normalization, which up to now always requires the adiabatic
limit \cite{bk:scharf}, \cite{pap:ep-gl}. For the first two of them
this should be possible by a suitable condition on the measure that
one can derive from the Jost-Lehmann-Dyson representation of our (time
ordered) two point function.


\chapter{Conclusion and Outlook}
\label{chap:conclusion}
We have provided for an explicit Poincar\'e covariant normalization in
the Epstein-Glaser approach to renormalization theory. Beside its
meaning as a closed loophole in the inductive construction it can
be useful for the practitioner. Especially for massless theories
where the central solution does not exist this is an advantage.

The local conservation of translation invariance by means of 
conservation of the \emt\ is likely to be generalized to an arbitrary
coupling containing also derivated fields by the same
method. Discussing these (and all other) symmetries locally by suitable
Ward identities moreover shows the advantage that massive and massless
fields can be treated on the same footing. 

An open problem still is a local definition of the $\beta$-function.
The validity of the equation ${{\Theta_{\mathrm{imp}\,g\Ll}}^\mu}_\mu
=2\beta\Llg$, as conjectured by Minkowski \cite{pap:minkowski}
and verified by Zimmermann's normal product quantization
\cite{proc:zimmermann} has to be examined in the local perturbative
approach. This might reveal how the free parameter which is still
present in the trace anomaly in $\vp^4$-theory as discussed by us has
to be chosen. A similar problem addresses the anomalous dimension. The
anomalous terms found by our procedure still lack the possibility to
derive the anomalous dimension of the interacting field as a real
parameter (in form of a formal power series).

The \ope\ has produced power series as algebraic structure
constants. These depend on the coupling only locally and may therefore
serve as the right objects for the discussion of wave function, mass and
coupling constant renormalization independent of the adiabatic
limit. 
\vfill
\hfill
\begin{minipage}{7cm}
\itshape \footnotesize
``\dots There is another theory which states that this has already
happened.''\\
  \normalfont \hfill -- Douglas Adams, "The Restaurant at the End of
  the Universe"
\end{minipage}
\label{pg:end}

\newpage
\begin{center}
{\large \textbf{Acknowledgements}}\\[\baselineskip]
\end{center}

First of all I thank my supervisor K.\ Fredenhagen for entrusting this
work to me. His guidance and support has contributed to the progress
of this thesis in many situations. He is a great teacher and I am
thankful for all I have learned in numerous discussions through the
course of the work.

Moreover I am very indebted to Michael D\"utsch. We spent a lot of time
at the blackboard and I profited from all of his explanations. 

My colleagues from the II. Institut have contributed to a
pleasant working atmosphere the whole time. Thanks to all of them.

I am very thankful to my parents Karin and
Rolf Prange whose support I could always rely on.

Last but not least I thank my wife Wiebke Kortum for her impayable
contribution to this work.

Financial support by the DFG Graduiertenkolleg ``Theoretische
Elementarteilchenphysik'' is gratefully acknowledged.


\appendix


\chapter{Representation of the symmetric groups}
\label{app:permrep}
Everything in this brief appendix should be found in any book about
representation theory of finite groups. We refer to
\cite{bk:simon,bk:boerner,bk:fulton}.  The group algebra
$\Acal_{S_p}$ consists of elements
\begin{align}
a&=\sum_{g\in S_p}\alpha(g)\cdot g,&
b&=\sum_{g\in S_p}\beta(g)\cdot g,
\end{align}
where $\alpha,\beta$ are arbitrary complex numbers. The sum of two 
elements
is naturally given by the summation in $\menge{C}$ and the product is
defined through the following convolution:
\begin{align}
ab&\doteq\sum_{g_1,g_2}\alpha(g_1)\beta(g_2)\cdot g_1 g_2
=\sum_g\gamma(g)\cdot g \text{ with} \\
\gamma(g)&\doteq\sum_{g_1 g_2=g}\alpha(g_1)\beta(g_2)
=\sum_{g_1}\alpha(g_1)\beta({g_1}^{-1}g)
=\sum_{g_2}\alpha(g{g_2}^{-1})\beta(g_2).
\end{align}
The group algebra is the direct sum of simple twosided ideals:
\begin{equation}
\Acal_{S_p}=I_1\oplus\dots\oplus I_k,
\end{equation}
and $k$ is the number of partitions of $p$. Every ideal $I_j$ contains
$f_j$ equivalent irreducible representations of $S_p$. $I_j$ is
generated by an idempotent $e_j\in\Acal_{S_p}$:
\begin{align}
I_j&=\Acal_{S_p}e_j & {e_j}^2&=e_j.
\end{align}
These idempotents satisfy the following orthogonality and completeness
relations:
\begin{align}
e_j e_i &= \delta_{ji} & \sum_{j=1}^k e_j &=\eins.
\end{align}
The center of $\Acal_{S_p}$ consists of all elements $\sum_j^k\alpha_j
e_j, \alpha_j\in\menge{C}$.

Every permutation of $S_p$ can be uniquely (modulo order) written as a
product of disjoint cycles. Since two cycles are conjugated if and
only if their length is the same, the number of conjugacy classes is
equal to the number of partitions of $p$. Denoting the $j$'th
conjugacy class by $c_j$ we build the sum of all elements of one class
\begin{equation}
k_j\doteq\sum_{\pi\in c_j}\pi \in \Acal_{S_p}
\end{equation}
which is obviously in the center of $\Acal_{S_p}$, too. So we can
expand $k_i$ in the basis $e_j$:
\begin{equation}
k_i=h_i\sum_{j=1}^k\frac{1}{f_j}\chi_j(c_i)e_j,
\label{eq:decompki}
\end{equation}
where $\chi_j(c_i)$ is the character of the class $c_i$ in the
representation generated by $e_j$ and $h_i$ is the number of elements
of $c_i$. The dimension of that representation is equal to the
multiplicity $f_j$.

The construction of the idempotents can be carried out via the 

\subsection*{Young tableaux}
A sequence of integers $(m)=(m_1,\dots,m_r), m_1\geq m_2\geq\dots\geq
m_r$ with $\sum_{j=1}^r=p$ gives a partition of $p$. To every such
sequence we associate a diagram with
\[
\begin{array}{cl}
m_1\text{ boxes}&\yng(3)\dots\yng(1) \\[-.5pt]
m_2\text{ boxes}&\yng(2)\dots \\
\vdots &
\end{array}
\]
called a \emph{Young frame} $(m)$. Let us take $p=5$ as an example:
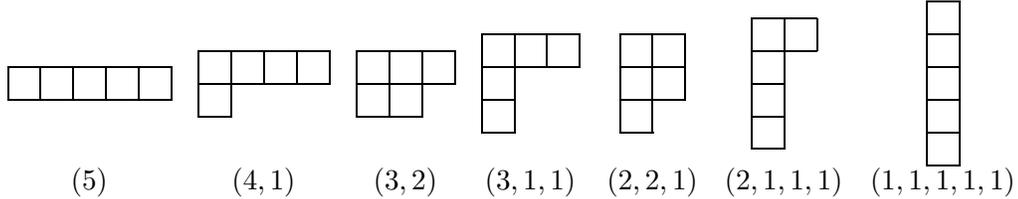
\[
\begin{array}{ccccccc}
\yng(5) & \yng(4,1) & \yng(3,2) & \yng(3,1,1) & \yng(2,2,1) &
\yng(2,1,1,1) & \yng(1,1,1,1,1) \\
(5) & (4,1) & (3,2) & (3,1,1) & (2,2,1) & (2,1,1,1) & (1,1,1,1,1) 
\end{array}
\]
An assignment of numbers $1,\dots,p$ into the boxes of a frame is
called a \emph{Young tableau}. Given a tableau $T$, we denote 
$(m)$ by $(m)(T)$. If the numbers in every row and in every column
increase the tableau is called \emph{standard}. The number of standard
tableaux for the frame $(m)$ is denoted by $f_{(m)}$. It is equal to the
dimension of the irreducible representation generated by the
idempotent $e_{(m)}$. We now answer the question 

\subsubsection*{How to construct $e_{(m)}$}
Set
\begin{align*}
\Rcal(T)&\doteq
\{\pi\in S_p|\pi\text{ leaves each row of $T$ set wise fixed}\}, \\
\Ccal(T)&\doteq
\{\pi\in S_p|\pi\text{ leaves each column of $T$ set wise fixed}\},
\end{align*}
and build the following objects:
\begin{align}
P(T)&\doteq\sum_{p\in\Rcal(T)}p, & \label{def:PQ}
Q(T)&\doteq\sum_{q\in\Rcal(T)}\sgn(q) q,
\end{align}
then
\begin{equation}
e(T)\doteq\frac{f_{(m)}}{p!}P(T)Q(T) \label{def:eT}
\end{equation}
is a minimal projection in $\Acal_{S_p}$ (generates a minimal left
ideal). The central projection (generating the simple two sided ideal)
is given by
\begin{equation}
e_{(m)}\doteq\frac{f_{(m)}}{p!}\sum_{T|(m)(T)=(m)} e(T).
\label{def:eF}
\end{equation}

\subsection*{Example $p=3$}
The frame {\scriptsize\yng(3)} has only one standard tableau
{\scriptsize\young(123)}. All different tableaux in (\ref{def:eF})
lead to the same idempotent (\ref{def:eT}) which is just the sum of
all permutations.
\[
e_{(3)}=\frac{1}{6}
\bigl(\eins+(1\,2)+(1\,3)+(2\,3)+(1\,2\,3)+(1\,3\,2)\bigr).
\]
For the frame {\scriptsize\yng(1,1,1)} we only need the column permutations in
(\ref{def:eT}). We find 
\[
e_{(1,1,1)}=\frac{1}{6}
\bigl(\eins-(1\,2)-(1\,3)-(2\,3)+(1\,2\,3)+(1\,3\,2)\bigr).
\]
The frame {\scriptsize\yng(2,1)} has two standard tableaux. For the
tableaux {\scriptsize
\young(12,3),
\young(13,2),
\young(21,3),
\young(23,1),
\young(31,2),
\young(32,1)} we find:
\begin{align*}
e_{(2,1)}&=\frac{2^2}{(3!)^2}
\bigl\{
(\eins+(1\,2))(\eins-(1\,3))
+(\eins+(1\,3))(\eins-(1\,2))
+(\eins+(1\,2))(\eins-(2\,3))+ \\
&\qquad+(\eins+(2\,3))(\eins-(1\,2))
+(\eins+(1\,3))(\eins-(2\,3))
+(\eins+(2\,3))(\eins-(1\,3))
\bigr\} \\
&=\frac{1}{3}
\bigl\{2\eins
-(1\,2\,3)-(1\,3\,2)
\bigr\}.
\end{align*}
Up to order $p=4$ the central idempotents are given by the sum of
minimal projectors of the standard tableaux -- they are orthogonal. 

The characters in the irreducible $(m)$ representation can be computed through
\begin{equation*}
\chi_{(m)}(s)
=\frac{f_{(m)}}{p!}\sum_{T|(m)(T)=(m)}
\sum_{\substack{p\in\Rcal(T)\\ q\in\Ccal(T) \\pq=s}}\sgn(q).
\end{equation*}
Many other useful formulas can be derived from the Frobenius character
formula.  Interchanging rows and columns in a frame $(m)$ leads us to
the \emph{dual frame} $\widetilde{(m)}$. For the characters one finds:
$\chi_{\widetilde{(m)}}(s)=\sgn(s)\chi_{(m)}(s)$. There is a nice
formula for the characters of the transpositions in \cite{bk:fulton}:

Define the rank $r$ of a frame to be the length of the diagonal. Let $a_i$
and $b_i$ be number of boxes below and to the right of the $i$'th box, reading
from lower right to upper left. Call 
$\begin{pmatrix}a_1&\dots&a_r\\b_1&\dots&b_r\end{pmatrix}$ the
characteristics of $(m)$, e.g.
\[
\young(X\hfil\hfil\hfil\hfil\hfil,%
       \hfil X\hfil\hfil,%
       \hfil\hfil X,%
       \hfil\hfil,%
       \hfil\hfil)
\quad r=3,
\text{ characteristics} =
\begin{pmatrix} 0&3&4 \\ 0&2&5 \end{pmatrix}.
\]
Then
\[
\chi_{(m)}(\tau)=
\frac{f_{(m)}}{p(p+1)}\sum_{i=1}^r(b_i(b_i+1)-a_i(a_i+1)).
\]



\chapter{Notations and Abbreviations}
\label{app:convention}
\begin{tabular}{lp{8cm}}
$\eta_{\mu\nu}
=
\begin{pmatrix}
  1 & 0 & 0 & 0\\
  0 & -1 & 0 & 0\\
  0 & 0 & -1 & 0\\
  0 & 0 & 0 & -1\\
\end{pmatrix}$
& Minkowski metric\\
& \\
$\epsilon_{AB}=
\begin{pmatrix}
  0 & 1 \\
  -1 & 0
\end{pmatrix}$
& Spinor metric\\
&\\
$\NN,\RR,\CC$ & Numbers: positive integer, real, complex\\
$\MM$ Minkowski space\\
$\Dd,\Sd,\Cunend$ & Test function spaces: compact support, rapid
decrease, infinitely differentiable\\
$\Aa,\Ba,\Ga,\Ga_g$ & Algebras: free quantized field, Boas algebra of
symbols, sub algebras of generators and basic generators.\\
$\Ho,\Do,\Fo$ & Hilbert spaces: Hilbert space, dense domain of
definition, Fock space\\
$\Ll,\Kl$ & Lagrangian\\
$\widehat{f}(p)=\int\dif x\, f(x) e^{ipx}$
& Fourier transformation in $\RR^n$ \\
$f(x)=\frac{1}{(2\pi)^n}\int\dif x\, \widehat{f}(x) e^{-ipx}$ \\
$\Delta,\Delta^F$ & Commutator function, Feynman propagator \\
$D,D^F$ & Massless commutator function, massless Feynman propagator \\
$\phicl$ & Classical field $\in\Dd(\MM)$\\
$\vp_j$ & Symbol $\in\Ga\subset\Ba$\\
$T(N)(x_N)$ & Time ordered product \\
$R(M;N)(y_M;x_N)$ & Retarded product \\
$N=\{W_1,\dots,W_n\}$ & Set of symbols $W_i\in\Ba$\\
$x_N=(x_1,\dots,x_n)$ & $n$-fold coordinate vector $\in\MM^n$\\
$\Wg$ & Interacting field with local interaction $g\Ll, g\in\Dd(\MM),
\Ll \in \Ba$, in case of more couplings $g\Ll = \sum_i g_i\Ll_i$.
\end{tabular}


\bibliographystyle{amsalpha}
\bibliography{literatur}   


\end{fmffile}
\end{document}